\documentclass[a4paper,11pt]{article}
\pdfoutput=1 

\usepackage{jheppub} 
\usepackage{amsmath,amssymb}
\usepackage{graphicx}
\usepackage{subcaption}
\usepackage[percent]{overpic}
\usepackage[dvipsname]{xcolor}
\usepackage{multirow}
\usepackage{bbm}
\usepackage{enumitem}

\definecolor{purple}{rgb}{0.7,0,0.7}

\newtheorem{comment}{Comment}

\def\be{\begin{equation}}
\def\ee{\end{equation}}

\newcommand{\CD}{\mathcal{D}}
\newcommand{\CS}{\mathcal{S}}
\newcommand{\CO}{\mathcal{O}}
\newcommand{\CK}{\mathcal{K}}

\newcommand{\CH}{\mathcal{H}}
\newcommand{\CL}{\mathcal{L}}
\newcommand{\CM}{\mathcal{M}}
\newcommand{\CN}{\mathcal{N}}

\newcommand{\eul}{{\mathbf{e}}}

\newcommand{\IC}{\mathbb{C}}
\newcommand{\IP}{\mathbb{P}}
\newcommand{\IZ}{\mathbb{Z}}

\newcommand{\IR}{\mathbb{R}}

\newcommand{\Tr}{\mathrm{Tr}}

\newcommand{\fM}{{\mathfrak{M}}}
\newcommand{\fD}{{\mathfrak{D}}}
\newcommand{\fL}{{\Lambda}}

\renewcommand{\(}{\left(}
\renewcommand{\)}{\right)}

\title{%
A-branes, foliations and localization
}

\author[a,b]{Sibasish Banerjee}
\author[c]{Pietro Longhi}
\author[d]{Mauricio Romo}

\affiliation[a]{Weyertal 86-90, Department of Mathematics, University of Cologne, 50679, Cologne, Germany }
\affiliation[b]{Max Planck Institute for Mathematics, Vivatsgasse 7, 53111 Bonn, Germany}

\affiliation[c]{Institute for Theoretical Physics, ETH Zurich, 8093, Zurich, Switzerland}

\affiliation[d]{Yau Mathematical Sciences Center, Tsinghua University, Beijing, 100084, China}

\emailAdd{
sbanerje@mpim-bonn.mpg.de,
longhip@phys.ethz.ch, 
mromoj@tsinghua.edu.cn
}

\abstract{
This paper studies a notion of enumerative invariants for stable $A$-branes, and discusses its relation to invariants defined by spectral and exponential networks.
A natural definition of stable $A$-branes and their counts is provided by the string theoretic origin of the topological $A$-model.
This is the Witten index of the supersymmetric quantum mechanics of a single $D3$ 
brane supported on a special Lagrangian in a Calabi-Yau threefold.
Geometrically, this is closely related to the Euler characteristic of the $A$-brane moduli space.
Using the natural torus action on this moduli space, we reduce the computation of its Euler characteristic to 
a count of fixed points via equivariant localization.
Studying the $A$-branes that correspond to fixed points, we make contact with definitions of spectral and exponential networks. 
We find agreement between the counts defined via the Witten index, and the BPS invariants defined by networks. By extension, our definition also matches with Donaldson-Thomas invariants of $B$-branes related by homological mirror symmetry.\\[5pt]
\today
}

\begin{document} 

\maketitle
\flushbottom

\section{Introduction}

A recurring theme in research at the interface of physics and mathematics is the subject of BPS states.
In physics BPS states are associated with symmetry-protected sectors of a gauge or string theory, whereas in mathematics they arise in various guises in the domains of geometry, algebra, low-dimensional topology, and beyond.

In this work we focus on a class of BPS states modeled by Lagrangian $A$-branes in a class of Calabi-Yau threefolds. 
Our main goal is to define a notion of `counting' stable $A$-branes that is motivated by physics, and that is meaningful from the viewpoint of mathematics.

Let $X$ be a hypersurface in $\IC^2\times (\IC^*)^2$ defined by $uv = F(x,y)$ for a certain Laurent polynomial $F(x,y)$, and let $\Omega^{3,0}$ denote a normalized holomorphic three-form on $X$.
At zero string coupling, an $A$-brane is characterized by a choice of special Lagrangian $L\subset X$ calibrated by $\Omega^{3,0}$, together with a choice of flat abelian local system $\CL\to L$.
In this work we restrict attention to cases where $L$ is a primitive cycle in $H_3(X,\IZ)$.
Each of these geometric data has a moduli space: let $\fM$ be the moduli space of $L$, and let $T^{b_1(L)}$ be the moduli space of the local system. The $A$-brane moduli space $\CM$ is fibered by $T^{b_1(L)}$ over $\fM$, with fibers degenerating at points $L\in \fM$ where cycles in $H_1(L,\IZ)$ pinch. Let $\fD\subset \fM$ denote the locus where a maximal collection of cycles pinches, namely where the whole torus fiber shrinks to a point. Since $\dim_\IR\fM = b_1(L)$ by a theorem of \cite{mclean1998deformations}, it follows that $\fD$ is a finite collection of points.
Our definition for counting $A$-branes is given by the following formula\footnote{Here and throughout the paper it is understood that invariants are always defined on homology classes of Lagrangians cycles $[L]$, although we slightly abuse notation and omit the brackets. }
\be\label{eq:Omega-formula-intro}
	\Omega(L) = (-1)^{b_1(L)} |\fD| \,.
\ee

The proposal (\ref{eq:Omega-formula-intro}) is motivated by physical reasoning.
If we consider type IIB string theory on $X\times \IR^4$, then $A$-branes descend from $D3$ branes wrapping $L\times \IR$.
$L$ is assumed to be compact, and we consider the dimensional reduction of the $D3$ worldvolume theory to a 1d $\CN=4$ quantum mechanics on $\IR$. We argue that this theory factorizes into a free $U(1)$ gauge theory and a nonlinear sigma model.
The $U(1)$ degrees of freedom are identified with transverse motion of the particle in $\IR^4$, while the target of the nonlinear sigma model is a K\"ahler manifold $\CM$ corresponding to internal moduli of the $A$-brane.
Neglecting translational degrees of freedom, the Hilbert space of supersymmetric vacua for the nonlinear sigma model is given by a suitable notion of cohomology $\CH^{BPS}_L \simeq H^*(\CM)$, details of which are discussed in the main text.
The Witten index provides an invariant graded count of supersymmetric vacua, and corresponds to the Euler characteristic $\chi(\CM)$ up to a sign.

The fact that $\CM$ admits a Lagrangian torus fibration, implies that there is a natural torus action $T^{b_1(L)}$ rotating the fibers.
Equivariant localization with respect to this action allows to express the Euler characteristic as a sum over fixed points, counted without signs 
\be
	\chi(\CM) = \sum_{F} 1 = \text{(\# of fixed points)} \,.
\ee
Fixed points of the torus action coincide with degenerate loci $\fD$ in the moduli space $\fM$ of the underlying special Lagrangian. This leads to our formula (\ref{eq:Omega-formula-intro}) up to an overall shift by a sign.

We next consider how this definition of counting $A$-branes compares with known enumerative invariants in related contexts.
One reason for restricting to the class of Calabi-Yau hypersurfaces considered here, is that they admit a class of special Lagrangians fibered by two-spheres over paths $x(t)\subset \IC^*$ \cite{Klemm:1996bj}.
Calibration of $L$ translates into calibrations of the path $x(t)$ by a suitable abelian differential.
This enables to model $\fM$ by the space of foliations on $\IC^*$ defined by the differential, or a certain generalization that we introduce, which involves multiple foliations interacting with each other.
For $S^2$-fibered special Lagrangians of this sort, we develop a description of $\fM$ in the language of foliations.
We then identify the locus $\fD$ of degenerate Lagrangians with a certain class of leaves known as `critical leaves'.
Through this observation we make contact with work of \cite{Gaiotto:2009hg, Gaiotto:2012rg, Eager:2016yxd, Banerjee:2018syt} on spectral networks and exponential networks.
We argue that counts of BPS states performed via networks techniques correspond to counting fixed points $\fD$ of the torus action on the moduli spaces $\CM$ of $A$-branes. This argument is only expected to hold for primitive cycles, while a more involved relation to networks is foreseen in the non-primitive case.

The connection with spectral and exponential networks indicates that (\ref{eq:Omega-formula-intro}) should reproduce the `BPS index' (second helicity supertrace) of 4d $\CN=2$ gauge theory engineered by Type IIB on $X$.\footnote{In the case of exponential networks, this would be a Kaluza-Klein 4d $\CN=2$ theory in the sense discussed in \cite{Banerjee:2019apt, Closset:2019juk}.}
When $X$ is the mirror of a toric Calabi-Yau threefold our definition then coincides with
the generalized (rank-zero) Donaldson-Thomas invariants of $B$-branes in the mirror, as computed via exponential networks in \cite{Banerjee:2018syt, Banerjee:2019apt, Banerjee:2020moh}.

This observation fits naturally in the mathematical framework of homological mirror symmetry.
From this viewpoint the category of $B$-branes is described by the bounded derived category of coherent sheaves on the mirror toric Calabi-Yau.
This is a triangulated category which admits a notion of Bridgeland stability condition \cite{Douglas:2002fj, Douglas:2000gi, bridgeland2007stability}.
Generalizations of Donaldson-Thomas theory formulated in \cite{Joyce:2008pc, Kontsevich:2008fj} define numerical invariants counting stable objects in this category.
Homological mirror symmetry establishes an equivalence with the Fukaya-Seidel category of $X$, whose objects are precisely the $A$-branes considered here.
At present there does not seem to be a definition of stability conditions for Fukaya-Seidel categories formulated directly in terms of geometric data on the Calabi-Yau \cite{Thomas:2001ve, Thomas:2001vf, 2011arXiv1103.5010B, 2011arXiv1106.3430B,  2014arXiv1411.2772K, 2014arXiv1401.4949J, 2018arXiv180108286B, 2021arXiv211213623H}. However, triangulated categories always admit a notion of Bridgeland stability.
Similarly, there seems no existing definition of enumerative invariants for $A$-branes defined directly in terms of geometric data, but once again the high-level constructions of \cite{Joyce:2008pc, Kontsevich:2008fj} apply to the triangulated structure on a Fukaya-Seidel category, motivating the possibility that a `down to earth' definition of enumerative invariants may exist.
Our findings suggest that (\ref{eq:Omega-formula-intro}) would be a viable candidate, at least if restricted to primitive cycles.

\medskip

In conclusion, we hope this paper may fulfil two purposes.
First, we show that a natural definition of enumerative invariants of $A$-branes, the Euler characteristic of their moduli space of a special  Lagrangian with a flat abelian local system, is actually well-motivated and natural from a physics perspective. 
Although our definition readily applies to the case of primitive cycles, the evidence provided in support of it may serve as motivation to search for a broader, and more rigorous, definition.
One quality of this definition that we stress, is its close connection to classical geometric data, as opposed to the abstraction of categories.
A not too far-fetched parallel may be the contrast between categorical definitions of (generalized) Donaldson-Thomas invariants \cite{Joyce:2008pc, Kontsevich:2008fj}, and the original definition of \cite{Donaldson:1996kp} based entirely on geometric data.\footnote{In this vein, the passage from primitive to non-primitive cycles may involve the introduction of weighted sums of Euler characteristics, as argued in \cite{2005math......7523B} for the original Donaldson-Thomas invariants.}

The second purpose of this work is to show how the definition we propose is amenable to computations via localization, and how this leads naturally to the framework of spectral and exponential networks.
In particular we hope the simplicity of our definition may offer a useful reinterpretation of how networks capture Donaldson-Thomas type invariants.\footnote{Discussions of stability conditions and categorical Donaldson-Thomas invariants related to $A_1$ networks can be found in \cite{2013arXiv1302.7030B, 2014arXiv1409.8611H, 2021arXiv210406018H}. Analogous structures for higher-rank networks are studied e.g. in \cite{2016arXiv160705228G, Smith:2020fdf, 2021arXiv211213623H}.}
Again we stress that our definition only applies to primitive cycles, and in this sense our results only reproduce a `linearization' of the $\CK$-wall formula of \cite{Gaiotto:2012rg}. This is, in our opinion, a sign that we only scratched the surface, and that understanding higher orders of the networks $\CK$-wall formula from an enumerative-geometric point of view, may uncover yet more beautiful secrets.

\subsection*{Organization}
In the attempt to be reasonably self cointained, we start in Section \ref{sec:A-branes-stability} with a review of notions of stability  for $A$-branes that arise by embedding the topological $A$-model into string theory, and related definitions of BPS counting.
In Section \ref{eq:eunmerative-inv-A-branes} we motivate and present our proposal for enumerative invariants of $A$-branes.
Section \ref{sec:sLag-moduli-foliations} contains a discussion of the structure of moduli spaces of $A$-branes in terms of foliations and a certain generalization thereof.
These structures are illustrated with examples in Section \ref{sec:some-moduli-spaces}, which also includes a discussion of Lagrangian $A$-branes that are not generically described by foliations such as SYZ fibers.
In Section \ref{sec:localization} we make contact with equivariant localization, explaining how it leads to counting critical leaves of foliations.
Section \ref{sec:relation-to-networks} contains a discussion of how the count of critical leaves compares with counting of BPS states defined by spectral and exponential networks.

\subsection*{Acknowledgements}

We are grateful to Richard Eager, Greg Moore, Andy Neitzke and Johannes Walcher for correspondence and comments on a draft of this paper.
Part of the work by SB was supported by the Alexander von Humboldt foundation for researchers. 
MR acknowledges support from the National Key Research and Development Program of China, grant No. 2020YFA0713000, and the Research Fund for International Young Scientists, NSFC grant No. 11950410500.
The work of PL was supported by NCCR SwissMAP, funded by the Swiss National Science Foundation.

\section{Lagrangian $A$-branes and stability}\label{sec:A-branes-stability}

This section is  a review of known facts about Lagrangian $A$-branes. There are many excellent (and richer) expositions in the literature, we follow \cite{Aspinwall:2004jr, Aspinwall:2009isa}.
Our exposition will be aimed at presenting two existing notions of `counting' stable $A$-branes. 
The first one arises in physics, in the setting of geometric engineering and 4d $\CN=2$ field theories. 
The second one arises in mathematics through the setting of homological mirror symmetry and related constructions of stability conditions and generalized Donaldson-Thomas theory.

The existence (and agreement) of these definitions serves as motivation for this work. In later sections we propose a third and alternative way of defining ways to count $A$-branes, and argue that it agrees with the two well-known definitions reviewed here, under certain assumptions.

\subsection{$A$-branes and BPS branes}

The notion of $A$-branes arises in the context of the topological twists of 2d $(2,2)$ worldsheet superconformal field theories \cite{Witten:1988xj, Witten:1991zz}.
Unless otherwise stated we will assume vanishing string coupling in what follows.
As a consequence we will think of branes as modeled by classical geometric objects, namely submanifolds of the ambient space, endowed with certain vector bundles.\footnote{We restrict attention to Lagrangian $A$-branes. There are however other types of $A$-branes, see \cite{Kapustin:2001ij}.}
In order to preserve half of the worldhseet supersymmetry, it can be shown that $A$-branes must be supported on Lagrangian submanifolds and must carry an Abelian flat connection. In addition the Lagrangian needs to have vanishing Maslov class and 
must satisfy the tadpole vanishing condition.\footnote{
With a choice of holomorphic structure on the Calabi-Yau, and the holomorphic top form denoted by $\Omega^{3,0}$, a quick definition of Maslov class is as follows. Let $\zeta(p)$ be the phase of $\Omega^{3,0}$ at $p\in L$. Then consider a loop $c\in \pi_1(L)$, and follow $\zeta(p)$ along this loop. The Maslov index is $(\zeta(c(1))-\zeta(c(0)))/2\pi$. The vanishing of the Maslov class requires that the index vanishes for any loop in $\pi_1(L)$.
In a nutshell, this arises by the requirement of anomaly cancellation for the ghost number for the BRST complex of the twisted theory, see e.g. \cite[Chapter 40]{hori2003mirror}.
The tadpole condition asserts that all open worldhseet instanton contributions to tadpole expectation values should vanish, and ensures that the brane is stable when wrapped on the Lagrangian. 
}
In fact, from the viewpoint of the full (untwisted, type II) string theory, preserving half of the spacetime supersymmetry requires that the Lagrangian supporting an $A$-brane (rather its uplift to the full theory) should be a \emph{special Lagrangian} \cite{Becker:1995kb}. 
This means that the phase of the holomorphic top form restricted to $L$ must be constant\footnote{This implies the vanishing of the Maslov class as a consequence.} 
\be\label{eq:sLag-condition-review}
	\zeta^{-1} \iota^*\Omega^{3,0}(TL|_p) \in \IR_{>0}\,.
\ee
where $\iota:L\to X$ is the immersion of $L$ in the Calabi-Yau $X$, $\zeta\in \IC^*$ is a constant, and $p\in L$ is any point.
A special Lagrangian $L$ is a calibrated submanifold \cite{10.1007/BF02392726}, and therefore minimizes volume in the respective homology class $[L]\in H_3(X,\IZ)$. In this sense (\ref{eq:sLag-condition-review}), is a natural condition for BPS branes: for given brane tension, the special Lagrangian cycle minimizes the mass, and therefore provides a brane that is stable against decay into a lighter objects $L'$ with the same charge $[L]$.

Clearly the existence of a special Lagrangian will depend on the choice of complex structure through $\Omega^{3,0}$. In fact, for a given complex structure there may be a whole family $\fM$ of special Lagrangians in the same class $[L]$. 
Given a member of this family $L\in \fM$, it is known that it admits $b_1(L)$ integrable deformations \cite{mclean1998deformations}, implying that $\dim_\IR \fM=b_1(L)$.
Moreover, given a smooth $L\in \fM$, the flat Abelian connection carried by the $A$-brane is characterized by $b_1(L)$ holonomies. So the moduli space of Lagrangian $A$-branes $\CM$ in class $[L]$ has the form of a Lagrangian torus fibration 
\be\label{eq:CM-fibration-review}
	T^{b_1} \rightarrow \CM \rightarrow \fM \,.
\ee
Varying the complex structure of $X$ changes $\Omega^{3,0}$, and therefore the types of solutions to (\ref{eq:sLag-condition-review}). This may deform of $\fM$, possibly including changes of topology and altogether disappearance. When this happens, the spectrum of $A$-branes jumps, see \cite{Berkooz:1996km, Sen:1998sm, Joyce:2003yj, Taylor:2003gn} for discussions of the physics and geometry behind $A$-brane decay.

\subsection{A first look at BPS counting for $A$-branes: geometric engineering}\label{sec:geometric-engineering}

So far we have assumed vanishing string coupling, and the picture of $A$-branes has been firmly grounded into classical geometry. As soon as the coupling is turned on, this classical picture is replaced by a quantum one \cite{Denef:2002ru}, and the moduli space $\CM$ gets replaced by a Hilbert space. 
later we will return to a more detailed description of the relevant Hilbert spaces, while for the moment we note that 
their dimension provides a notion of the `number' of BPS states.
A more appropriate definition of counts of BPS states would actually involve an index that is invariant under small deformations of $\fM$ and only feels jumps in its topology.
Discussing and computing an appropriate notion of such an index will be one of the central points of this paper.

For now we observe that a closely related way of counting BPS states already appeared in physics.
Stable Lagrangian $A$-branes correspond to BPS branes of the full string theory on $X$.
In the geometric engineering limit \cite{Katz:1996fh}
the full string theory is described by an effective 4d $\CN=2$ theory on the directions transverse to $X$.
BPS $D3$ branes wrapping a special Lagrangian $L$ in $X$ then descend to BPS particles in 4d.
In the Seiberg-Witten description of 4d BPS states, the central charge of BPS particles is computed by the period of a meromorphic one-form $\lambda_{SW}$ on a cycle of a Riemann surface $\Sigma$ \cite{Seiberg:1994rs}. From the viewpoint of geometric engineering the central charge coincides with the period of the holomorphic top form on $L$ \cite{Klemm:1996bj}
\be
	Z_\gamma = \frac{1}{\pi} \oint_\gamma \lambda_{SW} =\frac{1}{\pi} \int_L\iota^*\Omega^{3,0}\,.
\ee
From the viewpoint of the 4d theory, it is the central charge that determines whether a BPS particle is stable or not.
In recent years, much progress has been made on understanding BPS states of 4d $\CN=2$ theories and their wall-crossing from several different angles \cite{Kontsevich:2008fj, Joyce:2008pc, Gaiotto:2009hg, Alim:2011ae, Manschot:2010qz}.

In particular, in the context of 4d $\CN=2$ theories there is a well-defined index that expresses invariant counts of BPS states, up to wall-crossing.
This is known as the \emph{BPS index}, or second helicity supertrace\footnote{See the appendices of \cite{Kiritsis:1997gu} for a definition and discussion of properties of helicity supertraces.}
\be
	\Omega(\gamma)\in \IZ \,.
\ee

\subsection{Another look at BPS counting: homological mirror symmetry}

As reviewed above, under certain conditions such as vanishing string coupling, $A$-branes have a well-defined geometric interpretation as special Lagrangian submanifolds with a flat local abelian system.
These objects come in families parameterized by a moduli space $\CM$ as described in (\ref{eq:CM-fibration-review}).
In physics there is a notion of counting $A$-branes that comes from counting BPS states reviewed in section \ref{sec:geometric-engineering}. It is then natural to ask whether this count is related in any way to enumerative invariants associated to $\CM$.
Even better, it would be desirable to categorify the count of BPS states, by finding a definition of vector spaces isomorphic to the Hilbert spaces of BPS states of 4d $\CN=2$ theories.
Ideally, the definition of these vector spaces (or the related enumerative invariants) would only rely on geometric properties of $A$-branes without any references to physics.

A natural setting where enumerative geometry of $A$-branes may be formulated is the Fukaya-Seidel category, see \cite{hori2003mirror, seidel2008fukaya, Aspinwall:2009isa, fukaya2009lagrangian, fukaya2010lagrangian, Auroux:2013mpa} for a sample of reviews.
We will not need to delve into definitions here, except for mentioning that $A$-branes correspond to objects, morphisms are generated by Floer complexes associated to pairs of $A$-branes, and the composition of morphisms is governed by an $A_\infty$ structure. Although the subject of Fukaya-Seidel categories is an active area of research, there is in fact a large volume of literature devoted to its study. 
It may then be not too hopeless to ask whether an enumerative theory tailored to $A$-branes has been developed.
At present, it seems that no such framework has been formulated in definitive form yet, although interesting work in this direction has appeared in several places \emph{e.g.} \cite{lau2018quantum, Thomas:2001ve, Thomas:2001vf, 2014arXiv1401.4949J, 2021arXiv211213623H}.

A less direct approach to enumerative invariants for $A$-branes goes through homological mirror symmetry \cite{Kontsevich:1994dn}.
In this setting the Fukaya-Seidel category of $A$-branes on a Calabi-Yau threefold $X$ is related to the bounded derived category of coherent sheaves on the mirror Calabi-Yau $X^\vee$.
This relation is relevant for our purpose, as the latter has been known for some time to admit both a notion of stability \cite{2000math......9209D, Douglas:2000gi, Aspinwall:2001dz, Douglas:2002fj, bridgeland2007stability}, and a notion of enumerative invariants `counting' stable objects \cite{Donaldson:1996kp, Joyce:2008pc, Kontsevich:2008fj}.
Both of these constructions may be interpreted, through homological mirror symmetry, as stability conditions for $A$-branes and as enumerative invariants to count stables ones.
For the purpose of this work, this line of reasoning provides strong motivation to expect that a well-defined notion of `counting' $A$-branes should exist. We will later attempt to provide a basic definition of these invariants without invoking mirror symmetry or the relation to $B$-branes.

With a view towards our definition, we mention here one particular way to model enumerative invariants for $B$-branes.
We keep details to a minimum, and refer to our previous work \cite{Banerjee:2020moh} for notation, further details, and references.
The bounded derived category of coherent sheaves is equivalent to the derived category of representations of the path algebra of a quiver $Q$ with potential $W$ \cite{Douglas:2002fr, Aspinwall:2004bs}
\be
	\CD^b Coh(X^\vee) \simeq \CD^b ({\rm Mod}-(Q,W))\,.
\ee
The quiver description provides a somewhat more manageble handle on the definitions of stability and enumerative invariants.
The relevant notion of stability for quivers corresponds to King's stability \cite{king1994moduli}, see \emph{e.g.} \cite{Denef:2002ru, Alim:2011ae} for a review of its physical interpretation.
Let $Q$ be a quiver with potential $W$, corresponding to a formal sum of loops in the path algebra.
Given a choice of stability condition, a BPS particle is characterized by a certain dimension vector $\vec d$.
Entries of this vector are positive integers, and correspond to dimensions of vector spaces $V_i\simeq \IC^{d_i}$ associated to each vertex of $Q$.
For each arrow of the quiver, starting from vertex $i$ and ending on vertex $j$, one considers the space of linear maps ${\rm Hom}(V_i, V_j)$. The representation variety $\CM_{\vec d}$ is obtained by considering the space of all such maps, subject to linear constraints arising from $\partial W = 0$, modulo $\prod_i GL(V_j)$.
Then the appropriate enumerative invariant is the Euler characteristic of $\chi(\CM_{\vec d})$.

It is worthwhile mentioning the invariants associated to quiver representation theory because, at least under certain circumstances, one may expect a certain correspondence between quiver moduli spaces and the moduli spaces of $A$-branes \cite{Denef:2002ru}. 
This already suggests that the invariants we are after may be related to Euler characteristics $\chi(\CM)$ of $A$-brane moduli spaces, which indeed comes close to the definition we will propose below. 
Later we will see several examples where indeed we find that the moduli space of $A$-branes coincides, at least topologically, with the moduli space of suitable quiver representations $\CM_L \simeq \CM_{\vec d}$, corroborating the hypothesis that a putative enumerative invariant $\Omega(L)$ counting $A$-branes in class $[L]$ would be
\be
	\Omega(L)  \sim \chi(\CM_{L})\,.
\ee

\section{Enumerative invariants for $A$-branes}\label{eq:eunmerative-inv-A-branes}

Having reviewed notions of stability for $A$-branes, we will now start over from scratch and build towards a working definition of the Hilbert space of $A$-branes and the associated enumerative invariants.
The starting point will be to lift $A$-branes, defined by the $A$-model on a certain class of local Calabi-Yau threefolds, to $D3$ branes in type IIB string theory.

\subsection{Supersymmetric quantum mechanics of $D3$ branes}

Type IIB string theory on a Calabi-Yau threefold $X$ features a spectrum of D3 branes wrapping compact special Lagrangian cycles $L \subset X$ and a worldline $\IR$ in the transverse~$\IR^4$.\footnote{More accurately, the classical picture of D3 branes as geometric objects is valid only in the limit $g_s\to 0$.}
A middle-dimensional homology class $[L]\in H_3(X,\IZ)$ may support stable BPS states only if there exist calibrated Lagrangian cycles $L$ in that class.
Calibration, as defined in~\cite{10.1007/BF02392726}, means that the holomorphic top form $\Omega^{3,0}$ has fixed phase $\theta$ when restricted to $L$, namely $\Omega^{3,0}(TL) = e^{i\theta}\, |\Omega^{3,0}(TL)|$ (pullback of $\Omega^{3,0}$ to $L$ is understood).
A D3 brane wrapping a calibrated cycle preserves four supercharges, thus carrying a worldvolume $4d$ $\CN=1$ field theory of a $U(1)$ vectormultiplet.
Dimensional reduction along $L$ yields a $1d$ $\CN=4$ theory on $\IR$. 
On the classical level, fields of this theory arise from modes of open-strings with both endpoints on the D3 worldvolume.
Details of the dimensionally reduced theory will depend on the geometry of $L$ and its embedding in $X$.
Switching on the string coupling $g_s$ quantizes the one-dimensional worldvolume theory, leading to a $\CN=4$ supersymmetric quantum mechanics on $\IR$ that will be denoted $T_L^{1d\, \CN=4}$.

BPS states of the bulk theory are defined by the condition of preserving certain supercharges. On the other hand, the preserved supercharges are precisely those that descend to the worldvolume theory on the D3 brane.
A well-known consequence of the 1d $\CN=4$ super-Poincar\'e algebra, is that if a state in the Hilbert space of the quantum mechanics is annihilated by one of its supercharges, it must be a groundstate \cite{Witten:1982df}.
This leads to a natural definition for the Hilbert space of BPS states corresponding to the D3 brane on $L\times \IR$: it is identified with the space of groundstates of this quantum mechanics
\be\label{eq:Hilbert-space-proposal}
	\tilde\CH^{BPS}_L := \CH_{0}[T_L^{1d\, \CN=4}] \,.
\ee

\subsection{Nonlinear sigma models on $A$-brane moduli spaces}

Having settled on a definition of $\tilde \CH^{BPS}_L$, the next question we address is what can be said about its structure on general grounds. This will depend on the theory $T_L^{1d\, \CN=4}$.

The simplest type of theory arises when $H_1(L,\IZ)=0$, since in this case there is a unique calibrated cycle $L$ in class $[L]$ for given complex moduli $\Omega^{3,0}\in H^{3,0}(X)$ \cite{Joyce:1999tz, ASNSP_1997_4_25_3-4_503_0, Strominger:1996it}.
The gauge theory features only a 1d $\CN=4$ vectormultiplet, whose bosonic degrees of freedom include a one-form and a triplet of scalars.\footnote{This theory is the simplest instance of a quiver gauge theory of the type discussed in \cite{Denef:2002ru}, corresponding to the case of a single-node quiver.}
The vacuum equations imply that the connection is flat, and since the quantum mechanics is on $\IR$, the connection must be pure-gauge and gives rise to no moduli.
The moduli space of vacua consists entirely of the Coulomb branch $\IR^3$, parameterized by the v.e.v.s of the scalar triplet. 
Values of the three scalars $x_i(t)$ parameterize the transverse position of the BPS particle with worldline $\IR\subset \IR^4$.

When $H_1(L,\IZ)\neq 0$  the theory $T^{1d\,\CN=4}_L$ is more interesting.
On a purely geometric level, there is now a nontrivial moduli space $\fM_{L}$ of special Lagrangians in class $[L]$.\footnote{While it would be more appropriate to denote this moduli space by $\fM_{[L]}$, in an effort to keep notation light we simply denote this by $\fM_L$.} 
Locally $\fM_{L}$ is modeled by the vector space of harmonic one-forms on $L$ \cite{mclean1998deformations}. 
Using the metric on $L$, this space can be identified with cohomology classes on $L$
\be
	T\fM_{L}\big|_{L}\simeq H^1(L,\IR) \,.
\ee
These new geometric moduli of the underlying Lagrangian yield new degrees of freedom in the theory $T^{1d\,\CN=4}_L$. 
In addition to the $U(1)$ vectormultiplet described previously, there will now be $b_1(L)$ chiral multiplets, arising as follows.

On the one hand, there are moduli for the flat connection on $L$, corresponding to periods of the flat connection
$\theta_i = \oint_{\omega_i} A$ along generators $\omega_i$ of $H_1(L,\IZ)$. 
Due to large gauge transformations, these moduli are periodic $\theta_i \in \IR / 2\pi \IZ \simeq S^1$ for $i=1,\dots, b_1(L)$.
On the other hand there are moduli corresponding to the decomposition of the (also flat) \emph{dual} connection $\tilde A$, along a basis of harmonic 1-forms on $L$, namely $\tilde A = \rho_i \omega^i$ with $\rho_i\in \IR$. 
Together they give rise to $b_1(L)$ complex-valued scalars $\sigma_j = \exp\(\rho_j + i\theta_j\) \in \IC^*$.
These chiral fields parameterize deformations of an $A$-brane on $L$: deformations of $L$ correspond to fluctuations of $\rho_i(t)$, while deformations of the abelian flat local system correspond to fluctuations of holonomies $\theta_i(t)$.

In conclusion, the 1d $\CN=4$ theory arising from a single D3 brane on a special Lagrangian $L$ consists of two non-interacting parts
\be\label{eq:1d-theory-factorization}
	\underbrace{\text{U(1) gauge theory}}_{\text{center-of-mass d.o.f.}}\quad \times\quad \underbrace{\text{sigma model with target $\CM_L$}}_{\text{internal d.o.f.}}
\ee
reflecting the separation between translational and internal degrees of freedom of the BPS particle.
The former are described by the adjoint (neutral) scalars of a 1d $\CN=4$ $U(1)$ vector multiplet.
On the other hand, internal degrees of freedom are described by a sigma model of $b_1(L)$ neutral chiral multiplets with target $\CM_L$, the moduli space of $A$-branes on~$L$.\footnote{All chirals are neutral under $U(1)$ since they all descend from the 4d adjoint vectormultiplet, or from strings with both endpoints on the same D3 brane.}
This moduli space fibers over the moduli space of the underlying calibrated cycle~$L$
\be
	T^{b_1(L)} \to \CM_L \to \fM_L \,,
\ee
where fibers $T^{b_1(L)}$ parameterize flat abelian local systems on $L$.
$\CM_L$ admits a K\"ahler metric~\cite{Strominger:1996it}, consistently with $\CN=4$ supersymmetry on the particle worldvolume.

\subsection{The Hilbert space of BPS states}

With a clearer picture of the worldvolume theory of a BPS particle engineered by a D3 brane on $L$, we return to the Hilbert space introduced in (\ref{eq:Hilbert-space-proposal}).
Since center-of-mass degrees of freedom and internal ones do not interact (\ref{eq:1d-theory-factorization}), we consider their quantization separately.

Quantization of the center-of-mass degrees of freedom, together with fermionic partners, gives rise to the \emph{half-hypermultiplet} 
$\rho_{hh}$.
As a representation of $spin(3)\oplus su(2)_R$ of transverse spacetime rotations and R-symmetry of the bulk theory, the half-hypermultiplet corresponds to $ ({\bf 1} , {\bf 2})  \oplus ({\bf 2} , {\bf 1})$, see e.g. \cite{MoorePITP} for more details.

For the sigma model with target $\CM_L$, the Hilbert space of supersymmetric vacua can be identified with Dolbeault cohomology $\bigoplus_{p,q}H^{p,q}_{\bar\partial}(\CM_L)$ \cite{Witten:1982im, hori2003mirror}. 
The Dolbeault $(p,q)$ bi-grading translates physically into Fermion number $F = p+q- \dim \CM_L$ ($p+q$ corresponds to $k$-form degree) and $R$-charge $R = p-q$. 
Here the overall shift of Fermion number by $-\dim \CM_L$ 
arises through a careful analysis of the fields involved in the sigma model, 
which leads to identifying vacuum configurations with sections of 
$K_{\CM_L} \otimes \wedge T_{\CM_L}$.\footnote{By $K_{\CM_L}$ we denote the canonical line bundle of $\CM_L$ and by $ T_{\CM_L}$ the holomorphic tangent bundle.} 
This bundle is indeed isomorphic to $\wedge T^*_{\CM_L}$, but only up to an overall shift of $k$-form degree by $-\dim \CM_L$ due to the factor $K_{\CM_L}$. A derivation with details can be found in \cite[eq. (2.72)]{Hori:2014tda}.
For simplicity we will suppress the grading by $R$-charge, and pass from Dolbeault to de Rham cohomology $\bigoplus_k H_{dR}^{k}(\CM_L)$.

This definition is still incomplete, since there are cases when the moduli space $\CM_L$ is noncompact, due to non-compactness of the underlying moduli space of calibrated cycles $\fM_L$.\footnote{An example of this is the moduli space of $L\simeq T^3$ the SYZ fiber, whose mirror dual is a D0 brane: in fact the moduli space of the D0 is the whole mirror Calabi-Yau $X^\vee$, which is always noncompact for the class of geometries we consider.}
When the target space is noncompact, there are several possible definitions of cohomology, potentially leading to different Hilbert spaces. A physically motivated choice would be to consider $L^2$-cohomology \cite{Harvey:1996gc, Hori:2014tda, Lee:2016dbm, Duan:2020qjy}.
On the other hand, if we wish to make contact with Donaldson-Thomas theory in mathematics, a more appropriate choice would be to consider cohomology with compact support \cite{Martinec:2002wg, Mozgovoy:2020has}.\footnote{Here we refer to Donaldson-Thomas theory on the mirror Calabi-Yau. By mirror symmetry the moduli space of stable B-branes on $X^\vee$ are expected to coincide with moduli spaces of $A$-branes on $X$. Then, roughly, Euler characteristics of compactly supported de Rham cohomology of moduli spaces of $B$-branes coincide with Donaldson-Thomas invariants. A more precise relation between Donaldson-Thomas invariants and Euler characteristics is discussed in \cite{2005math......7523B}.}
These choices do lead to different Hilbert spaces: for example if $\CM_L\simeq \IC$ one has
\be\label{eq:C-cohomology}
	\begin{array}{rrrr}
	H^0_{dR}(\IC) = \IC & \quad H^1_{dR}(\IC) = 0 & \quad H^2_{dR}(\IC) = 0 & \quad \text{(de Rham)}\\
	H^0_{c,dR}(\IC) = 0 &\quad  H^1_{c,dR}(\IC) = 0 & \quad H^2_{c,dR}(\IC) = \IC  & \quad \text{(compact support)}\\
	H^0_{L^2}(\IC) = 0 &\quad  H^1_{L^2}(\IC) = 0 & \quad H^2_{L^2}(\IC) = 0  &\quad  \text{($L^2$ cohomology)}\\
	\end{array}
\ee
In our approach, tailored to studying geometric properties of $A$-branes, it is natural to adopt the second option  \cite{Banerjee:2020moh}. 
We thus identify the Hilbert space of internal degrees of freedom with \emph{compactly supported de Rham cohomology} of the moduli space of $A$-branes 
\be\label{eq:H-BPS}
	\CH^{BPS}_{L} = \bigoplus_k H_{c,dR}^{k}(\CM_L)[-\dim \CM_L] \,,
\ee
where $[-\dim \CM_L]$ denotes the shift by $\dim \CM_L$ in the Fermion number, explained earlier.
The full Hilbert space $\tilde \CH^{BPS}_L$ in (\ref{eq:Hilbert-space-proposal}), of BPS states of a wrapped $D3$ brane on the cycle $[L]\times \IR$, is $\rho_{hh}\,\otimes \, \CH^{BPS}_{L} $.
But since $A$-branes know nothing about the transverse $\IR^4$, and any associated degrees of freedom, we drop  $\rho_{hh}$ and simply define $\CH^{BPS}_{L}$ as the Hilbert space of BPS states.

\subsection{Witten index}

The Hilbert space of BPS states (\ref{eq:H-BPS}) comes equipped with a natural enumerative invariant, the Witten index of the supersymmetric quantum mechanics on~$\CM_L$.
We take this as the definition of an enumerative invariant for $A$-branes
\be
	\Omega(L)  = \Tr (-1)^F e^{-\beta H} = \Tr_{\CH^{BPS}_{L}} (-1)^{2 J_3}\,.
\ee
Here $J_3$ is a Cartan generator of the spin algebra $spin(3)$, realized by the $su(2)$ Lefshetz action on cohomology \cite{Denef:2002ru}. 
Following the identification with compactly-supported de Rham cohomology in (\ref{eq:H-BPS}), the Witten index computes the Euler characteristic of the moduli space of $A$-branes, up to an overall sign due to the shift in Fermion number
\be\label{eq:omega-chi}
\begin{split}
	\Omega(L) 
	&=(-1)^{\dim \CM_L} \, \sum_{k} (-1)^k \dim H_{c,dR}^{k}(\CM_L)  \\
	& = (-1)^{\dim \CM_L} \,\chi(\CM_L) \,.
\end{split}
\ee

This is the main point of this section: supersymmetric quantum mechanics on the worldvolume of $D3$ branes provides a natural definition for enumerative invariants of $A$-branes, corresponding to the Euler characteristic of their moduli spaces up to a sign.

\medskip

We close this section with a remark on the two-fold role of string theory for the definition of enumerative invariants (\ref{eq:omega-chi}) and their categorification (\ref{eq:H-BPS}).
On the one hand, embedding $A$-branes into string theory naturally leads to a supersymmetric quantum mechanics of D3 branes, whose Witten index corresponds to the invariants considered here.
On the other hand, mirror symmetry further relates these to D4-D2-D0 boundstates in the mirror Calabi-Yau $X^\vee$, whose own enumerative invariants are the rank-zero (generalized) Donaldson-Thomas invariants \cite{Joyce:2008pc,Kontsevich:2008fj}. 
Recall from section \ref{sec:A-branes-stability} that the category of $A$-branes admits a definition of generalized DT invariants. 
Homological mirror symmetry further relates those to the categorical DT invaraints for $B$-branes on $X^\vee$.
This web of relations suggests that our definition of enumerative invariants should agree with the categorical one, providing an alternative definition based entirely on classical geometric data.

\begin{comment}
\label{comm:DT-quivers}
{
In relation to the definition of DT invariants based on BPS quivers, it should be noted that $\CM_L$ can be viewed as the Higgs branch of a supersymmetric quiver quantum mechanics.
The relevant theory is a 1d $\CN=4$ GLSM with superpotential \cite{Douglas:1996sw, Denef:2002ru, Fiol:2000wx}. There is an overall $U(1)$ parameterizing the center of mass degrees of freedom, which decouples.
The moduli space of vacua features both Higgs and Coulomb branches. The Higgs branch describes the fusion of parton branes corresponding to nodes of the quiver, which in the limit $g_s\to 0$ should correspond precisely to the $A$-brane on $L$ that we discuss above \cite{Denef:2002ru}.
Then the Higgs branch is identified with the moduli space $\CM_L$ of an $A$-brane in class $[L]$. The definitions of BPS Hilbert space (\ref{eq:H-BPS}) and of enumerative invariants (\ref{eq:omega-chi}) then coincide with the definitions based on quiver representation theory in the context of generalized DT theory \cite{Kontsevich:2008fj}.
}
\end{comment}

\section{Moduli spaces of $S^2$-fibered special Lagrangians}\label{sec:sLag-moduli-foliations}

The definition of enumerative invariants for $A$-branes in (\ref{eq:omega-chi}) and their categorification (\ref{eq:H-BPS}) 
are based on the notion of a moduli space $\CM_L$ of $A$-branes in class $[L]$.
In this section we begin discussing the structure of $\CM_L$, starting from the observation that it admits a natural fibration (\ref{eq:CM-fibration-review}). 
Here we study the base $\fM_L$ of this fibration, which parameterizes moduli of the underlying special Lagrangians, and will later return to the discussion of moduli of local systems on $L$.
In this section we focus entirely on a class of special Lagrangians characterized by the fact that they admit fibrations by $S^2$, although in later sections we will also discuss the case of SYZ fibers.

\subsection{A class of Calabi-Yau hypersurfaces}

Let $X$ be a Calabi-Yau hypersurface described as the vanishing locus of a function
\be
	H = uv - F(x,y)  \quad  \subset \ \IC^2 \times ({{\IC}^*})^2
\ee 
for some Laurent polynomial $F$ of variables $(x,y)$.
This class of geometries arises in at least two distinct, though overlapping, settings.
On the one hand, varieties like $X$ may arise as Hori-Vafa mirrors of toric Calabi-Yau threefolds, with toric data encoded by the polynomial $F(x,y)$ \cite{Hori:2000kt}.
On the other hand, the curves $F(x,y)=0$ also appear as semiclassical moduli spaces of $A$-branes on noncompact special Lagrangians in toric Calabi-Yau threefolds, such as toric Lagrangians \cite{Aganagic:2000gs, Aganagic:2001nx} and knot conormals \cite{Ooguri:1999bv, Aganagic:2013jpa, Aganagic:2012jb, Ekholm:2019yqp}.

The holomorphic three form on the Calabi-Yau is given by the pull-back of
\begin{equation}
	\Omega^{3,0} = \frac{1}{2\pi} \, \oint_{\gamma_H}\frac{du \wedge dv \wedge d\log x\wedge d\log y}{H}
\end{equation}
where $\gamma_H$ is a small loop around the locus $H=0$.
By an application of the Poincar\'e residue theorem in the $v$-plane, this reduces to 
\begin{equation}\label{eq:top-form-log}
	\Omega^{3,0} = i\, \frac{du\wedge d\log x\wedge d\log y}{\partial H/\partial v}
\, = \, i \, d \log u\wedge d\log x \wedge d\log y.
\end{equation}
A special Lagrangian cycle $L$ immersed via $\iota :L\to X$ into $X$ is defined by the condition 
\be\label{eq:sLag-condition}
	\zeta^{-1} \iota^*\Omega^{3,0}(TL|_p) \in \IR_{>0}
\ee
when evaluated on the fiber of $TL|_p$ for all $p\in L$. Here $\zeta$ is a constant phase, which coincides by definition with the phase of the BPS central charge\footnote{Normalization is chosen to agree with \cite{Banerjee:2018syt} where the central charge of $D0$ branes is $2\pi/R$  ($R=1$ here).}
\be
	Z_L = \frac{1}{4\pi^2} \int_L \iota^*\Omega^{3,0} \quad \in \zeta \,\IR_{>0} \,.
\ee

\subsection{$S^2$-fibered Lagrangians and graded lifts}\label{sec:S2-fibers}

We now restrict attention to a specific class of calibrated Lagrangian cycles, having the property that they admit fibrations by two-spheres. One motivation for studying these cycles is the direct connection to cycles on Seiberg-Witten curves in 4d $\CN=2$ theories  \cite{Klemm:1996bj}.

\begin{figure}[h!]
\begin{center}
\includegraphics[width=0.85\textwidth]{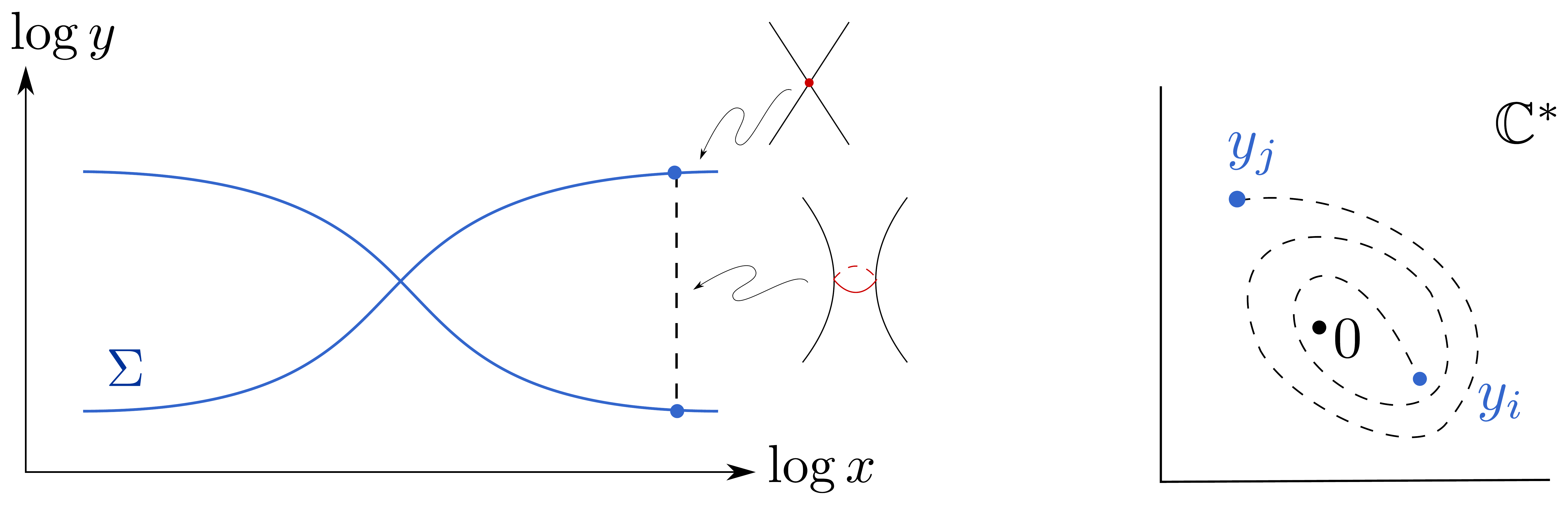}
\caption{Left: The conic fibration over the universal cover of $(\IC^*)^2$. 
An $S^2$ fibered over the dashed segment $\tilde I_{ij,M,N}(x)$ is shown, stretching between $\log y_i+2\pi i M$ and $\log y_j+2\pi i N$.
Right: the segment maps to a path in $\IC^*$ winding $N-M$ times around $y=0$.}
\label{fig:conic-bundle}
\end{center}
\end{figure}

Let us start with generic Lagrangians, temporarily postponing a discussion of calibration.
To describe the structure of a generic Lagrangian $L$ of this type, it will be convenient to view the ambient space of $X$ as a fibration of $\IC^2$ over $(\IC^*)^2$. At each $(x,y)\in (\IC^*)^2$ there is a complex conic in $\IC^2$ described by $uv = c$, with $c = F(x,y)\in \IC$. 
If $c\neq 0$ the conic is a one-sheeted hyperboloid, with a noncontractible $S^1$.
This circle shrinks when $c=0$, which happens when $(x,y)$ lie on the complex curve $\Sigma \subset (\IC^*)^2$ defined by $F(x,y)=0$. The geometry is sketched in Figure \ref{fig:conic-bundle}.

We consider a segment ${\tilde I}_{ij,M,N}(x)$ parameterized\footnote{%
At this point the segment is defined up to relative homotopy, and needs not have the geometry described in (\ref{eq:y-of-s}). However, later we will derive this precise shape from the study of the special Lagrangian condition.%
} 
by $s\in [0,1]$  
\be\label{eq:y-of-s}
	x(s) = x
	\qquad
	\log y(s) = (1-s)(\log y_i(x) +2\pi i M)+ s (\log y_j(x) + 2\pi i N)
\ee
for any two roots $y_i, y_j$ of $F(x,y)=0$ and any choice of logarithmic branches $M,N\in \IZ$.
Here and in the following, a choice of trivialization for the covering $\Sigma\to\IC^*$ over the $x$-plane is understood, so that we may unambiguously assign labels $i,j,\dots$ etc to roots $y_i, y_j,\dots$ of $F(x,y)=0$.
Note that ${\tilde I}_{ij,M,N}(x)$ is defined on the universal covering of the $\IC^*$ $y$-plane, namely $\IC$ with local coordinate $\log y$.
This plane is divided into horizontal strips of height $2\pi i$ corresponding to different branches of the logarithm. 
The segment (\ref{eq:y-of-s}) stretches between branches labeled by $M,N$. Let
\be
	{I}_{ij,n}(x) = \tilde\pi({\tilde I}_{ij,M,N}(x)) 
\ee
be the projection down to $\IC^*$ by the exponential map $\tilde\pi$. This will be a path from $y_i$ to $y_j$ with winding number around $y=0$ (rounded to)
\be\label{eq:winding-number-graded-lift}
	n = N-M\,.
\ee
A sketch of this projection is provided in Figure \ref{fig:conic-bundle}.

We define a two-sphere ${\tilde S}^2_{ij,M,N}(x)$ by considering a circle on the complex conic, fibered over $\tilde I_{ij,M, N}(x)$\footnote{We consider the pullback of the bundle of complex conics over $(\IC^*)^2$ to a family of complex conics over the universal cover.}.
This two-sphere lives in the universal cover, and maps to a two-sphere on the base
\be\label{eq:S2-fibers}
	{S}^2_{ij,n}(x) = \tilde\pi({\tilde S}^2_{ij,M,N}(x)) \,.
\ee
Of course, the latter is fibered by circles over the path $I_{ij,M,N}(x)$ winding $n$-times around $y=0$, above a fixed $x\in \IC^*$. 

At this point it is important to observe that ${S}^2_{ij,n}(x)$ contains slightly less information than its preimage ${\tilde S}^2_{ij,M,N}(x)$, having traded two integer labels $M,N$ for the single $n$.
This is because any simultaneous shift $(M,N)\to (M+k,N+k)$ by $k\in \IZ$ would leave invariant the projection by $\tilde\pi$ to a two-sphere on the base.
The information that gets lost corresponds to a choice of \emph{graded lift} for $S^2$ to a two-sphere in the universal cover.

\begin{figure}[h!]
\begin{center}
\includegraphics[width=0.5\textwidth]{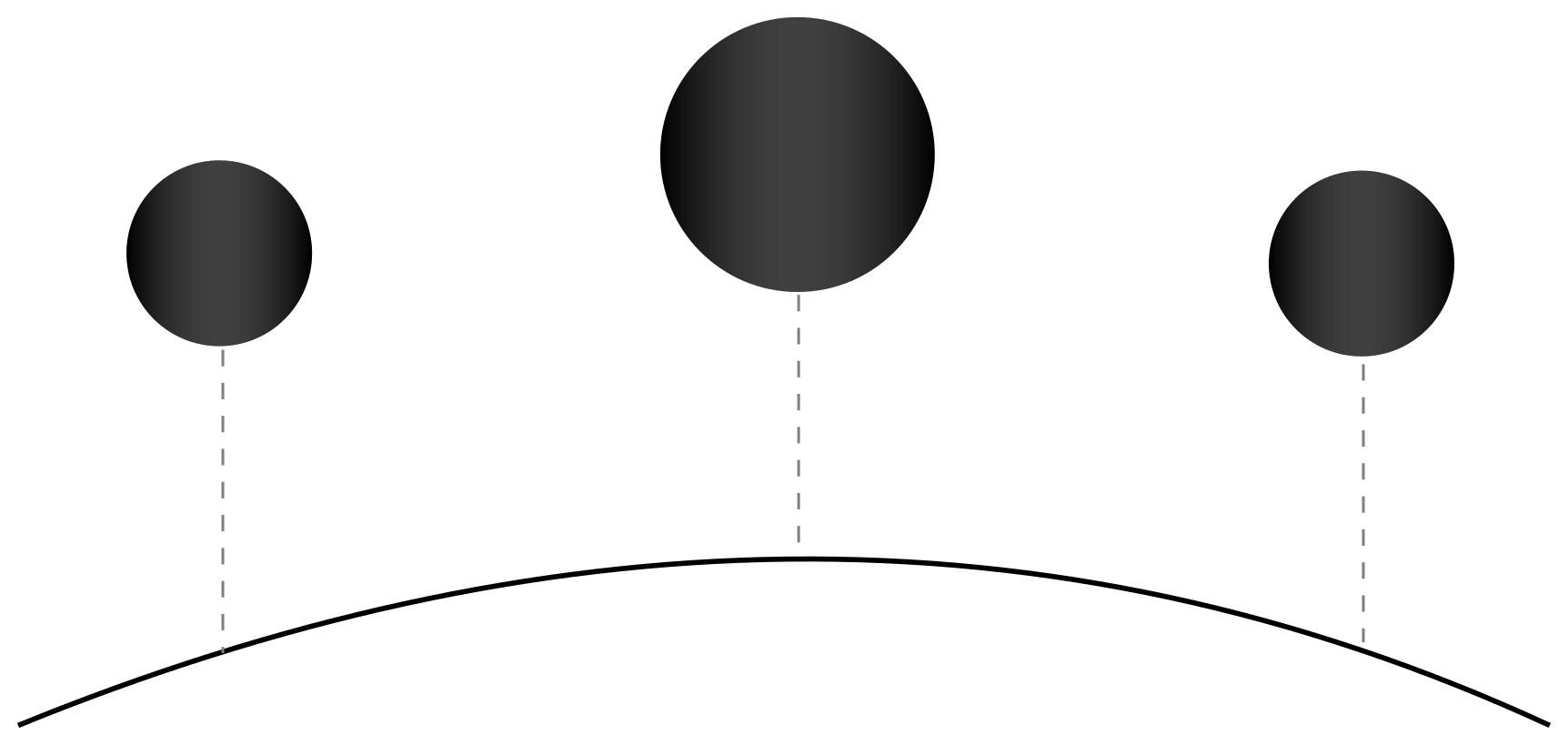}
\caption{A three-manifold locally fibered by $S^2$ over a segment.}
\label{fig:3-manifold-piece}
\end{center}
\end{figure}

Now an $S^2$-fibered three-manifold can be obtained by choosing a path $x(t)$ on $\IC^*$, for $t\in [0,1]$, and considering the family of $S^2_{ij,n}(x)$ fibered over such a path, see Figure \ref{fig:3-manifold-piece}. 
The three-manifold obtained in this way will generically have a boundary at the endpoints of the path $x(t)$, namely $\partial M = S^2_{ij,n}(x(1)) - S^2_{ij,n}(x(0))$.
To obtain a compact three-manifold one needs to impose conditions on what happens to these boundaries. 
There are essentially three types of conditions:

\begin{enumerate}[label=c\arabic{enumi}.,ref=(c\arabic{enumi})]
\item \label{item:S2xS1}
The most natural option is to pick a periodic path, namely $x(0) = x(1)$.
However, not every such path will produce a closed three-manifold. 
The essential requirement is that, as we proceed along the path and come back to the initial point, the shifts in logarithmic branches of $\log y_i \to \log y_i + 2\pi i k$ and $\log y_j\to \log y_j + 2\pi i k'$ are equal $k=k'$. This ensures that the two-sphere $S^2_{ij,n}(x(0))$ gets transported back to itself $S^2_{ij,n+k'-k}(x(1)) \equiv S^2_{ij,n}(x(1))$.
In this case, we get a closed Lagrangian with topology 
\be
	L\simeq S^2\times S^1\,.
\ee

\item \label{item:S3}

The second option is that an endpoint, say $x(0)$, corresponds to a branch point of $\Sigma\to \IC^*$ (a point where $y_i(x) = y_j(x)$). If we take the 2-sphere $S^2_{ij,n=0}$, it shrinks at the endpoint and the three-manifold locally has the topology of a three-ball. 
In particular if \emph{both} endpoints lie on branch points (where the \emph{same} two-sphere shrinks, as in Figure \ref{fig:S3-fibered}), the overall topology is 
\be
	L\simeq S^3\,.
\ee

\item \label{item:fancy}

The third option is more sophisticated. We may allow the path $x(t)$ to have an endpoint at generic $x\in \IC^*$, as long as it joins \emph{two} other paths at a \emph{junction}, see Figure \ref{fig:junction}. 
At junctions we require certain compatibility conditions on the types of $S^2$ fibered over each adjoining path, to ensure that they can all be glued consistently. We will discuss these conditions shortly.
With junctions in the game, all sorts of interesting topologies can arise for $L$, we will later encounter  examples with 
\be
	b_1(L) = 0, 1, 2, 3\dots
\ee

\end{enumerate}

Having constructed different types of closed Lagrangian 3-manifolds fibered by 2-spheres, we return to the issue of \emph{graded lifts}.
As explained above, the 2-sphere $\tilde S^2_{ij,M,N}$ in the universal cover represents a choice of graded lift of the two-sphere $S^2_{ij,n}$ on the base, with $n=N-M$.
The Lagrangian $L$ is locally fibered by $S^2_{ij,n}$ over a path $x(t)$ in $\IC^*$, and \emph{locally} can be lifted to a \emph{graded} Lagrangian fibered by $S^2_{ij,M,N}$ over the same path $x(t)$.
The consistency conditions \ref{item:S2xS1}-\ref{item:S3}-\ref{item:fancy} enforced at endpoints of $x(t)$ ensure that the grading determined by $S^2_{ij,M,N}$ extends \emph{globally}.
In conclusion, we have described a construction of $S^2$-fibered Lagrangians in $X$, endowed with the notion of a graded lift to the universal cover of the $\IC^*$ $y$-plane.

The data of graded lifts plays an important role in the definition of Fukaya-Seidel categories \cite{seidel2008fukaya, Aspinwall:2009isa, Auroux:2013mpa}.
In fact our definition agrees with the one arising in that context.
The $\IZ$-grading of Lagrangians is defined provided that the Maslov class of $L$ vanishes, which is automatically the case for special Lagrangians, which is the case we restrict to.
In that case, the grading is defined as a lift of the phase of the holomorphic top form $\Omega^{3,0}$ from values in $S^1$ to values on the covering $\IR$.
In our case, the top form restricted to $X$ is in local coordinates given by $\Omega^{3,0}=i\, \frac{dx\wedge dy\wedge du}{x y u}$ (recall (\ref{eq:top-form-log})), therefore its phase is linearly related to the phases of the $x,y,u$-coordinates.
Since $N$ is defined as a choice of branch for $\log y$, it indeed coincides with the $\IZ$-grading induced by trivializing the phase of the top form. 
Another interpretation of $N$, arising from a spacetime point of view, was discussed in \cite{Banerjee:2018syt}.

\subsection{Junctions}\label{sec:junctions}

\begin{figure}[h!]
\begin{center}
\includegraphics[width=0.5\textwidth]{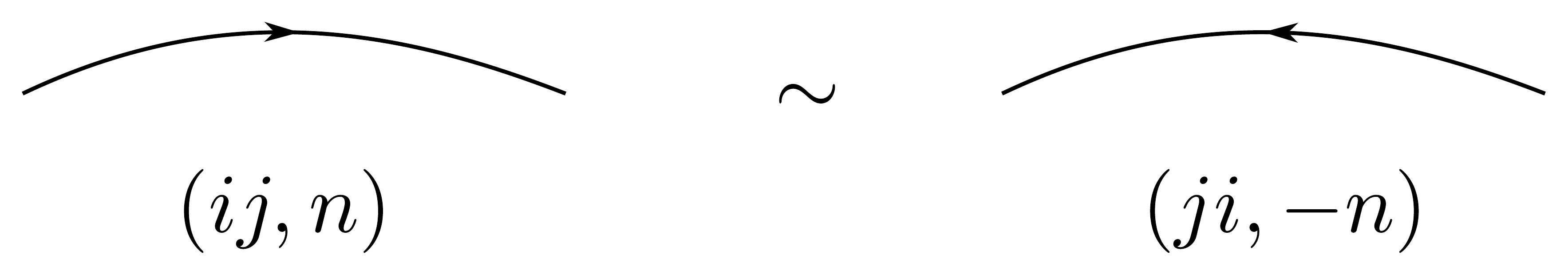}
\caption{Segments with opposite orientations and flipped labels are equivalent.}
\label{fig:orientation-reversal}
\end{center}
\end{figure}

A junction is a point $x\in\IC^*$ where three distinct segments can end.
Recall that we consider a two-sphere $S^2_{ij,n}$ fibered above each segment, we may keep track of this data by attaching a label $(ij,n)$ to the segment itself.
Moreover, reversing the orientation of the segment while keeping the orentation of the three-manifold unchanged requires reversing the orientation of the two-sphere. Therefore an $(ij,n)$ segment  is equivalent to a $(ji,-n)$ segment with the opposite orientation, see Figure \ref{fig:orientation-reversal}.
By convention, all segments attached to a junction will be understood to be incoming, when referring to their labels.

The main issue we need to deal with, is that the three-manifolds fibered over the three incoming segments have disjoint boundaries, which need to be glued together in some way in order to produce a closed three-manifold.
We propose the following construction. 
The boundary of a segment ending at the junction $x$ is a two-sphere $S^2_{ij,n}$. Recall that this sphere is fibered over an interval $I_{ij,n}$ in the $\IC^*$ $y$-plane above $x$. 
We may choose a graded lift to a segment $\tilde I_{ij,M,N}$ in the covering $\IC$-plane with coordinate $\log y$, for an arbitrary $M\in \IZ$. Correspondingly there is a lift of the two-sphere to $\tilde S^2_{ij,M,M+n}$ as described previously.
In choosing a graded lift, we require that the endpoints of lifted segments match together, bounding a polygon in the $\IC$-plane. We then take a circle fibration over such polygon,  this produces a three-manifold whose boundary matches precisely with the boundaries of the three incoming pieces that we wish to glue. We illustrate this construction with a few concrete examples.

\begin{figure}[h!]
\begin{center}
\includegraphics[width=0.85\textwidth]{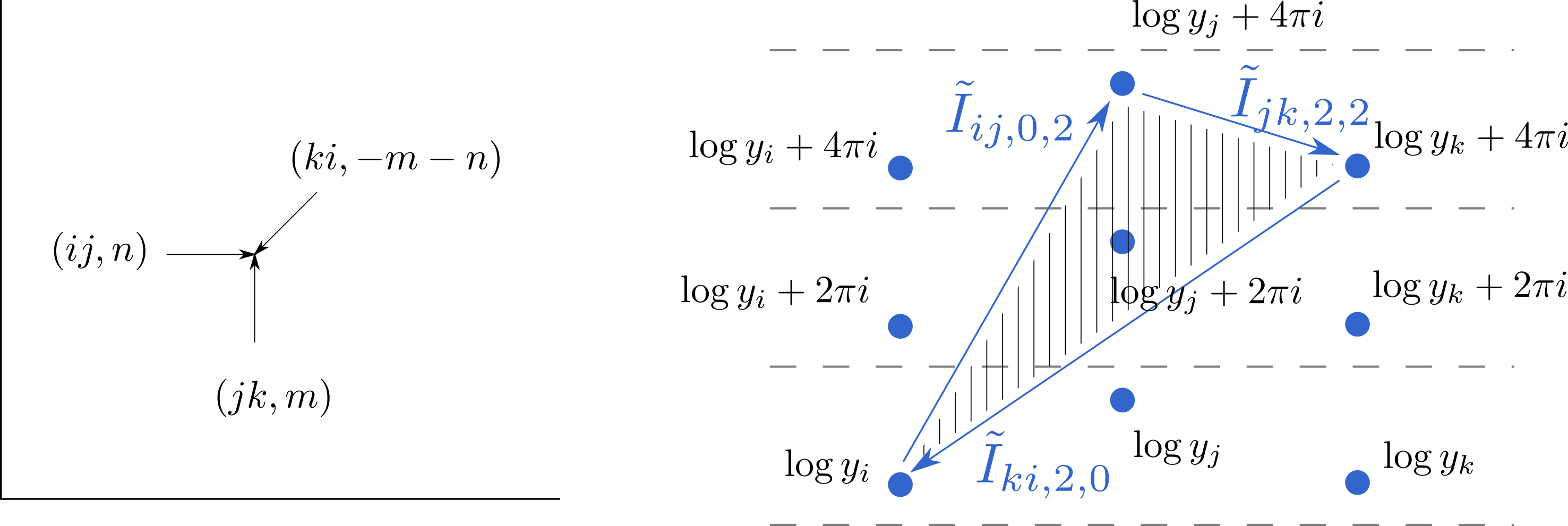}
\caption{Left: the junction of three paths in the $\IC^*$ $x$-plane. Right: the $\IC$ plane with local coordinate $\log y$, above the junction point. The two-spheres $\tilde S^2_{ij,0,2}$ etc are represented by segments $\tilde I_{ij,0,2}$ etc (we fixed $n=2,m=0$ for illustration). The junction three-manifold is obtained by fibering the geodesic circle of the conic $uv=F(x,y)$ over the shaded triangle.}
\label{fig:junction}
\end{center}
\end{figure}

\begin{figure}[h!]
\begin{center}
\includegraphics[width=0.85\textwidth]{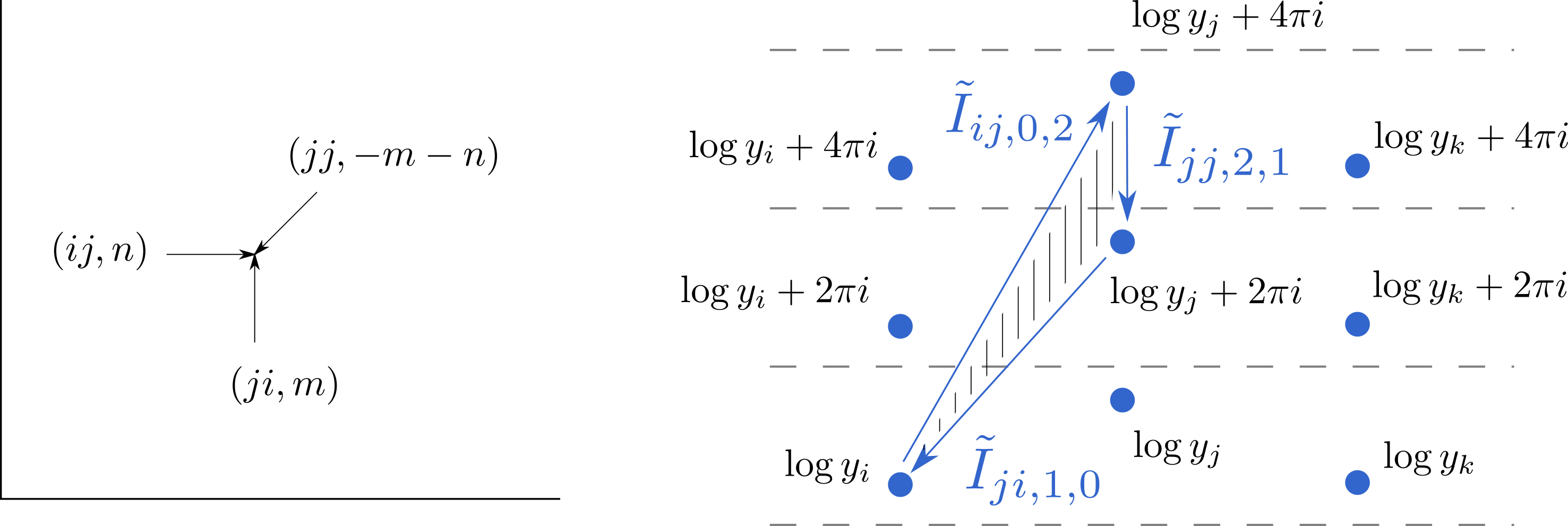}
\caption{A variant of the junction considered in Figure \ref{fig:junction} with $j=k$. (Here we fixed $n=2,m=-1$ for illustration).}
\label{fig:junction-bis}
\end{center}
\end{figure}

\paragraph{Example 1.}
First consider the case of three paths of types $(ij,n), (jk,m)$ and $(ki,-m-n)$ with $i,j,k$ not necessarily all distinct.
Let us choose graded lifts for each path as follows
\be
\begin{split}	
	(ij,n)& \to (ij,M,M+n)\,,\\
	(jk,m)& \to (jk,M+n,M+n+m)\,,\\
	(ki,-m-n)& \to (ki,M+n+m,M)\,,
\end{split}
\ee
for arbitrary $M\in \IZ$.
We take two-spheres fibered above these paths in the corresponding logarithmic branches on the $\IC$-plane of $\log y$. Figure \ref{fig:junction} shows these spheres represented by the respective segments $\tilde I_{ij,M,M+n}$ etc.
The segments bound a triangle in $\IC$, and we build a three-manifold by fibering a circle on the complex conic $uv=F(x,y)$ over the triangle. By construction this is a three-manifold whose boundary are the two-spheres associated to edges of the triangle.
A variant of this example, is when two labels coincide. The corresponding picture is shown in Figure \ref{fig:junction-bis}

\begin{figure}[h!]
\begin{center}
\includegraphics[width=0.85\textwidth]{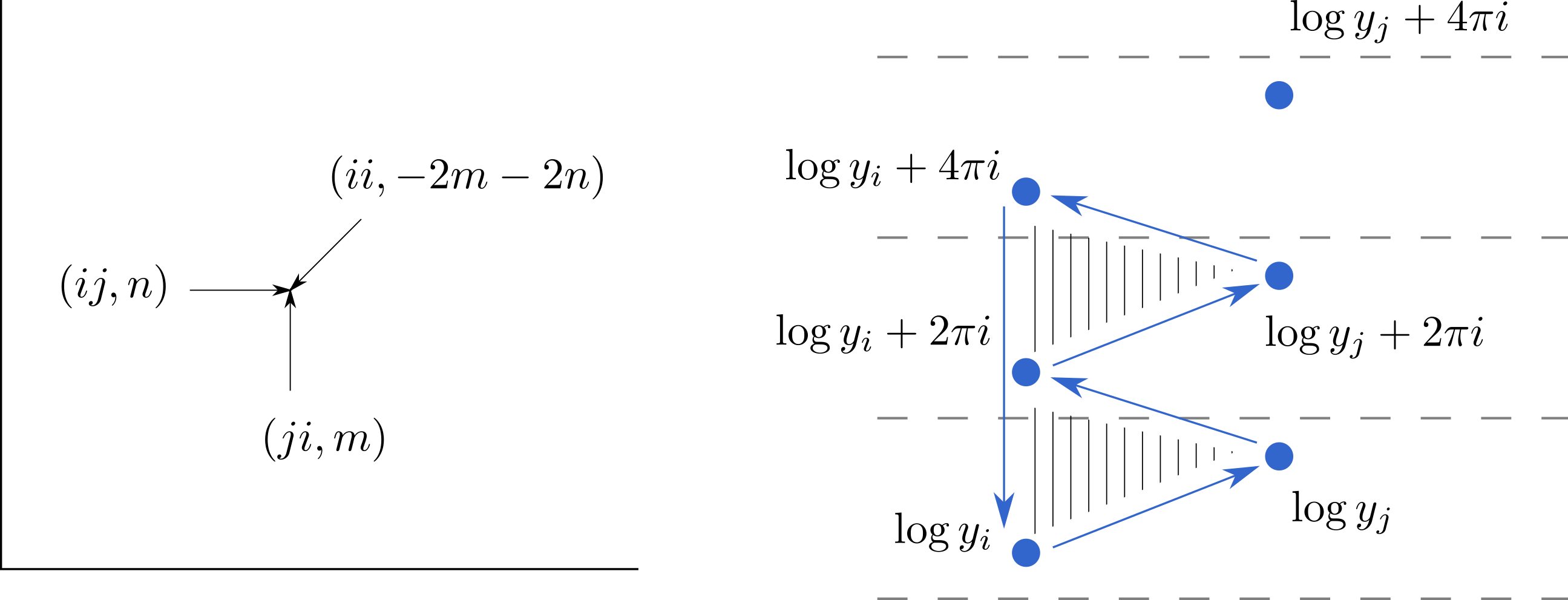}
\caption{Left: the junction of three paths in the $\IC^*$ $x$-plane. Right: the $\IC$ plane with local coordinate $\log y$, above the junction point. (Here we fixed $n=0,m=1, p=2$ for illustration).}
\label{fig:junction-ex-2}
\end{center}
\end{figure}

\paragraph{Example 2.}
Next let us consider the case of three paths of types $(ij,n), (ji,m)$ and $(ii,s)$ with $p(m+n)+  s = 0$ for some $p\in \mathbb{N}$.
In this case we take $p$ lifts of both $(ij,n)$ and $(ji,m)$ $S^2$-fibered 3-manifolds to the universal cover, graded as follows
\be
\begin{split}	
	(ij,n)& \to (ij,M,M+n), \ (ij,M+m+n,M+m+2n)  \dots \\
		& \qquad\qquad\qquad \dots (ij,M+(p-1)(m+n),M+(p-1)m+ pn) \,,\\
	(ji,m)& \to (ji,M+n,M+n+m), \ (ji, M+m+2n, M+2m+2n) \dots\\
		& \qquad \qquad\qquad\qquad\dots  (ji, M+(p-1)m+ pn, M+p(m+n))\\
	(ii,s)& \to (ii,M+p(m+n),M)\,,
\end{split}
\ee
for arbitrary $M\in \IZ$.
These lifts mean that we take two-spheres fibered above these paths in the corresponding logarithmic branches on the $\IC$-plane of $\log y$. Figure \ref{fig:junction-ex-2} shows these spheres represented by the respective segments $\tilde I_{ij,M,M+n}$ etc.
The segments bound a polygon in $\IC$. We then fiber a circle over this polygon to obtain the three-manifold to be glued at the junction.

\begin{figure}[h!]
\begin{center}
\includegraphics[width=0.85\textwidth]{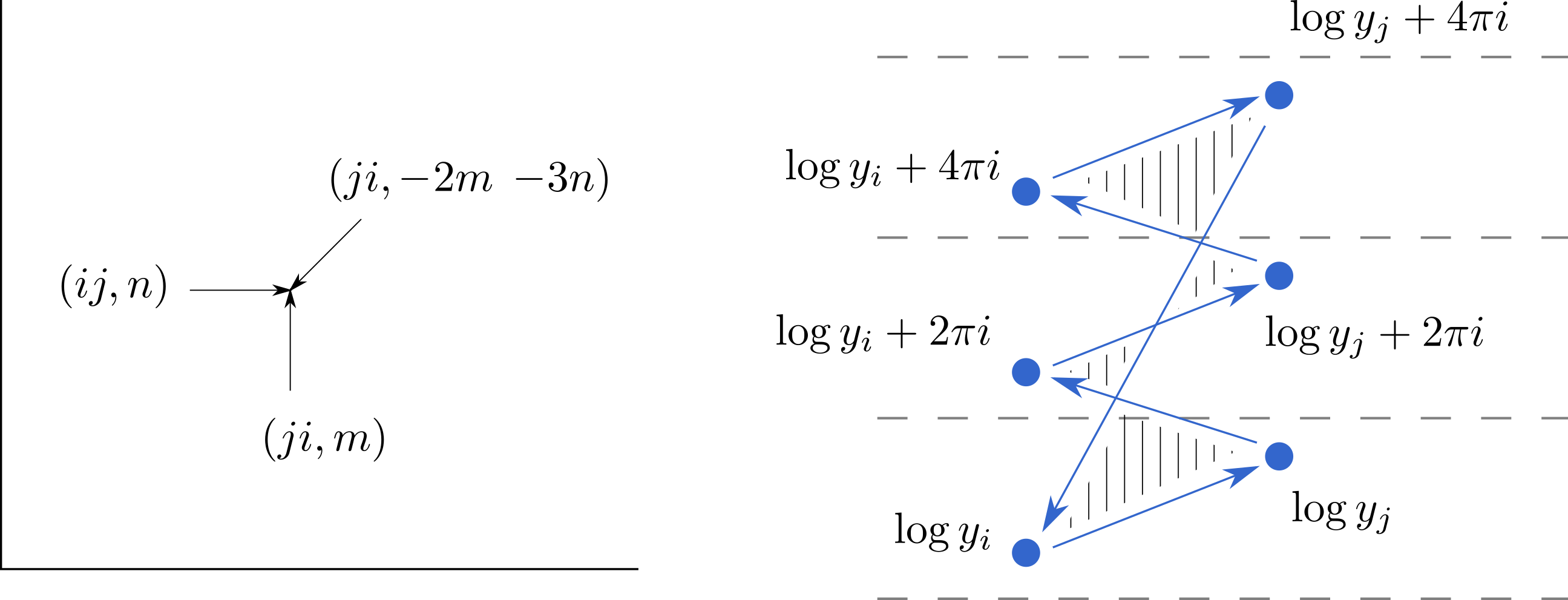}
\caption{Left: the junction of three paths in the $\IC^*$ $x$-plane. Right: the $\IC$ plane with local coordinate $\log y$, above the junction point. (Here we fixed $n=0,m=1,p=2$ for illustration).}
\label{fig:junction-ex-3}
\end{center}
\end{figure}

\paragraph{Example 3.}
The third and last example that will be relevant to us is when the three paths have types $(ij,n), (ji,m)$ and $(ji,-n-p(m+n))$ for some $p\in \mathbb{N}$.
In this case we take $p+1$ lifts of $(ij,n)$ and $p$ lifts of $(ji,m)$ segments, graded as follows
\be
\begin{split}	
	(ij,n)& \to (ij,M,M+n), \ (ij,M+m+n,M+m+2n)  \dots \\
		& \qquad\qquad\qquad \dots (ij,M+p(m+n),M+pm+ (p+1)n) \,,\\
	(ji,m)& \to (ji,M+n,M+n+m), \ (ji, M+m+2n, M+2m+2n) \dots\\
		& \qquad \qquad\qquad\qquad\dots  (ji, M+(p-1)m+ pn, M+p(m+n))\\
	(ii,s)& \to (ii,M+p(m+n),M)\,,
\end{split}
\ee
for arbitrary $M\in \IZ$.
The corresponding two-spheres $\tilde S^2_{ij,M,M+n}$ etc., fibered over the $\IC$-plane of $\log y$ are shown in Figure \ref{fig:junction-ex-3}, where they are represented by segments $\tilde I_{ij,M,M+n}$ etc..
These segments now bound a collection of polygons. 
We take a circle in the complex conic $uv=F(x,y)$ fibered over these fragments to build a three-manifold, and use this three-manifold with boundary to glue at the junction.\footnote{The polygon resulting from this construction may be non-convex in certain cases, as shown in figure. In this case we consider a circle fibered over the shaded triangles, and fibered above the segments $\tilde I_{ij,N,N+n}$ in the $\log y$ plane. Note that, at the intersection of segments of types $\tilde I_{ij,N,N+n}$ and  $\tilde I_{ji,M,M+m}$, the $u$-plane circles that fiber above each segment respectively need not coincide. 
If they do not, then there is a hole in the $u$-plane above the intersections of these segments, that has the form of an annulus stretching between the $u$-circle of one segment, and the $u$-circle of the other segment (As we show in Section \ref{sec:calibration}, the $u$-plane circle must be round and centered at $u=0$). In this case, we define the full $3$-manifold by filling in this hole by an annulus in the $u$-plane. }

\bigskip

As a closing remark on junctions, we note that the polygons that we introduce for gluing have null area. This is simply because $\Omega^{3,0}$ restricts to zero along the 3-manifold that we build, since it has no extension along $x$. 
In particular, since the complex number $Z=0$ can be given any phase, the circle fibrations over these polygons may be regarded as a convex hull that satisfies trivially the condition (\ref{eq:sLag-condition}). This is important since in the end we wish to construct calibrated cycles using junctions as building blocks. We turn to a discussion of calibration next.

\subsection{Calibration and graded lifts}\label{sec:calibration}

So far we described a class of compact Lagrangians in $X$, with the property that they admit $S^2$ fibrations over paths $x(t)$ in $\IC^*$.
Enforcing the condition that these are \emph{special} Lagrangians imposes precise restrictions on the shape of the path $x(t)$.

For concreteness let us introduce a local parametrization of a three-cycle $L$ by
\be
	(t,s,\theta) \in [0,1] \times [0,1]\times S^1 \,,
\ee
where
\begin{itemize}
\item $x(t)$ only depends on $t$
\item for fixed $t$, $y(t,s)$ traces a segment in the $y$-plane connecting two (possibly coincident) sheets of $\Sigma$ above $x(t)$

\item at fixed $(t,s)$, both $x,y$ are fixed, and $L$ traces a circle on the $u$-cylinder, by initial assumption on its topology.
This $u$-circle fibers over the segment parameterized by $s$, shrinking at the endpoints to yield a two-sphere at each $t$.

\end{itemize}

The special Lagrangian condition fixes the dependence of coordinates $(x,y,u)$ on the local parameters $(t,s,\theta)$ as follows.

We start from the dependence of $u$ on $\theta$ for fixed $(t,s)$.
It is easy to see that this circle must be round, since 
the holomorphic top form $\Omega^{3,0}$ restricts to $du/u$, and the special Lagrangian condition 
\be\label{eq:u-theta-dependence}
	\iota_{\partial_\theta} \frac{du}{u} = \frac{\partial \log u}{\partial \theta} \in \zeta' \IR_{\geq 0}
\ee 
is solved by $u = u_0 \cdot e^{\zeta' \theta}$. Now this is a periodic function of $\theta$ only if $\zeta'= k\, i$ for some $k\in \IZ$, and this
gives a round circle of radius $|u_0|$.
Primitive cycles will have $k=\pm1$, with the sign determined by their orientation.
At this point we have not yet determined the dependence of the $u$-circle radius on the $(x,y)$ coordinates.  
We will return to this in a moment.

Next we consider the $s$-dependence of $y(t,s)$ at fixed $t$.
By our assumptions on the topology of $L$, this must trace a line in the $y$ cylinder $\IC^*$, connecting two sheets of $\Sigma$ at $x(t)$.
The special Lagrangian constraint restricted to the $y$-cylinder $\IC^*$ is then
\be
	\iota_{\partial_s} \frac{dy}{y} = \frac{\partial\log y}{\partial s} \in \zeta' \IR_{\geq 0}
\ee 
for some choice of $\zeta'$.
This is the equation of a straight line in coordinate $\log y$, which is uniquely fixed by the choice of two sheets of $\Sigma$ together with a choice of logarithmic branch for each.
Without loss of generality, let these sheets be labeled $\log y_i+2\pi i M$ and $\log y_j+2\pi i N$ (it could be that $i=j$ as long as $M\neq N$ in that case), then the 
explicit solution is exactly (\ref{eq:y-of-s}). We rewrite this as
\be\label{eq:y-shape}
	y(s) = y_i^{1-s}\, y_j^s\, e^{2\pi i\, (N-M) s}\,.
\ee
From here it is manifest that the shape of the Lagrangian $L$ only depends on the relative winding number $n=N-M$ defined in (\ref{eq:winding-number-graded-lift}), at least in the $y$-plane. Only the graded lift of $L$ keeps track of $M,N$.

Now we come back to the question of how the $u$-circle radius depends on $y$, at fixed~$x$.
The main point we wish to make here, is that the functional dependence of the radius on $y$ is uniquely fixed by the special Lagrangian condition. This is important because we wish to study moduli spaces of special Lagrangians, and our claim implies that the $u$-circle radius is not a modulus.
The argument is simple and goes as follows. 
At fixed $x$, we have a slice $X_{x}$ of the Calabi-Yau $X$, which consists of the $u$-cylinder $\IC^*$ fibered over the $y$-cylinder $\IC^*$, with degenerations over the sheets of $\Sigma$ at $y_i(x)$ where the $u$-cylinder $\IC^*$ is replaced by a $u$-plane $\IC$.
Let $X_x^*$ denote the complement of the degenerate loci, \emph{i.e.} where we remove the fibers at $y_i$.
On $X_x^*$ we the holomorphic top form restricts to $\Omega^{2,0} = \frac{du}{u}\frac{dy}{y}$, and the special Lagrangian $L$ restricts to a special Lagrangian $S^2$ in $X_x$, calibrated by $\Omega^{2,0}$.\footnote{This follows because $L$ is fibered by $S^2$ over a segment in the $x$-plane, by assumption, and the fact that $\Omega^{3,0}$ is compatible with this fibration.}
The $S^2$ has the north and south poles at the singular loci $(y_i,u=0)$ and $(y_j,u=0)$, but elsewhere it is calibrated by $\Omega^{2,0}$.
McLean's theorem asserts that this must be a rigid special Lagrangian cycle in $X_x$, since $\pi_1(S^2)$ is trivial. 
The absence of moduli implies that the $u$-circle radius must be a uniquely determined function of $s$, as claimed.\footnote{By contrast in the case of SYZ fibers where $L\simeq T^3$, the radius of the $u$-plane circle will be a true modulus, see Section~\ref{sec:SYZ-fibers}.
The crucial difference lies precisely in the fact that the north/south poles of $S^2$ lie on singular fibers in $X_x$: here the $u$-circle shrinks to a point, and this is what makes $\pi_1(S^2)$ trivial, freezing the deformation associated to the $u$-circle.}

Having completely fixed the $u$ and $y$ dependence on $\theta$ and $s$, the only moduli of $L$ can be encoded in the $x(t)$ dependence. 
This observation is very important, because it means that the whole moduli space of $L$ can be captured by studying the dependence of $x$ on $t$, to which turn next.

The pullback of the holomorphic top form to $L$ can now be evaulated explicitly
\be
\begin{split}
	\iota^*\Omega^{3,0}(\partial_t,\partial_s,\partial_\theta) 
	& = i\, \iota^*\(\frac{du}{u} \frac{dx}{x} \frac{dy}{y}\) (\partial_t,\partial_s,\partial_\theta)  \\
	& = i\(k \,i\, d\theta \, \frac{d\log x}{dt} dt\,  \frac{d\log y}{ds} ds\) (\partial_t,\partial_s,\partial_\theta)  \\
	& = -k \, \frac{d\log x}{dt} \,  (\log y_j(x) - \log y_i(x) + 2\pi i \, n) \,,
\end{split}
\ee
where we used the solution to (\ref{eq:u-theta-dependence}) discussed previously, and (\ref{eq:y-shape}). 
The special Lagrangian constraint (\ref{eq:sLag-condition}) translates\footnote{This is based on the fact that the volume form on $L$ is a positive real function of $(s,t,\theta)$. We just demand that restriction of $\Omega^{3,0}$ is proportional to the volum form up to a phase.} into the first-order differential equation
\be\label{eq:E-wall}
	\frac{d\log x}{dt} \,  (\log y_j(x) - \log y_i(x) + 2\pi i \, n) 
	\qquad \in \zeta\IR_{+}
\ee
as first observed in \cite{Klemm:1996bj} (here we assumed $k<0$ for the orientation of the $u$-circle, we fix this choice of convention).
This is the equation describing Exponential Networks, see \cite{Eager:2016yxd, Banerjee:2018syt}, or Spectral Networks \cite{Gaiotto:2012rg} in the case $n=0$.

As a check, the period of $\Omega^{3,0}$ along such a three-manifold is
\be
\begin{split}
	& \int_L \iota^* \Omega^{3,0} 
	=i \int_L \frac{dx}{x} \frac{dy}{y} d\log u 
	= \int_L \frac{dx}{x}  \frac{dy}{y}  \, d\theta 
	= 2\pi \, \int_{\IR}  \frac{d\log x}{dt} \int_{[0,1]}  \frac{d\log y}{ds}  ds  \\
	& = 2\pi \, \int_{\IR}  \frac{d\log x}{dt} (\log y_j(x) - \log y_i(x) + 2\pi i \, n)  \,.
\end{split}
\ee
There are three points we wish to stress about this computation. 
First, the phase of this period is $\zeta$, as expected by the special Lagrangian condition. Second, the computation of periods of $\Omega^{3,0}$ reduces to integration of an abelian differential (in the second line) along the path $x(t)\subset \IC^*$.
And last, but not least, the calibrating equation (\ref{eq:E-wall}) for the path $x(t)$ only depends on $n$, and \emph{not} on the choice of graded lift, corroborating earlier observations on the geometry over the $y$-plane.
This means that a compact special Lagrangian $L\subset X$ constructed locally from a solution to (\ref{eq:E-wall}) and glued together globally with boundary conditions of types \ref{item:S2xS1}-\ref{item:S3}-\ref{item:fancy} admits a whole $\IZ$-worth of graded lifts 
\be
	\tilde\pi^{-1}(L)  = \bigcup_{N\in \IZ} \tilde L_N 
\ee
to special Lagrangians on the  universal cover.

\subsection{Foliations}\label{sec:foliations}

In previous parts of this section we have discussed how a special Lagrangian $L$ fibered by $S^2$ can be built out of certain building blocks. Each building block consists of a segment $x(t)$ in $\IC^*$, above which we consider a family of two-spheres $S^2_{ij,n}$. The calibrating equation (\ref{eq:E-wall}) governs the shape of the segment, and we have argued above that the whole Lagrangian can be represented by this collection of segments.
In other words, the geometry of $L$ is rigid along $S^2$ and the whole moduli space of $L$ coincides with the moduli space of the system of segments $x(t)$.

In order to build a compact special Lagrangian, all that remains to be done is to find suitable pieces that glue together consistently, according to the rules outlined above.
This step involves passing from a \emph{local} description of $L$ to a  \emph{global} one: while the calibrating equation (\ref{eq:E-wall}) admits solutions for generic choices of $i,j,n,\zeta$ and of boundary conditions, only certain choices will lead to integral solutions $x(t)$ that close up globally into compact trajectories (possibly with junctions).

A systematic way to approach this problem goes as follows. First, we fix a choice of cycle $[L]\in H_3(X,\IZ)$. 
Then we define $\zeta = \exp ( i \arg \int_{[L]}\Omega )$ as the phase of the period of $\Omega$.
We then consider \emph{foliations} of $\IC^*$ induced by abelian differentials as in (\ref{eq:E-wall}) for all possible combinations of $i,j,n$. Concretely, we integrate the calibrating equation with boundary conditions $x_0$ taken at generic points in $\IC^*$ and study the resulting trajectories $x(t)$ with $x(t)=x_0$.\footnote{A global assignment of $i,j,n$ labels for trajectories is subject to a choice of trivialization over $\IC^*$, which we always assume fixed in this paper. See \cite{Banerjee:2018syt} for an extensive discussion of trivializations.}
Let $\phi_{ij,n}(\zeta)$ denote the foliation described by (\ref{eq:E-wall}).
In general, leaves of $\phi_{ij,n}(\zeta)$ will tend to $x=0,\infty$ or other singularities of the abelian differential (if present). 
Several examples will be given in Section \ref{sec:some-moduli-spaces}.

For the purpose of studying compact special Lagrangians, the interesting leaves are those that do \emph{not} end on any puncture. There are several possitiblities.
\begin{itemize}
\item 
Foliations of types $\phi_{ij,0}$ are special, since they admit \emph{critical leaves}, characterized by the fact that they run into branch points $x_b$ where $y_i(x_b)=y_j(x_b)$.
If a critical leaf has both endpoints at branch points, we obtain a Lagrangian of the type discussed in \ref{item:S3}, also shown in Figure \ref{fig:S3-fibered}.
\item 
Likewise, foliations of type $\phi_{ii,0}$ are also special since they admit compact leaves that run in circles around $x=0$. Indeed if $\zeta\in \IR$ the parametric form of these leaves is $x(t) = x_0\, \exp(\frac{\zeta}{2\pi i n}t)$. In this case we have a Lagrangian of the type discussed in \ref{item:S2xS1}.
\item
More generally, one may allow leaves to end at \emph{junctions}. A junction is a generic point $x \in \IC^*$ which we take as the boundary condition for three leaves of three different foliations $\phi_{ij,n}, \phi_{jk,m}$ and $\phi_{ki,s}$. Constraints on the labels appearing in these foliations have been discussed above in section \ref{sec:junctions}, also see \cite{Banerjee:2018syt}.
\end{itemize}

If complex moduli of $X$ are generic enough, then any compact special Lagrangian arising in this way will be of class $[L]$. If $\fM_L$ is nontrivial, the foliation will feature families of compact leaves, possibly including junctions.
In this way, studying foliations provides a direct handle on the moduli space of calibrated special Lagrangians fibered by two-spheres. As we will see, we will be able to deduce basic facts about the topology of $\fM_L$, and ultimately of $\CM_L$, by studying compact leaves of appropriate foliations. We illustrate this with a few examples next.

\section{Some moduli spaces}\label{sec:some-moduli-spaces}

We will now analyze directly the moduli spaces of special Lagrangians encoded by certain types of foliations. 
Here we will focus on topological properties of foliations and of their moduli spaces. 
For convenience we skip the process of finding numerical solutions of (\ref{eq:E-wall}) and discuss abstract  topological toy examples. Nevertheless, each of the examples we will discuss corresponds to an actual foliation by abelian differentials, we include relevant references for interested readers.

\subsection{The bi-critical leaf}

\begin{figure}[h!]
\begin{center}
\includegraphics[width=0.5\textwidth]{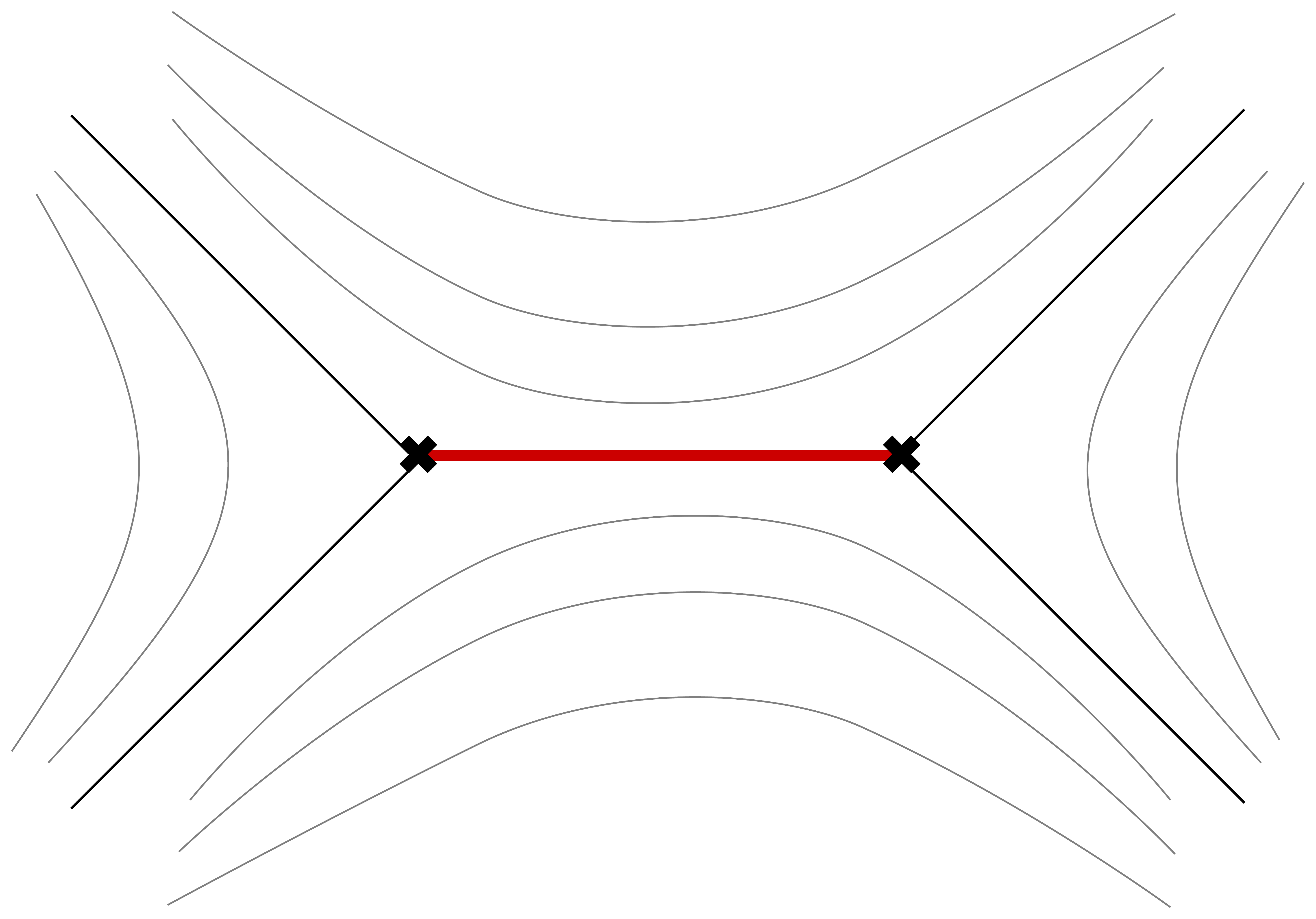}
\caption{Foliation with a bi-critical leaf.}
\label{fig:bicritical-leaf}
\end{center}
\end{figure}

Consider a foliation $\phi_{ij,0}$, with at least two branch points where $y_i=y_j$. An example is shown in Figure \ref{fig:bicritical-leaf}. 
There is a closed cycle $[L]\in H_3(X,\IZ)$ obtained by fibering a two-sphere $S^2_{ij,0}$ from one branch point to the other. 
A three-cycle in this class projects down to a segment of type $(ij,0)$ running from one branch point to the other, and the calibrating equation (\ref{eq:E-wall}) implies that this path must be a leaf of $\phi_{ij,0}$ with endpoints on \emph{both} branch points. We call this a bi-critical leaf.

\begin{figure}[h!]
\centering
\includegraphics[width=.4\textwidth]{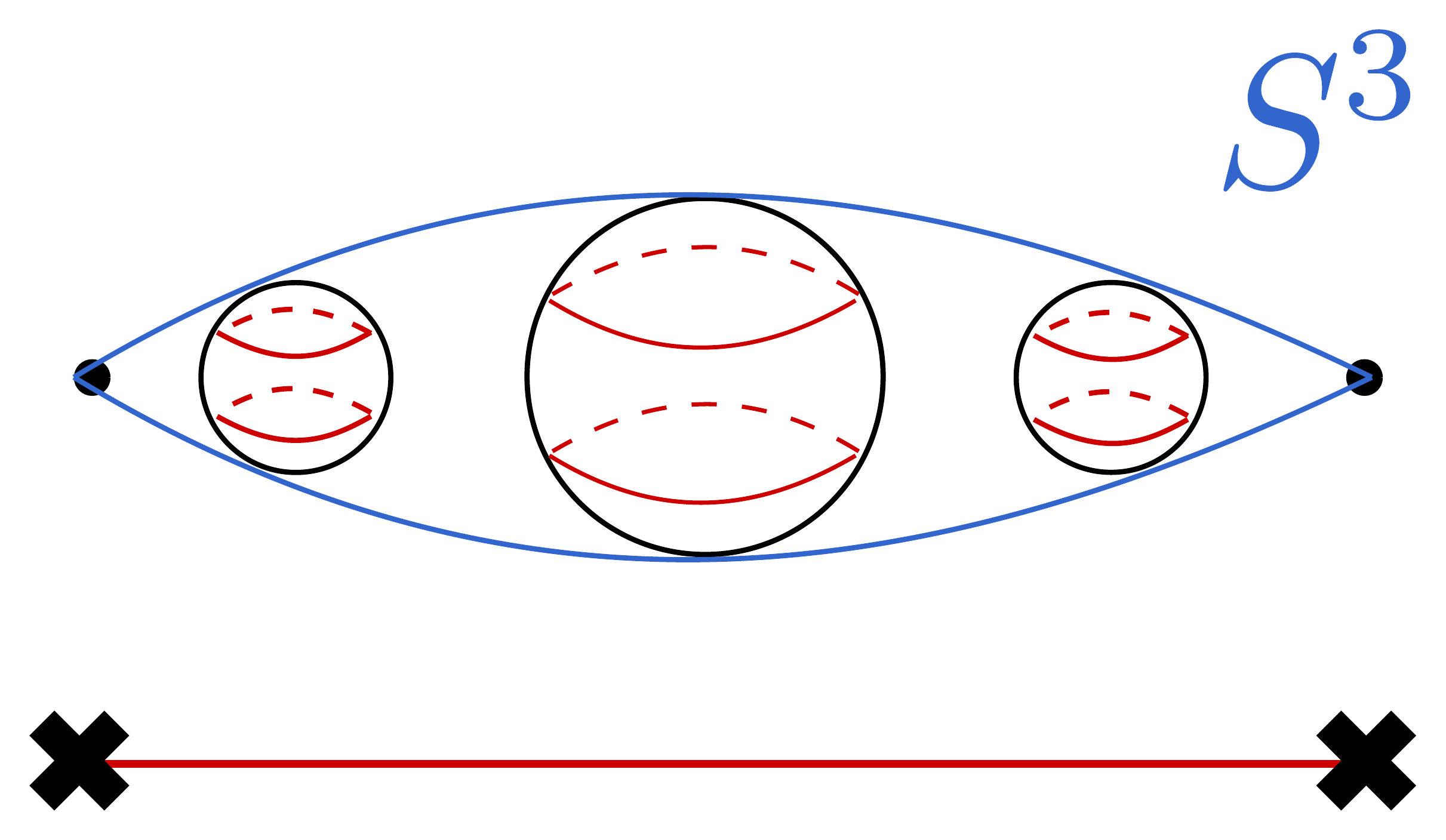}
\caption{The bi-critical leaf corresponds to a special Lagrangian $S^3$ (in blue) fibered by $S^2$  (in black) over the compact leaf. Each $S^2$ is itself fibered by an $S^1$ (red) in the $u$-plane, over a segment in the $\log y$-plane.}
\label{fig:S3-fibered}
\end{figure}

There is a unique leaf in this foliation that supports a compact special Lagrangian. Therefore the moduli space $\fM_L$ is a point in this case.
As a check, the topology of $L$ is also easy to read off: it is an $S^3$ fibered over a segment by $S^2$ that shrink at the endpoints, see Figure \ref{fig:S3-fibered}. Indeed $b_1(L)=0$ matches the dimension of the moduli space.
It follows that $H_1(L)$ is trivial, and therefore 
\be
	\CM_L = \{\text{pt}\}\qquad\Rightarrow\qquad \Omega(L) = (-1)^{\dim \CM_L} \chi(\CM_L) = 1
\ee

This kind of saddle, and the corresponding special Lagrangian, appears commonly in relation to hypermultiplets of $4d$ $\CN=2$ theories theories \cite{Klemm:1996bj, Gaiotto:2009hg, Mikhailov:1997jv, Shapere:1999xr}.
It also appears in the study of mirrors of $D2$ branes wrapping rigid $\IP^1$'s in toric Calabi-Yau threefolds, such as the conifold, and as hypermultiplets of  $5d$ $\CN=1$ theories \cite{Eager:2016yxd, Banerjee:2019apt}.

\subsection{Unbounded compact leaves}\label{sec:unbounded-circles}

\begin{figure}[h!]
\begin{center}
\includegraphics[width=0.5\textwidth]{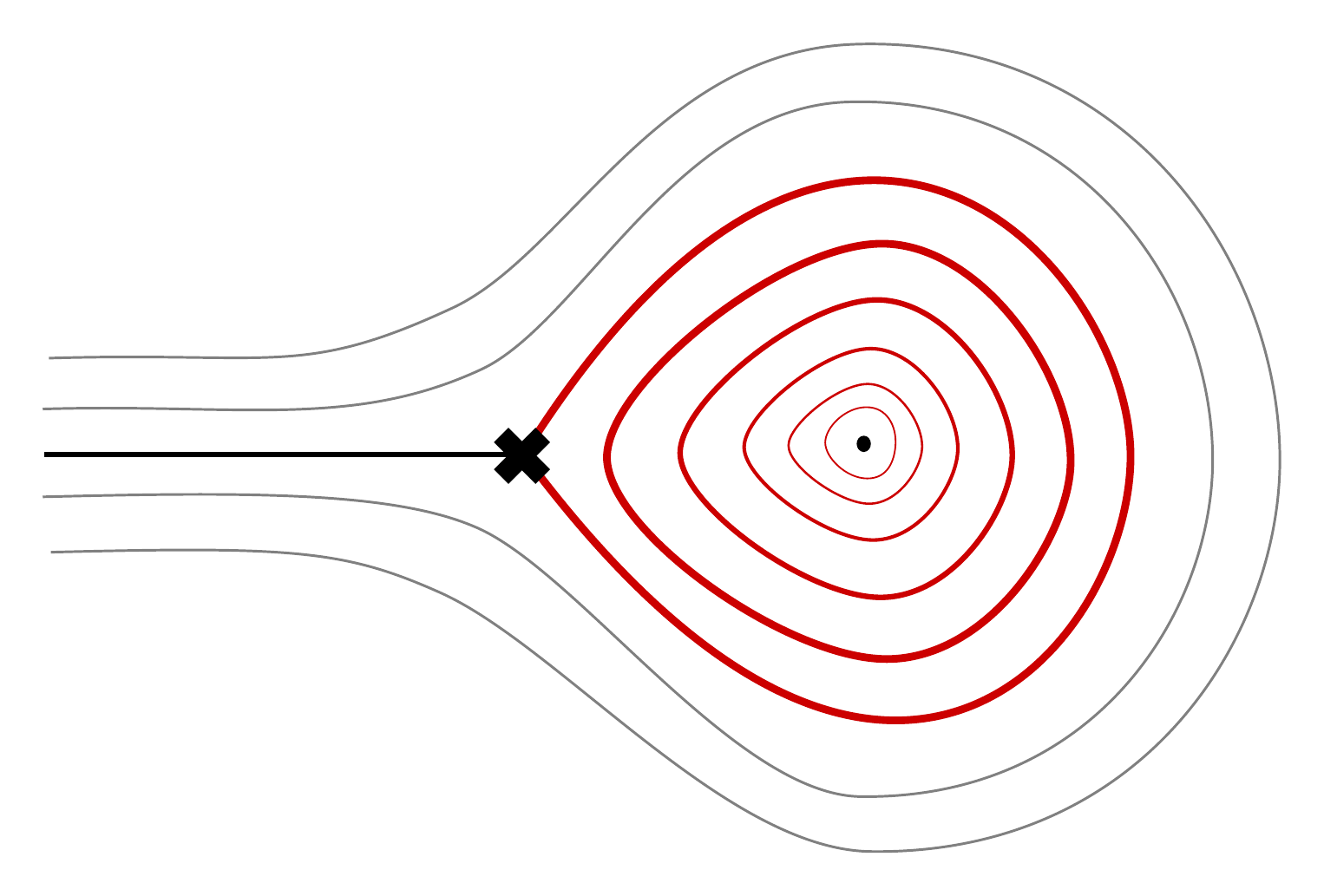}
\caption{Foliation with an unbounded family of compact leaves.}
\label{fig:unbounded-circles}
\end{center}
\end{figure}

Consider a foliation $\phi_{ij,0}$, with at least one branch point where $y_i=y_j$, and a puncture nearby where $\log y_i - \log y_j \sim 1/(x-x_0)$. An example is shown in Figure \ref{fig:unbounded-circles}. 
There is a closed cycle $[L]\in H_3(X,\IZ)$ obtained by fibering a two-sphere $S^2_{ij,0}$ along a closed path surrounding the puncture. 
A three-cycle in this class projects down to closed path of type $(ij,0)$ running around the puncture.
The calibrating equation (\ref{eq:E-wall}) implies that this path must be a leaf of $\phi_{ij,0}$.

There is a whole family of leaves in this foliation that support a compact special Lagrangian in class $[L]$, corresponding to circles of radius $0<r\leq r_0$. The circle of maximal radius corresponds to a bi-critical leaf of $\phi_{ij,0}$, with both endpoints on the branch point.
The moduli space is therefore
\be
	\fM_L \simeq \IR_{\geq 0}
\ee
with coordinate $\xi = \log r_0/r$.
We may check that $\dim_\IR \fM_L = b_1(L)$, in fact the generic leaf corresponds to a topology $L\simeq S^2\times S^1$. It follows that $H_1(L,\IZ)\simeq \IZ$ is generated by the class of the circle path on the $x$-plane. This generator disappears when $r=r_0$.
At $r_0$ the base circle attaches to the branch point, where 
the sphere fibered over the base circle $S^2_{ij,0}$ collapses. 
The topology of $L$ changes from $S^2\times S^1$ to $S^3$ with north and south pole identified.
Despite the identification of the poles, there is no well-defined holonomy for the flat Abelian connection on $L$ however.
This is because the tangent (and cotangent) space to the Lagrangian at the north pole of $S^3$ does not glue with the one at the south pole, as can be seen by the different slopes of segments attaching to the branch point.\footnote{A flat connection with holonomy would be gauge equivalent to $A\sim c\, dt$ for some constant $c\in \IR$, where $t\in 2\pi \IR/\IZ$ is the local coordinate on the circle. But now $dt$ is ill-defined at the branch point, therefore the connection and its holonomy become ill-defined.}

The $A$-brane moduli space is then an $S^1$-fibration over $\fM_L$, with $S^1$ shrinking at $r=r_0$ where the holonomy is ill-defined. This gives a manifold homeomorphic to $\IC$ with coordinate $z=\xi \, e^{i\oint A}$, from which we deduce the enumerative invariant
\be\label{eq:CP1-BPS-index}
	\CM_L \simeq \IC \qquad\Rightarrow\qquad \Omega(L) = (-1)^{\dim \CM_L} \chi(\CM_L) = -1\,.
\ee
Here we used compactly supported de Rham cohomology (\ref{eq:C-cohomology}) to compute $\chi(\IC)$.
Note that, had we chosen, say, $L^2$ cohomology, the result would have been different.

This kind of foliation and the corresponding family of special Lagrangians appears in the study of the $\IC\IP^1$ sigma model, and other 2d BPS states in coupled 2d-4d systems \cite{Gaiotto:2009hg, Gaiotto:2011tf}.
Our result, based on compactly supported de Rham cohomology, agrees with results from a field theoretic analysis for the $\IC\IP^1$ sigma-model \cite{Dorey:1998yh, Gaiotto:2011tf}.
This foliation also appears in the context of mirror symmetry: this family of Lagrangians is mirror to $D2$ branes supported on non-rigid $\IP^1$'s in toric Calabi-Yau threefolds, such as $\CO(0)\oplus \CO(-2) \to \IP^1$ \cite{Banerjee:2019apt}.

\subsection{Bounded compact leaves}\label{sec:bounded-circles}

\begin{figure}[h!]
\begin{center}
\includegraphics[width=0.5\textwidth]{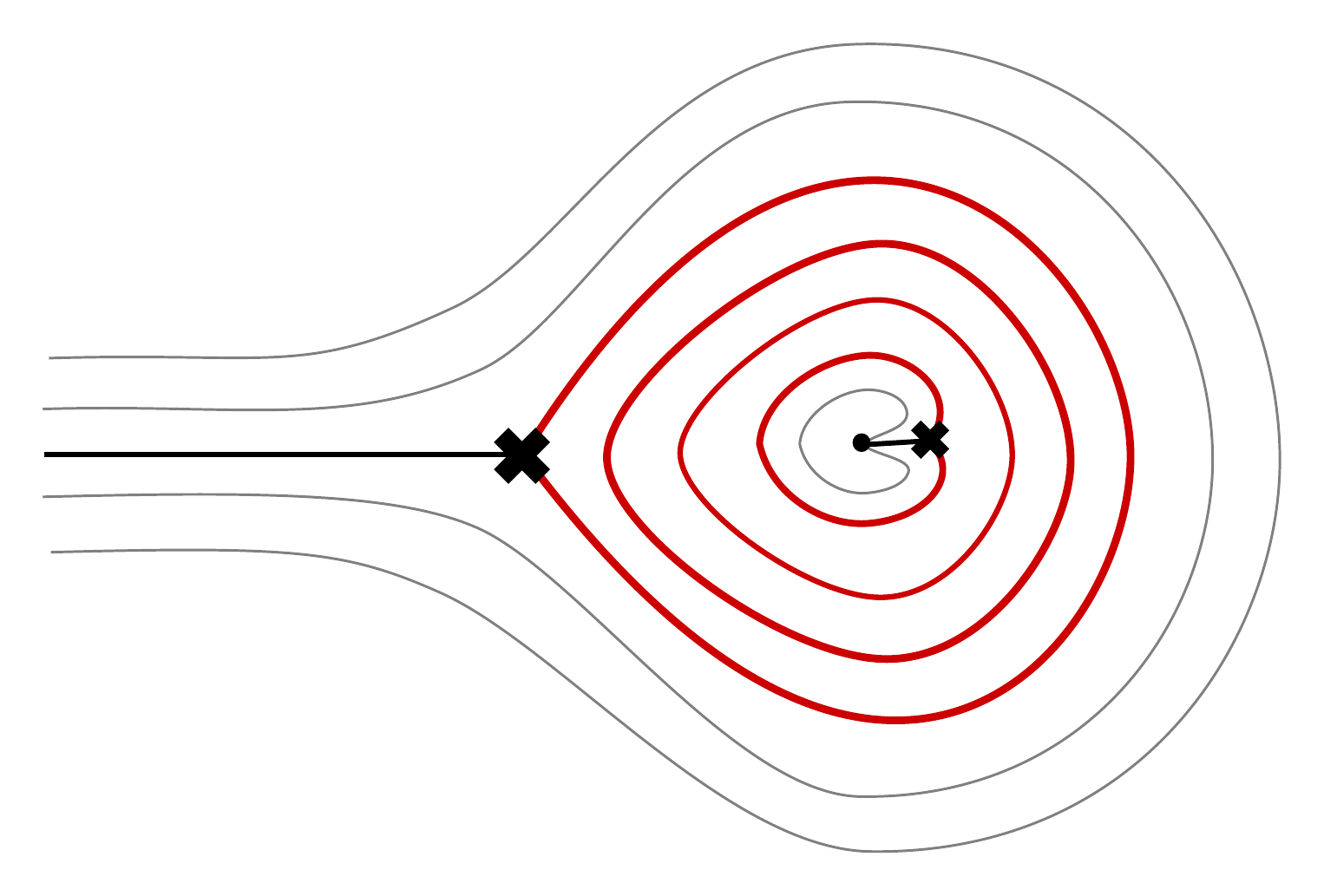}
\caption{Foliation with a bounded family of compact leaves.}
\label{fig:bounded-circles}
\end{center}
\end{figure}

A closely related type of foliation, also of type $\phi_{ij,0}$, involves two branch points where $y_i=y_j$, and a higher-order puncture somewhere between them, see Figure \ref{fig:bounded-circles}. 
Just as in the previous example, there is a closed cycle $[L]\in H_3(X,\IZ)$ obtained by fibering a two-sphere $S^2_{ij,0}$ along a closed path of type $(ij,0)$ surrounding the puncture. 
The calibrating equation (\ref{eq:E-wall}) implies that this path must be a leaf of $\phi_{ij,0}$. 

Again one finds a whole family of compact leaves in this foliation supporting a compact special Lagrangian in class $[L]$. However this time the circles have radius $r_0\leq r\leq r_1$. 
The circle of maximal radius corresponds to a bi-critical leaf attached to one branch point, and the one of minimal radius corresponds to a bi-critical leaf attached to the other branch point.
The moduli space is therefore
\be
	\fM_L \simeq [r_0,r_1]
\ee
This again matches with the expectation $\dim_\IR \fM_L = b_1(L)$, since the topology is again $L\simeq S^2\times S^1$. 
The homology generator disappears when $r=r_0$ and $r=r_1$, therefore the $A$-brane moduli space is
\be\label{eq:VM-BPS-index}
	\CM_L \simeq \IP^1 \qquad\Rightarrow\qquad \Omega(L) = (-1)^{\dim \CM_L} \chi(\CM_L) = -2
\ee

This kind of saddle, and the corresponding special Lagrangian, appears commonly in relation to vector multiplets of $4d$ $\CN=2$ theories theories \cite{Klemm:1996bj, Gaiotto:2009hg, Mikhailov:1997jv}.
It also appears in the study of mirrors of $D2$ branes wrapping  $\IP^1$'s in certain toric Calabi-Yau threefolds, such as $\CO(-2,-2)\to\IP^1\times \IP^1$ \cite{Banerjee:2020moh}.

\subsection{A junction with critical leaves}

\begin{figure}[h!]
\begin{center}
\includegraphics[width=0.5\textwidth]{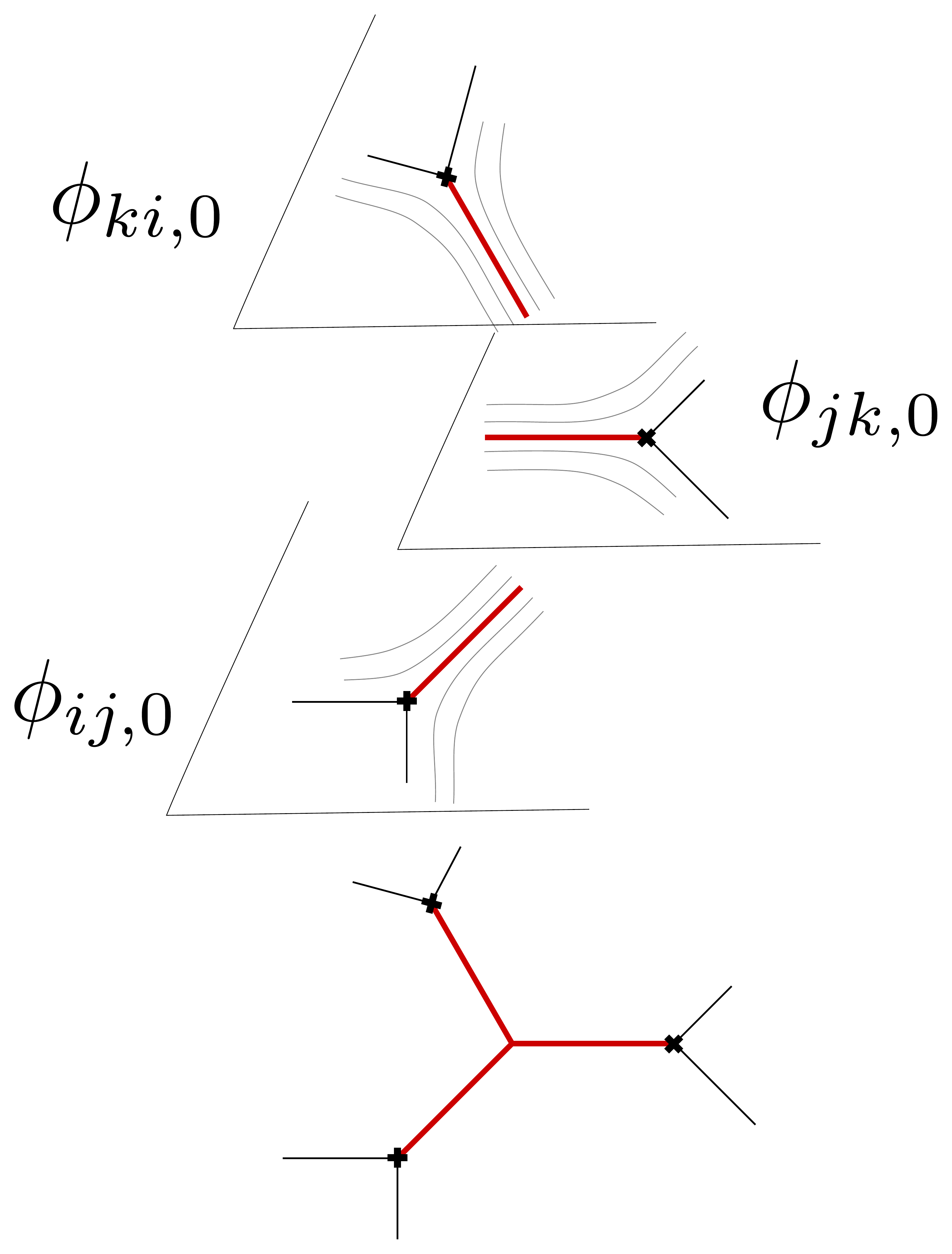}
\caption{A junction of three critical leaves of three different foliations.}
\label{fig:tri-junction}
\end{center}
\end{figure}

As the first example with a junction, let us consider three foliations $\phi_{ij,0}, \phi_{jk,0}$ and $\phi_{ki,0}$, with at least one branch point each (namely, a branch point where $y_i=y_j$ for $\phi_{ij,0}$ and so on).
An example is shown in Figure \ref{fig:tri-junction}. 
There is a closed cycle $[L]\in H_3(X,\IZ)$ obtained by as the union of three 3-balls glued along a junction like the one from Figure \ref{fig:junction}.
Consider a path of type $(ij,0)$ starting from the $ij$ branch point and ending at the junction, with an $S^2_{ij,0}$ fibered over it. Since the two-sphere shrinks at the branch point, the resulting 3-manifold has topology $B^3$.
Similarly consider three-balls fibered over paths from the two other branch points to the junction.
The calibrating equation (\ref{eq:E-wall}) implies the three paths must be leaves of respective foliations $\phi_{ij,0}, \phi_{jk,0}$ and $\phi_{ki,0}$.

There is a unique leaf in each of the three foliations that combines with the others to support a compact special Lagrangian in class $[L]$. These are the critical leaves that emanate from the respective branch points.
The moduli space is therefore a point $\fM_L = \{\text{pt}\}$. This agrees with the fact that the three $B^3$'s glue together into a three-sphere topology $L\simeq S^3$, which has $b_1=0$.
The $A$-brane moduli space is also a point
\be
	\CM_L \simeq \{\text{pt}\} \qquad\Rightarrow\qquad \Omega(L) = (-1)^{\dim \CM_L} \chi(\CM_L) = 1
\ee
This kind of foliation can be found in the study of hypermultiplets in higher-rank 4d $\CN=2$ theories of class $\CS$ \cite{Gaiotto:2012rg}.

\subsection{Sliding junctions}\label{sec:sliding-junctions}

\begin{figure}[h!]
\begin{center}
\includegraphics[width=0.85\textwidth]{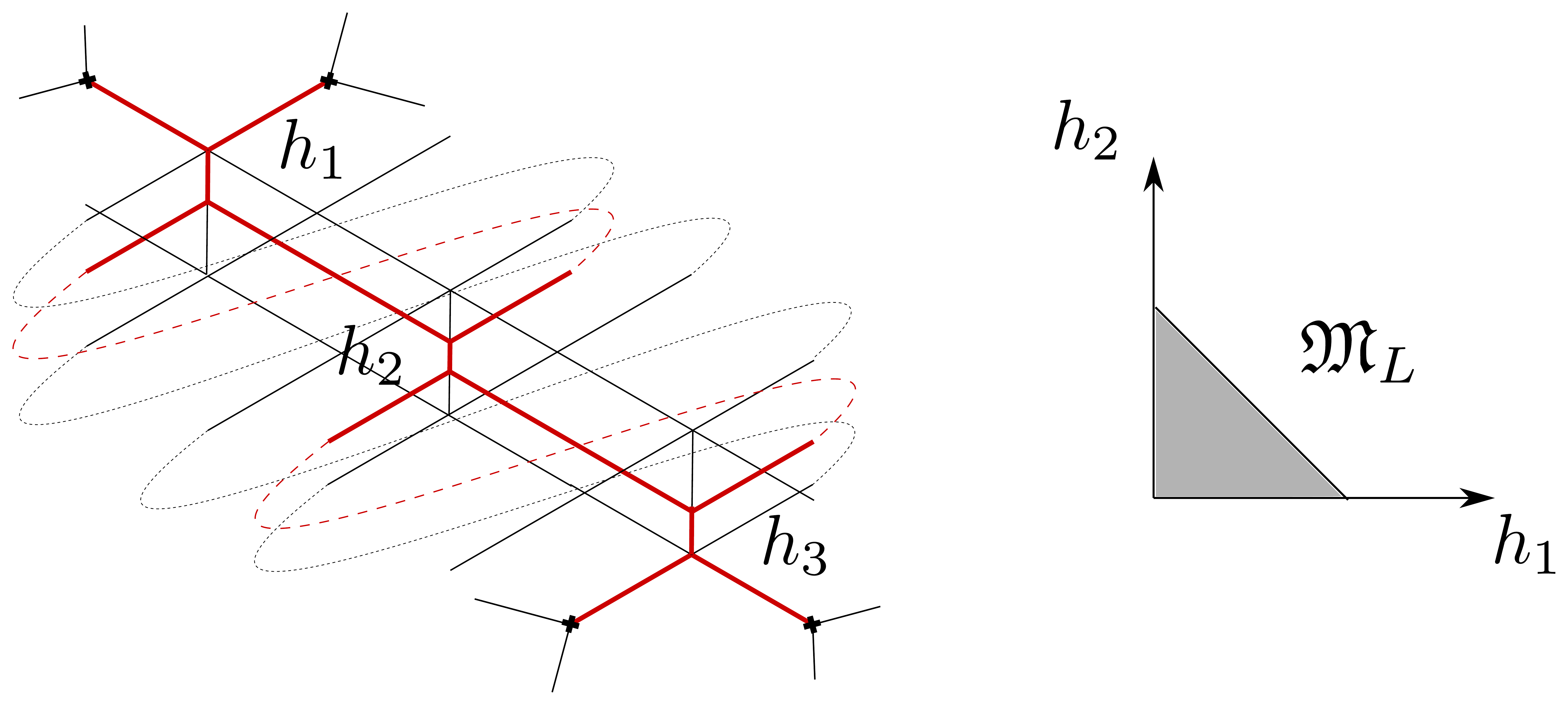}
\caption{Multiple junctions joining leaves of three different foliations. }
\label{fig:herds}
\end{center}
\end{figure}

A more interesting example with junctions involves considering multiple ones at the same time.
Again let us consider three foliations $\phi_{ij,0}, \phi_{jk,0}$ and $\phi_{ki,0}$, with at least two branch points of type $ij$ and at least two branch points of type $jk$.
An example is shown in Figure \ref{fig:herds}, the picture is drawn on a cylinder. 

Above the red paths coming into each junction, we have spheres of three different types: $S^2_{ij,0}, S^2_{jk,0}$ and $S^2_{ki,0}$ respectively.
If a path ends on a branch point, the resulting 3-manifold has the topology of a three-ball $B^3$. Instead if a path has both endpoints on junctions, the resulting 3-manifold has topology $S^2\times I$. All pieces glue together at junctions of the type shown in Figure \ref{fig:junction}. The result is a closed three-manifold in class $[L]\in H_3(X,\IZ)$.
The calibrating equation (\ref{eq:E-wall}) implies each of the paths underlying $L$ must be leaves of respective foliations $\phi_{ij,0}, \phi_{jk,0}$ and $\phi_{ki,0}$. 

There is a whole family of (systems of) leaves that join together at junctions in this way.
The family is parameterized by the heights $h_1, h_2, h_3$ of the three vertical segments in Figure \ref{fig:herds}.
Each segment can extend or shrink, while keeping angles of attaching segments unchanged.
The latter condition implies that when a segment shrinks, the others must extend, and vice versa.
Overall the moduli space is described by the condition that, in a suitable normalization
\be
	h_1+h_2+h_3 = 1\,, \qquad h_i \geq 0\,.
\ee
This describes a 2-simplex $\Delta^2$, also shown in Figure \ref{fig:herds}.
The moduli space is therefore  $\fM_L = \Delta^2$.
As a check, it is not hard to see that $b_1(L)=2$: there are two non-trivial cycles stretching (roughly) horizontally in Figure \ref{fig:herds}. The first cycle bounces off the lower end of segment $h_1$, and off the upper end of $h_2$, going from left to right, and then back on the other side of the cylinder. 
The second one does the same, bounding off the lower end of $h_2$ and the upper end of $h_3$.
The first cycle degenerates when $h_1=0$ or $h_2=0$. The second cycle degenerates if $h_2=0$ or if $h_3=0$.
 
The $A$-brane moduli space is a $T^2$ fibration over $\Delta^2$, with $T^2$ degenerating to a circle on the edges, and to a point at the vertices. This is the toric description of $\IC\IP^2$
\be\label{eq:3-herd-BPS-index}
	\CM_L \simeq \IC\IP^2 \qquad\Rightarrow\qquad \Omega(L) = (-1)^{\dim \CM_L} \chi(\CM_L) = 3
\ee

This kind of foliation can be found in the study of wild BPS states in higher-rank 4d $\CN=2$ theories of class $\CS$ \cite{Galakhov:2013oja}.\footnote{As often happens, there may be different families of foliations with isomorphic moduli spaces. In fact the same moduli space as for the 3-herd, namely $\IC\IP^2$ was observed to arise in the context of exponential networks for the mirror of local $\IP^2$ in \cite[Figure 30]{Eager:2016yxd}.}
In fact, it is closely related to the construct of $k$-herds, to which we will return later. There is a generalization of the above family of Lagrangians, parameterized by $h_1,\dots, h_k$ internal vertical segments. The moduli space $\fM_L$ in that case if the $k-1$-simplex $\Delta^{k-1}$. The $A$-brane moduli space is then a $T^{k-1}$ fibration over $\Delta^{k-1}$, where $T^{k-1}$ degenerates to $T^{k-1-l}$ on boundaries of codimension $l$. This is the toric description of $\IC\IP^{k-1}$. Therefore for an example with $2k$ junctions we have
\be\label{eq:k-herd-BPS-index}
	\fM_{L} \simeq \Delta^{k-1}\,,
	\qquad
	\CM_{L} \simeq \IC\IP^{k-1}\,,
	\qquad
	\Omega(L) = (-1)^{k-1}k \,.
\ee
As a check, we observe that BPS states of $k$-herds correspond to representation of Kronecker quivers with $k$ arrows, and dimension vectors $(1,1)$ \cite{Galakhov:2013oja}. These quiver representation varieties coincide exactly with $\CM_{L}$, for appropriate choice of stability data \cite{Denef:2002ru, 2003InMat.152..349R}.

\subsection{SYZ fibers}\label{sec:SYZ-fibers}

The last example we are going to discuss, is the first one where we consider foliations with nontrivial shifts of the logarithmic branch in the abelian differential, namely $\phi_{ij,n}$ with $n\neq 0$.
These arise, for instance, in the study of special Lagrangians arising as fibers of the SYZ-fibration, namely $L\simeq T^3$.
Note that a smooth $T^3$ does not admit an $S^2$-fibration, therefore the upcoming discussion will require a certain extension of the ideas from section \ref{sec:sLag-moduli-foliations}.
By SYZ mirror symmetry, the moduli space of an $A$-brane wrapping a $T^3$ fiber should correspond to the moduli space of a $D0$ on the mirror $X^\vee$, namely we expect $\CM_L \simeq X^\vee$.

\subsubsection{Smooth fibers}

Consider a Lagrangian cycle parameterized by $(t,s,\theta)$ as follows
\be\label{eq:smooth-T3}
	x(t) = x_0 e^{i t}\,,
	\qquad
	y(s) = y_0 e^{i s}\,,
	\qquad
	u(\theta) = u_0 e^{i \theta}\,.
\ee
Let $y_i(x)$ denote roots of $F(x,y)=0$ at $x\in \IC^*$. Also let $\tilde x_j$ denote any punctures in the $x$-plane, corresponding to roots of $F(x,0)=0$ or $F(x,\infty)=0$.
Choose $|x_0|< \min_j |\tilde x_j|$ to define a sufficiently small circle in the $x$-plane. Likewise choose $|y_0| < \min_{t,i} |y_i(x(t))|$ fixing a sufficiently small circle in the $y$-plane.
Then the $T^2$ parameterized by $(x(t), y(s))$ never crosses the locus $F(x,y)=0$, and the conic $uv=F(x,y)$ never degenerates, when restricted to this $T^2$. 
Thus the $u$-circle never shrinks, and gives $L$ the overall topology of $T^3$.
It is straightforward to check that $\iota^*\Omega = dt \wedge ds\wedge d\theta$, and therefore the special Lagrangian condition (\ref{eq:sLag-condition}) is satisfied with $\zeta=1$.
By McLean's theorem \cite{mclean1998deformations} the moduli space of $L$ has dimension $\dim_\IR \fM_L = b_1(L) = 3$. The three moduli correspond to the radii $|x_0|, |y_0|, |u_0|$.

\subsubsection{Degenerate fibers}

The Lagrangian $T^3$ that we just described is only one possible choice of special Lagrangian in class $[L]$.
It does not belong to the class of examples discussed in section \ref{sec:sLag-moduli-foliations}, since $T^3$ does not admit an $S^2$-fibration. 
However varying the moduli of $L$ we may run into a locus on $\fM_L$ where $L$ degenerates and admits an $S^2$-fibration. When this happens, that sub-locus of $L$ can be sometimes studied using foliations.

\begin{figure}[h!]
\begin{center}
\includegraphics[width=0.5\textwidth]{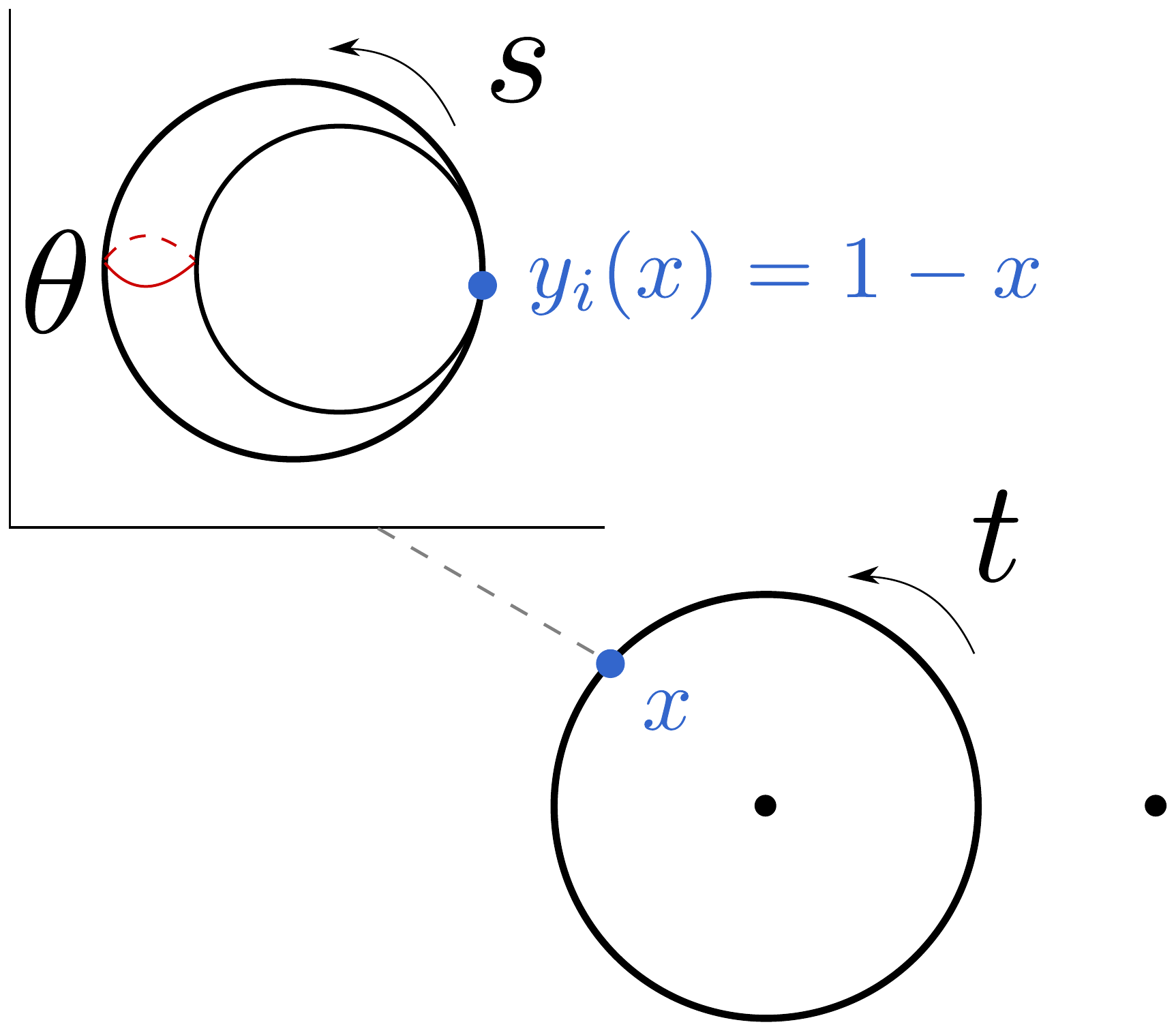}
\caption{Compact leaves of $\phi_{ii,1}$ in the $x$-plane are parameterized by $t$. 
Black dots denote punctures at $x=0$ and $x=1$.
Shown in the top-left is the $y$-plane above $x(t)$. The Lagrangian projects to a circle parameterized $y(s)$ that goes through $y_i(x)$.
There is an additional circle in the $u$-plane parameterized by $\theta$ that fibers over the $y$-circle, which pinches in correspondence of $y_i(x)$.}
\label{fig:ii-1-S2}
\end{center}
\end{figure}

To illustrate this with a simple example, consider $F(x,y)=1-y-x$, corresponding to the Hori-Vafa mirror of $\IC^3$.
Here we can take a Lagrangian parameterized as follows
\be\label{eq:ii-1-sLag}
	x(t) = x_0 e^{i t}\,,
	\qquad
	y(t,s) = (1-x(t)) e^{i s}\,,
	\qquad
	u(s,t,\theta) = u_0(t,s) e^{i \theta}\,.
\ee
Now we have a fixed circle in the $x$-plane parameterized by $t$, but over it we fiber a whole family of $y$-circles with varying radius $|1-x(t)|$. Such $y$-circles are paramterized by $s$, and moreover the $y$-circle intersects $\Sigma$ precisely at $s=0$, for each $t$.
In turn this means that the $u$-circle fibers continuously over $s\neq 0$ shrinking at $s=0$. 
So at fixed $t$, coordinates $(s,\theta)$ parameterize a $T^2$ with a cycle pinching above point $y(s=0)$. 
This can be viewed as a two-sphere with north and south poles identified, see Figure \ref{fig:ii-1-S2}.
This topology has $b_1(L)=2$, with one circle parameterized by $t$ and one by $s$, while the $u$-circle has become contractible on $S^2$.
Correspondingly \cite{mclean1998deformations}, the radius $|u_0(t,s)|$ is not a deformation modulus, but is determined by the special Lagrangian constraint as a function of $(t,s)$.

Noting that $y(s=0)= 1-x$ is the (only) sheet $y_i(x)$ of $\Sigma$, we identify the degenerate $T^2$ precisely with the two-sphere $S^2_{ii,1}$ introduced in (\ref{eq:S2-fibers}).
Thus the $x(t)$-circle must correspond to a leaf of the foliation $\phi_{ii,1}$. Indeed such a foliation is characterized by the differential equation (\ref{eq:E-wall}) which reduces to
\be\label{eq:ii-1-foliation-generic}
	\phi_{ii,1} : \qquad \frac{d\log x}{dt} \cdot 2\pi i \in \IR^+\,,
\ee
of which (\ref{eq:ii-1-sLag}) is indeed a solution.

The two deformations corresponding to $b_1(L)=2$ are the sizes of circles $x(t)$ and $y(t,s)$.
However the $y$-radius deformation is frozen by the choice to restrict to degenerate special Lagrangians of the specific form (\ref{eq:ii-1-sLag}). 
Going back to our derivation of (\ref{eq:E-wall}), recall that it was crucial that $\log y$ depended linearly on $s$, this is what we have chosen in (\ref{eq:ii-1-sLag}). A more general choice would have allowed for $s$-dependence of the $y$-circle radius (still demanding that $y(t,s=0)=1-x(t)$), but this would \emph{not} have led to a circle like (\ref{eq:ii-1-foliation-generic}) in the $x$-plane.
Taking into account the freezing of both the $y$-radius and the $u$-radius, we conclude that this type of foliations actually sees a codimension-two subspace of the moduli space of SYZ fibers, parameterized uniquely by radius $|x_0|$.

\subsubsection{SYZ fibers without a leaf representative}

The degenerate Lagrangian (\ref{eq:ii-1-sLag}) is closely related to the smooth fiber (\ref{eq:smooth-T3}). The main difference is that 
we have frozen two of the moduli, namely the freedom to shift the $y$-circle and $u$-circle radii.
Recall that $\CM_L\simeq X^\vee$ for SYZ fibers, and note that $\fM_L$ must therefore resemble the base of the $T^3$-fibration of $\IC^3$, namely the positive octant $(\IR_{\geq 0})^{\times 3}$ spanned by $\rho_i  =|z_i|$ for $i=1,2,3$.

Fixing two moduli may lead to a degeneration of the $T^3$ fiber of $\IC^3$ to a $T^2$ or to an $S^1$, constraining us respectively onto a 2-dimensional or a 1-dimensional slice of the base. 
When we considered the degenerate Lagrangian we fixed $|y|$ and $|u|$ moduli to certain functions of $(t,s,\theta)$, 
which moreover changed the topology of $L$ from $T^3$ to $S^1\times S^2/S^0$ ($S^2/S^0$ denotes a two-sphere with poles identified). 
The $u$-circle got pinched at $s=0$, inducing $b_1(L)$ to decrease from $3$ to $2$.
In the language of a $D0$ on $X^\vee\simeq \IC^3$ it means we have `hit a wall' where, say $\rho_1$ corresponding to variations of $|u|$, got fixed to zero.
In the example above we further imposed (by hand) a restriction on the $y$-radius, and for this reason we only saw a one-dimensional slice of this 2-dimensional subspace.

One may ask whether it is possible to explore more of the moduli space $\fM_L$, and perhaps see the other walls too. In particular, what about the locus corresponding to the boundary for the modulus $|x|$?
Given the symmetry of the curve $F(x,y)=1-y-x$ under exchange of $x,y$ we may simply consider
\be\label{eq:ii-1-sLag-flipped}
	x(t) = (1-y(s)) e^{i t}\,,
	\qquad
	y(s) = y_0 e^{i s}\,,
	\qquad
	u(\theta) = u_0(t,s) e^{i \theta}\,.
\ee
Now the $y$-circle has fixed radius, while it is the $x$-circle whose radius depends on $y$. 
This is a special Lagrangian, a point in $\fM_L$, with a pinched cycle corresponding to the $S^1$ paramterized by $\theta$. Hence $b_1(L)=2$ and this choice of Lagrangian corresponds to another wall in moduli space.

But one should note that this Lagrangian is not represented by a single leaf of the foliation in the $x$-plane: in fact for each $s\in S^1$ we have a different $x$-circle with radius $|1-y(s)|$.
This shows that single leaves of foliations in the $x$-plane cannot model special Lagrangians at generic points in the moduli space, when it comes to $A$-branes wrapping SYZ fibers.
This should not come as as surprise, since after all SYZ fibers have topology~$T^3$ and therefore evade the framework developed in section~\ref{sec:sLag-moduli-foliations} to study $S^2$-fibered special Lagrangians.

\subsection{Codimension-one strata for SYZ fibers}\label{sec:codimension-one-strata}

Despite the fact that foliations in the $x$-plane cannot capture the whole moduli space of $A$-branes wrapping SYZ fibers, foliations can still detect a real-codimension one (or higher) stratum, as illustrated by the example (\ref{eq:ii-1-sLag}).
Surprisingly, despite the lack of a global picture of the whole $\fM_L$, nonetheless foliations still contain enough information to compute the correct Euler characteristic of $\CM_L$. 
We will later explain this fact through the localization principle.

To set the stage for the main example in support of this claim, we consider a curve defined by the vanishing of
\be
	F(x,y) = y^2+y+x \,.
\ee
This is a two-sheeted cover of the $x$-plane $\IC^*$, and corresponds to the mirror curve of a toric brane in $\IC^3$ is a specific choice of framing, see \cite{Banerjee:2018syt} for a detailed analysis of its trivialization.
It is important to note that the puncture at $x=0$ lifts to two punctures on the curve, while the puncture at infinity lifts to a single puncture on the curve: this means there is a square-root branch cut starting from a branch point at $x=1/4$ and landing at $x=\infty$.

We label the two sheets by $i=\pm$. Then we consider foliations $\phi_{+-,n}$ and $\phi_{++,n}$. (The calibrating equation for $\phi_{--,n}$ coincides with the one for $\phi_{++,n}$) 
There is one nontrivial three-cycle in this geometry, corresponding the the SYZ fiber $[L]$, or equivalently the mirror of a D0 brane in $\IC^3$. 
Since the D0 central charge is real and positive, we shall study foliations defined by (\ref{eq:E-wall}) with $\zeta=1$.

As explained previously, foliations cannot probe the whole moduli space of SYZ fibers $\fM_L$, but only a codimension-one stratum. Here we will discuss this stratum following and expanding upon an analysis sketched in \cite{Eager:2016yxd}. 
Since we can only see part of the moduli space through foliations, we will not be able to compute $\Omega(L)$ here by applying the definition of Euler characteristic. We will explain in the next section how to overcome this difficulty without the need for any additional data.

\subsubsection{Circular leaves}

\begin{figure}[h!]
\begin{center}
\includegraphics[width=0.6\textwidth]{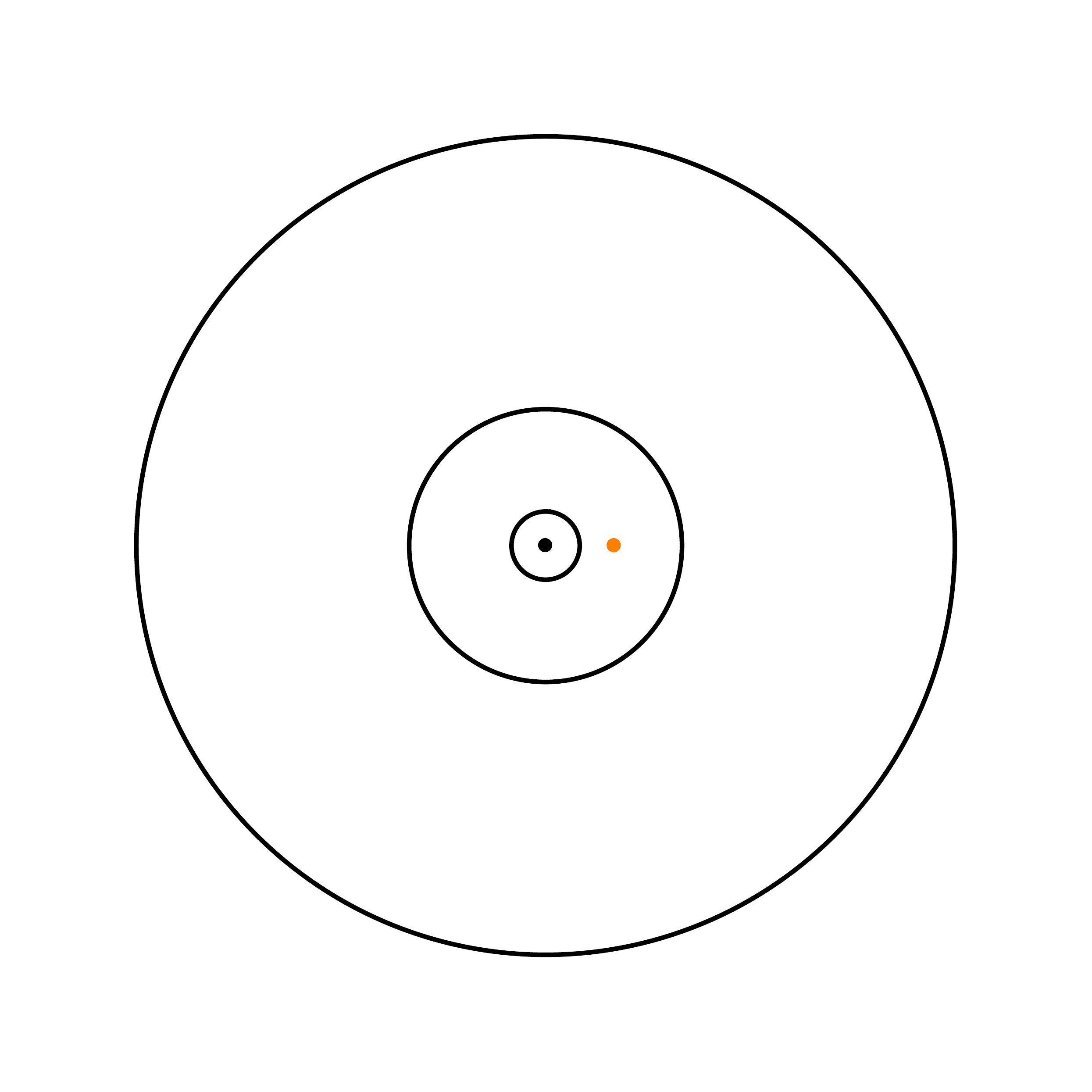}
\caption{Circular leaves of the foliation $\phi_{++,1}$. The black dot is the puncture at $x=0$, the yellow dot is the branch point at $x=1/4$. There is a square-root branch cut exchanging $\phi_{++,1}$ and $\phi_{--,1}$ running from $x=1/4$ to $\infty$. Circular leaves with radius $>1/4$ must wrap twice around to produce a closed 3-manifold fibered by $S^2_{\pm\pm,1}$.}
\label{fig:C3-circles}
\end{center}
\end{figure}

It is natural to begin with the degeneration of $T^3$ already discussed in (\ref{eq:ii-1-sLag}).
Here we study the foliation (\ref{eq:ii-1-foliation-generic}), which is independent of the specific form of $F(x,y)$. Leaves are circles centered at $x=0$, see Figure \ref{fig:C3-circles}

There is here one subtlety to take into account: if the radius of the circle $x(t)$ is $0<r<1/4$ then a leaf of type $(++,1)$ comes back to a leaf of type $(++,1)$ after a full turn.
On the other hand, if $r>1/4$ the leaf will cross the branch cut running between $x=1/4$ and $\infty$, and the leaf will come back of type $(--,1)$. Thus, for circles of radius $r>1/4$ the leaf must go around twice to come back to the same type. This is essential in order to obtaine a closed 3-cycle $L$, obtained by fibering an $S^2$ over the circle $x(t)$.

\subsubsection{Junction bubbling}

As explained before, the topology of the degenerate Lagrangians captured by $(ii,1)$-foliations, such as those in Figure \ref{fig:C3-circles}, is such that $b_1(L)=2$. One deformation is obviously the freedom to choose the radius of the $x$-circle. The second one is more subtle.
As it turns out, it is possible to turn on a topology-changing deformation on the $x$-plane, involving junctions. Here we discuss the relevant topologies and explain how they connect to the circular leaves of $\phi_{++,1}$ discussed so far.

\begin{figure}[h!]
\begin{center}
\includegraphics[width=0.3\textwidth]{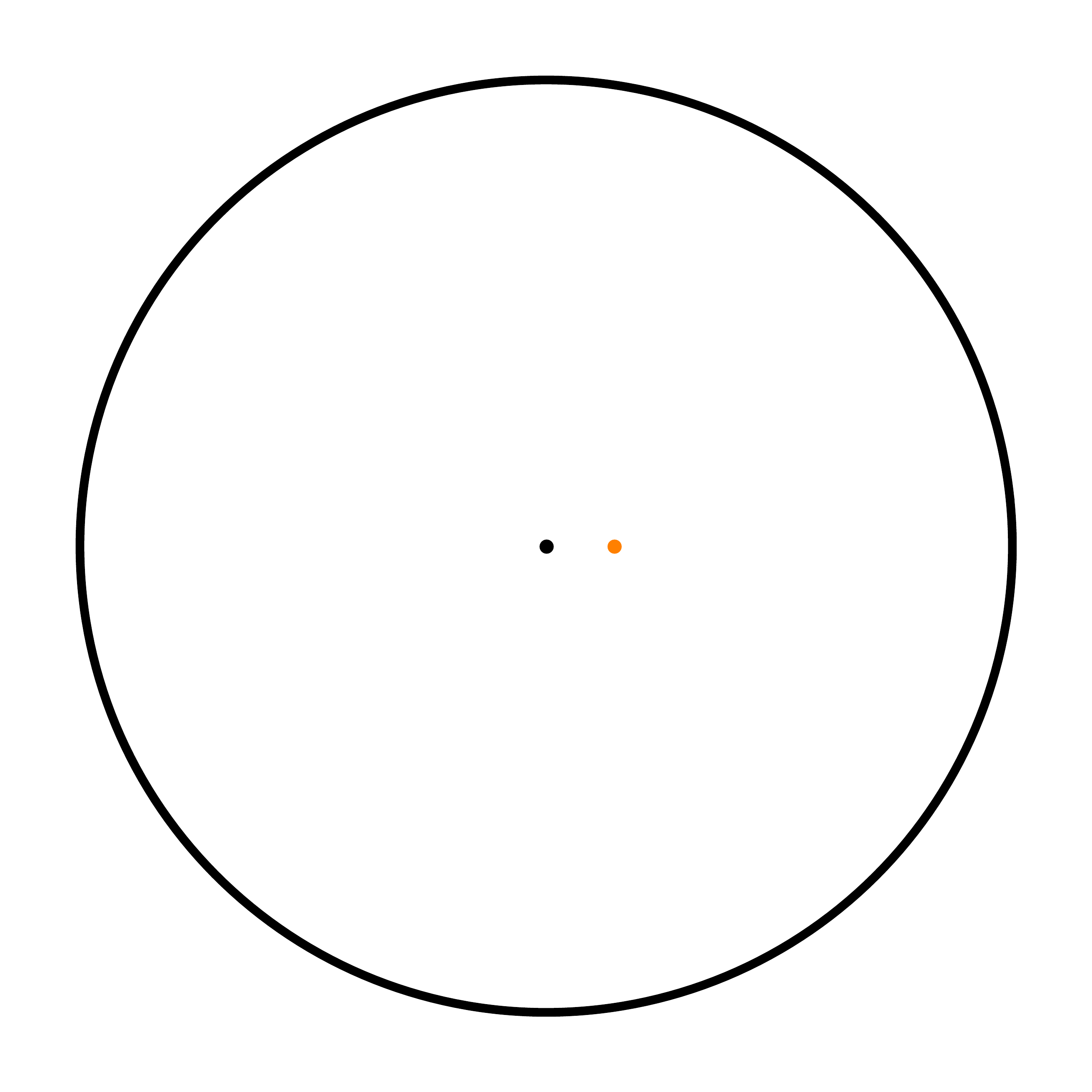}\hfill
\includegraphics[width=0.3\textwidth]{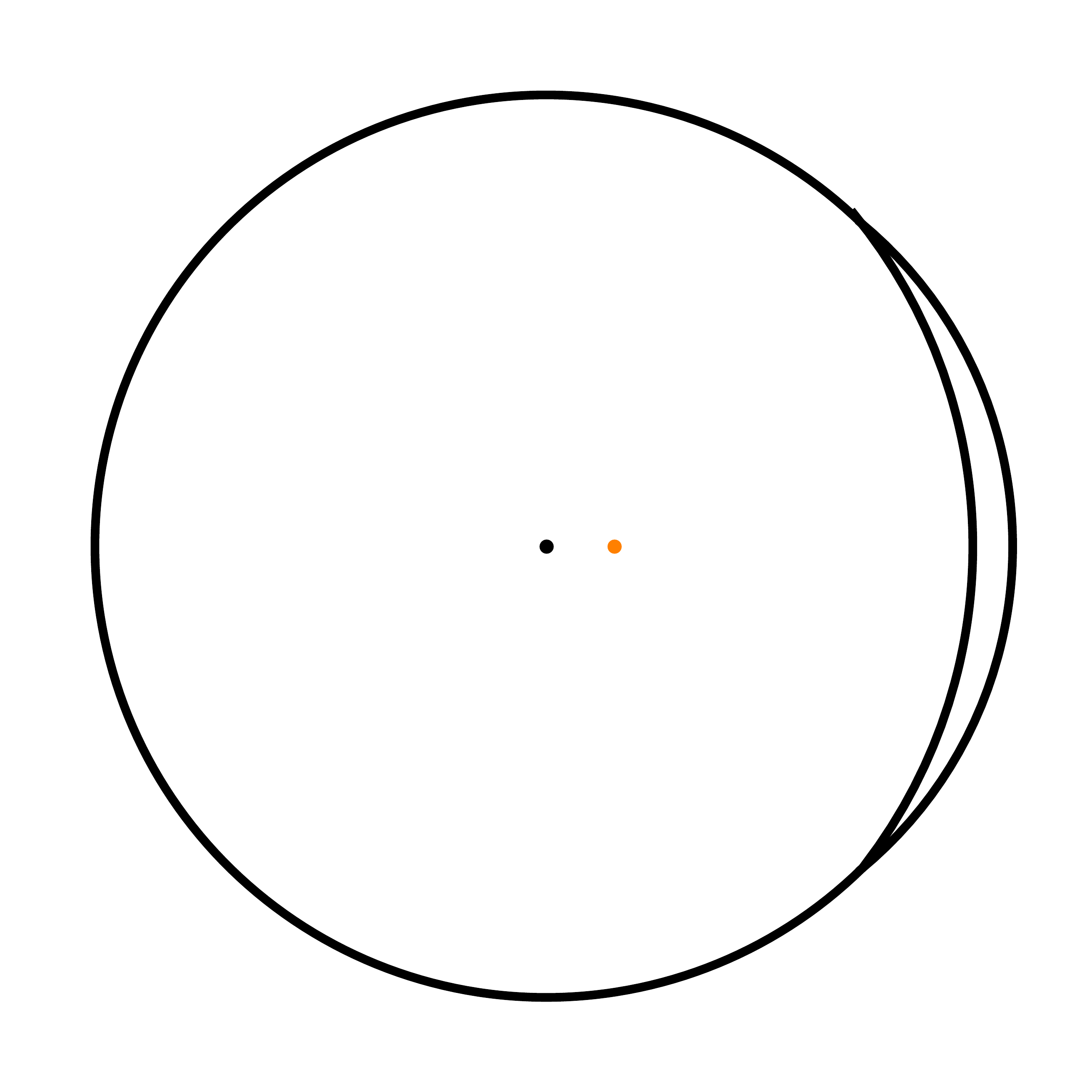}\hfill
\includegraphics[width=0.3\textwidth]{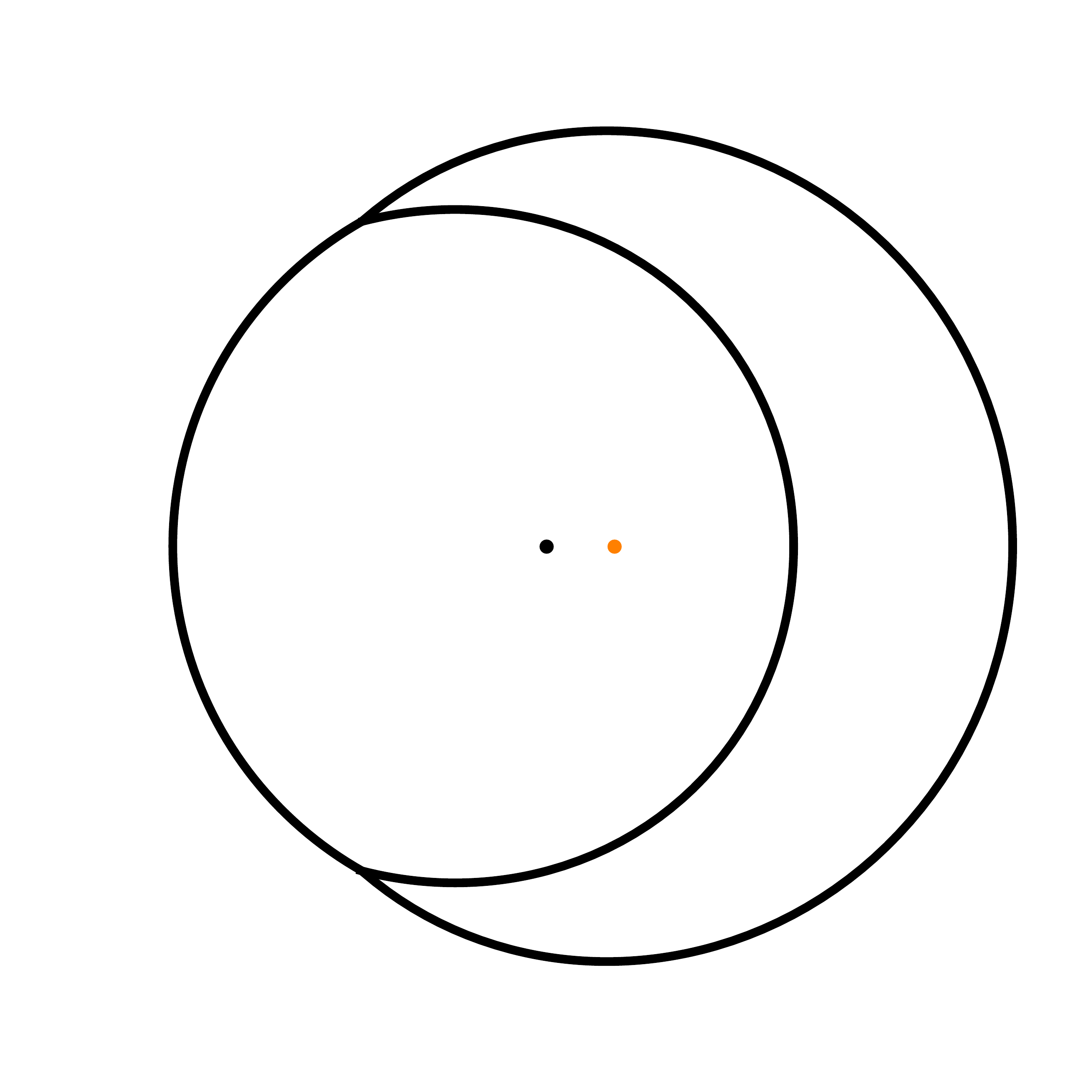}\\
\includegraphics[width=0.3\textwidth]{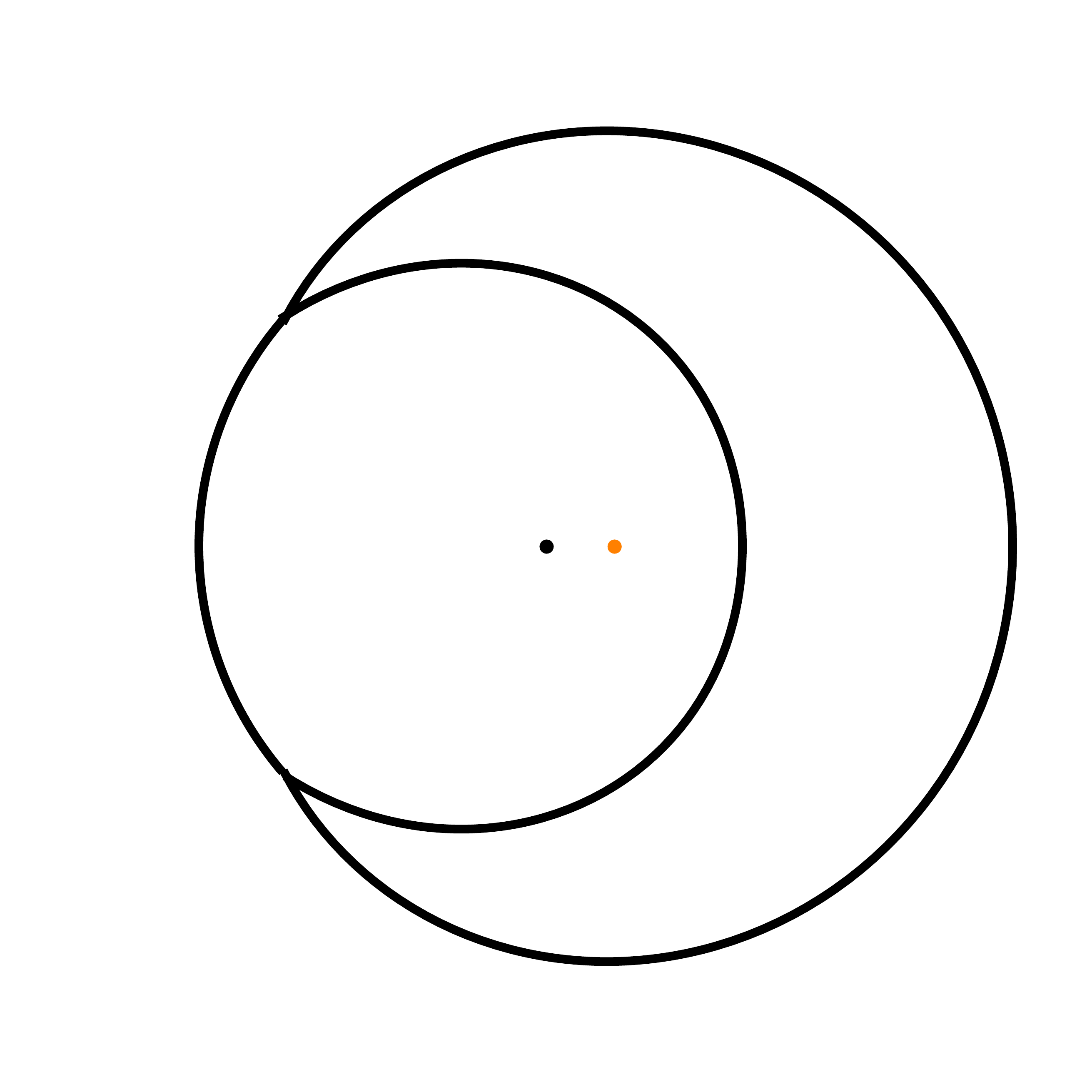}\hfill
\includegraphics[width=0.3\textwidth]{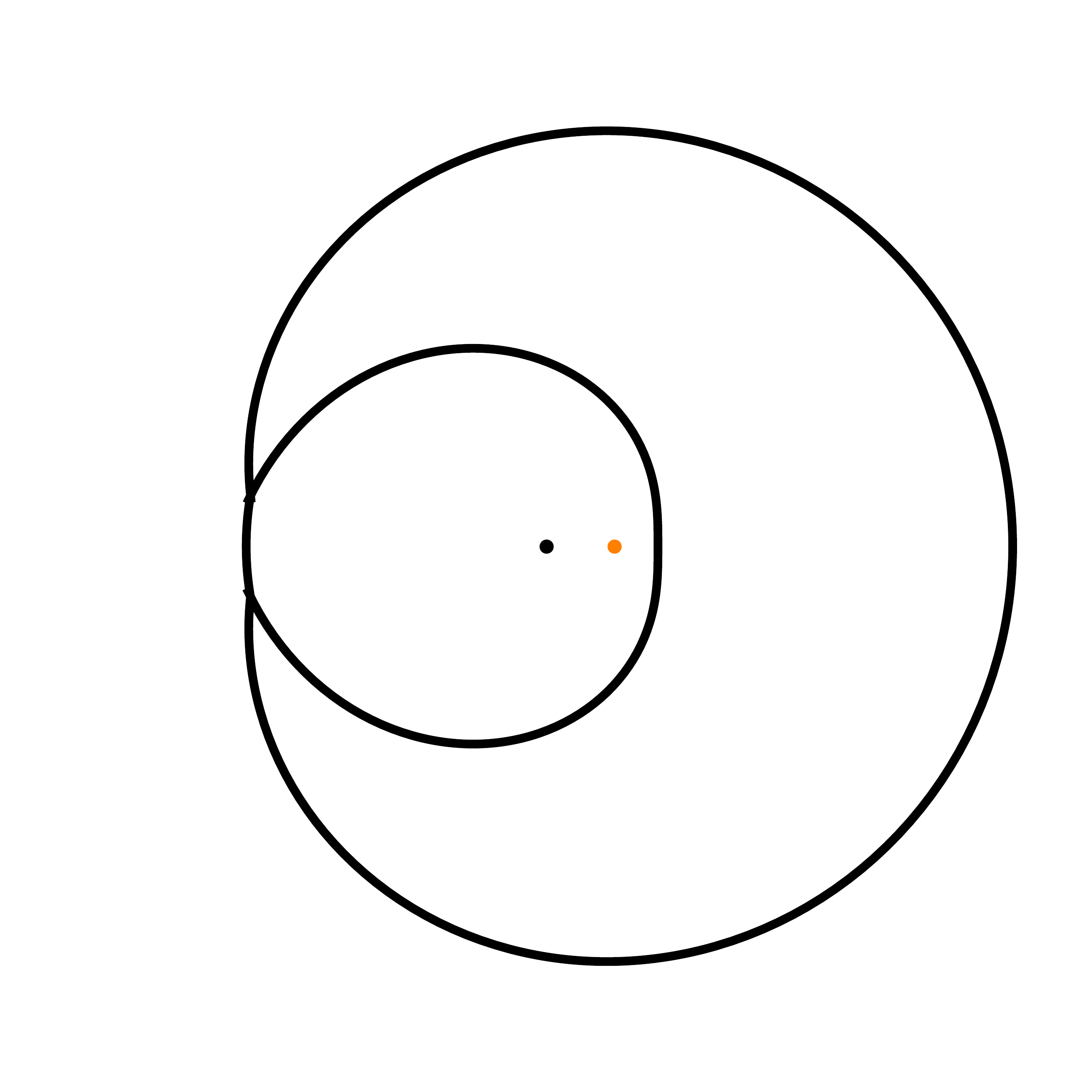}\hfill
\includegraphics[width=0.3\textwidth]{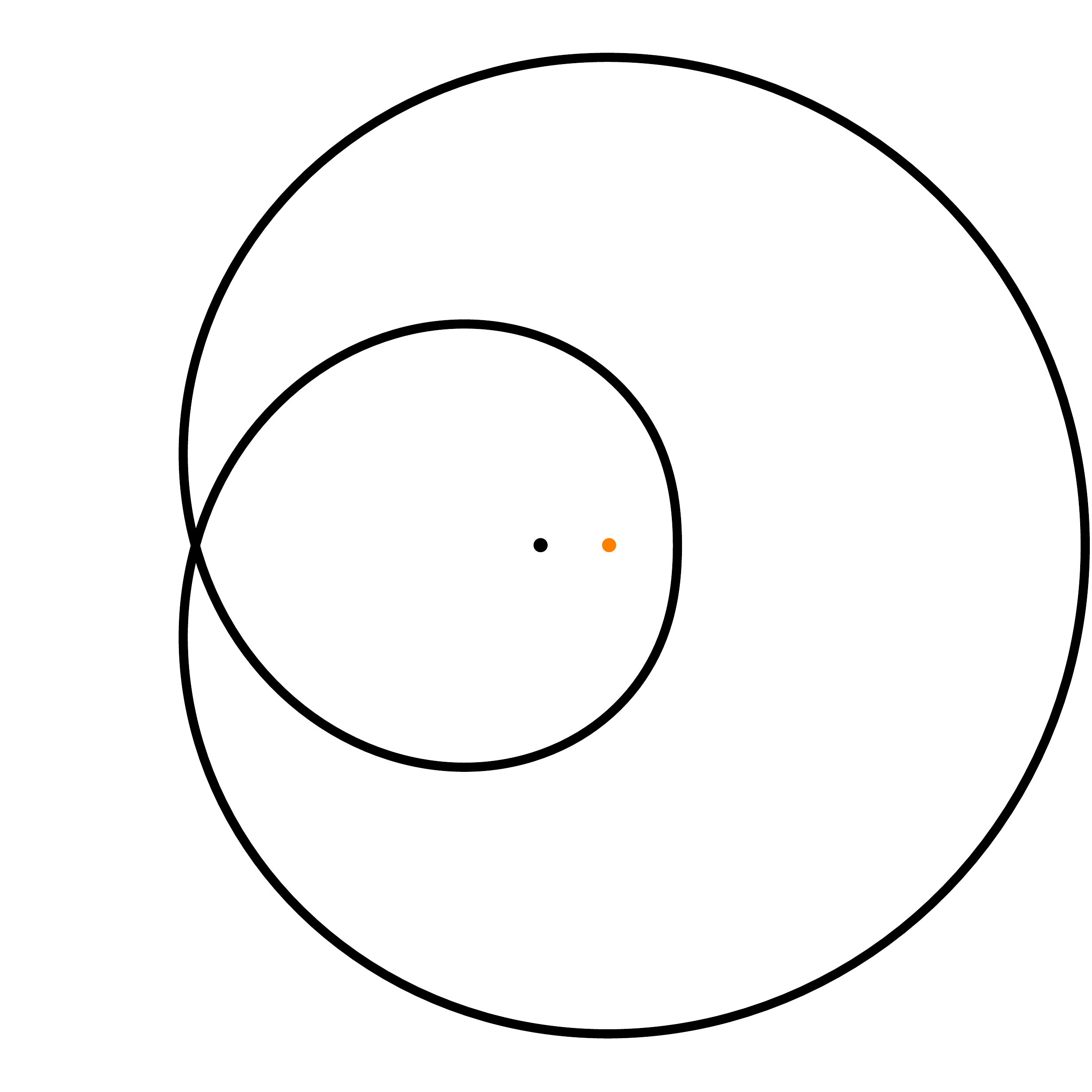}
\caption{A path in a codimension-one stratum of $\fM_L$, connecting different calibrated Lagrangians in class $[L]$. 
The top-left and bottom-right belong to different codimension-two strata. The bottom-right stratum is parameterized by overall size, and contains the codimension-three fixed point shown in Figure \ref{fig:C3-D0-fixed-point}.
}
\label{fig:circle-decay}
\end{center}
\end{figure}

The essential process, identified in \cite{Eager:2016yxd}, is the phenomenon by which a circular leaf of $\phi_{ii,1}$ may develop a pair of junctions. 
The process is detailed in Figure \ref{fig:circle-decay}.
When the pair of junctions bubbles up, we have a system with new leaves of types  $\phi_{ij,0}$ and $\phi_{ji,1}$. The moduli space is 2-dimensional, in agreement with $b_1(L)=2$.

The two moduli can be described as follows. Consider the $\phi_{ij,0}$ leaves, there is a modulus corresponding to the coordinate in leaf-space. There is another leaf-space modulus for $\phi_{ji,1}$. For any choice of these moduli, those leaves will intersect (they have different angles on $\IC^*$ according to (\ref{eq:E-wall})). The intersections always lie on the same circle centered at $x=0$, therefore one may always connect the intersections with a leaf of $\phi_{ii,1}$.

\section{Localization}\label{sec:localization}

In the previous section we have illustrated the use of foliations for the purpose of exploring the global topology of the moduli space of special Lagrangians $\fM_L$ and the associated moduli space of $A$-branes $\CM_L$.
While in certain cases it is possible to capture the global topology of $\CM_L$, and therefore compute the Euler characteristic $\chi(\CM_L)$, in other cases this is not possible.
A notable counterexample that we encountered in sections \ref{sec:SYZ-fibers} and \ref{sec:codimension-one-strata} is provided by $A$-branes wrapped on SYZ fibers.

Equivariant localization offers a way to sidestep the need to see the global structure of a manifold. The computation of topological invariants, such as the Euler characteristic, are reduced to the study of a finite number of points in the moduli space $\CM_L$, corresponding to fixed points of a certain $G$-action. In this section we explore how this idea can be applied to moduli spaces of $A$-branes.

\subsection{Equivariant fixed point formula for the Euler characteristic}

Let $M$ be a smooth manifold endowed with a smooth $G$-action by a Lie group $G$. 
In our setting we will be only concerned with torus actions by $G=T^m$.
Existence of a nontrivial $G$ action on a manifold $M$ carries implications on the topology of $M$.
Of particular interest is the space of $G$-orbits and their type, as classified by the respective stabilizer group.
One way to access this information is provided by equivariant cohomology, which we briefly review.\footnote{Here we follow \cite{hori2003mirror} in reviewing the Borel model of equivariant cohomology, which is suitable for stating the main result that we will need: the localization formula of Atiyah-Bott-Berline-Vergne \cite{duistermaat1982variation, berline1982classes, atiyah1984moment}. See \cite{Cordes:1994fc, Szabo:1996md, 2006math......7389V, Pestun:2016qko, Pestun:2016zxk} for pedagogical accounts of some other models of equivariant cohomology, and their applications to geometry and quantum field theory.}

The universal bundle of $G$ is a contractible space carrying a free $G$-action. This space is denoted $EG$ and is unique up to homotopy \cite{milnor1956construction, milnor1956construction-II}. The quotient space $EG/G$ is a smooth manifold called the classifying space of $G$, and denoted $BG$.
The name derives from the fact that any $G$-bundle on $M$ can be pulled back from $EG\to BG$ via a map $i:M\to BG$.
For the case $G=S^1$ the universal bundle is the infinite-dimensional sphere, and the classifying space is $\IC\IP^\infty$. For $T^m$ we simply take $BT^m = (BS^1)^{\times m}$.
We introduce the homotopy quotient
\be
	M_G = EG \times_G M 
\ee
defined as the quotient of the direct product $EG\times M$ by $(e,gm)\sim (eg,m)$ for all $g\in G$.
An important property of $M_G$ is that it admits a fibration over $M$, with \emph{different} fibers depending on the $G$-orbit.
More speficically, given a $G$-orbit $[Gm]$ through $m$, the corresponding fiber is the quotient $EG/\{g\in G; gm=m\}$ by the stabilizer group of the orbit. The two extreme cases are: $i)$ if $m$ belongs to a free orbit, the fiber is the whole $EG$, and $ii)$ if $m$ is a fixed point with stabilizer the whole $G$, then the fiber is $BG$. 
The $G$-equivariant cohomology of $M$ is defined as ordinary de Rham cohomology of $M_G$.
\be
	H_G^*(M) = H^*(M_G) \,.
\ee

Observe that if $G$ acts freely on all of $M$, then $M_G \simeq EG \times M/G$ fibers trivially over the space of $G$-orbits. Since $EG$ is contractible it follows that $H_G^*(M) = H^*(EG \times M/G) = H^*(M/G)$ in this case.
Conversely, for a trivial $G$-action we obtain $M_G\simeq BG\times M$, implying $H^*_G(M) = H^*(M)\times H^*(BG)$.
More generally there are orbits with different stabilizers, and equivariant cohomology may have a richer structure.
For example if $M=S^2$ and $G=S^1$ acts by rotations around an axis, then $M_G$ is fibered over the segment $[0,1]$ with generic fiber $EG$, except at the endpoints where the fiber is $BG$. In this case $H_G^*(M)\simeq H^*(\IC\IP^\infty)^{\oplus 2}$ is generated by the $G$-fixed points.

More generally, let $F\subset M$ denote the fixed locus of the $G=T^m$-action. 
By pullback through the inclusion map $i: F\to M$ we have\footnote{Functoriality of equivariant cohomology is used in application of pullback. Although we didn't discuss this property here, it is well-known to hold.}
\be
H_{T^m}^*(M) \mathop{\to}^{i^*} H_{T^m}^*(F) =  H^*(F) \times H^*(BT^m)\,.
\ee
The localization theorem asserts that this map is in fact an isomorphism, up to torsion.
In other words, the whole $G$-equivariant cohomology of $M$, except for the torsion part, is captured by equivariant cohomology of the fixed locus $F$. Furthermore this corresponds simply to cohomology of the fixed locus itself, times the cohomology of the classifying space $H^*(BT^m,\IC) \simeq H^*((\IC\IP^\infty)^{\times m})\simeq \IC[\epsilon_1,\dots, \epsilon_m]$.

There is a natural $G$ action on the restriction of $TM$ to $F$, which induces a splitting $TM|_F \simeq  TF\oplus N_{F/M}$ into the `fixed' and `moving-away' parts with respects to the action of $G$. 
Given the inclusion map $i:F\to M$, it can be used to push forward cohomology $i_*:H^*(F)\to H^*(M)$. 
\footnote{The careful reader will notice here a technical subtlety: This operation involves using Poincar\'e duality on $F$, then pushing forward a dual homology cycle, then using again Poincar\'e duality on $M$. 
As we will be interested, occasionally, in cases where $M$ is noncompact, one must take case of restricting to compactly supported de Rham cohomology in order for Poincar\'e duality to be applicable.
}
This operation is accomplished by using the Thom form $\Phi_F$ associated to $F$, by taking $\omega \in H^*(F)$ to $ \Phi_F\wedge\omega$ as a form on the normal bundle $N_{F/M}$. 
Viewing $N_{F/M}$ as a neighbourhood of $F$ and extending the Thom form by zero as usual, this gives a map to $H^*(M)$.
In particular, pushing forward the trivial cohomology class gives the Thom class in $H^*(M)$. 
Its pullback by $i^*$ is the Euler class of the normal bundle $i^*i_* 1 = \eul(N_{F/M})$. 
This construction extends to the equivariant setting, producing the $G$-equivariant Euler form on the normal bundle to $F$.

In the case when $F$ is a point, or a finite collection of points, $N_{F/M}\simeq TM|_F$ is just the restriction of $TM$ to the fixed locus. Representations of $G=T^m$ on $TM$ are labeled by $m$ weights $a_i\in \IR$. In this case the Euler form is a top form on $M$, in fact $\eul(N_{F/M}) = \prod_{i=1}^{m} a_i \, dx_1\wedge\dots\wedge dx^d$ for a given fixed point $F$. 
Since $a_i\neq 0$, the Euler form is invertible.
The Atiyah-Bott-Berline-Vergne localization theorem \cite{berline1982classes, atiyah1984moment} asserts that the Euler class  of $N_{F/V}$ is always invertible if $F$ is the fixed locus of a $G$ action.
This implies that $\sum_F(\eul(N_{F/M}))^{-1} \, i_*i^*$, where the sum runs over components of the fixed locus, acts as the identity on $H^*(M)$. It follows that the integral of an equivariant form $\phi \in \Omega_G^*(M)$ over all $M$
\be
	\int_M \phi = \sum_F\int_F\frac{i^*\phi}{\eul(N_{F/M})}
\ee
reduces to contributions at the fixed points.
This is the celebrated localization formula.

We will be interested in a basic application of this formula, corresponding to taking $\phi = \eul(TM)$ the $G$-equivariant Euler class of the tangent bundle to $M$.
\be\label{eq:localization-Euler}
	\chi(M) = \int_M \eul(TM) 
	= \sum_F\int_F\frac{i^*\eul(TM)}{\eul(N_{F/M})}
	= \sum_F 1  = \text{(\# of fixed points)} \,.
\ee
When $M$ has a finite number of fixed point under $G=T^m$, its Euler characteristic is just the number of isolated fixed points.

\subsection{Fixed points from foliations}\label{eq:foliation-fixed-points}

We now return to the study of moduli space of $A$-branes. 
Since $\CM_L$ admits a Lagrangian fibration by tori $T^{b_1(L)}$, it carries a natural $G=T^{b_1(L)}$ action rotating the fibers.
The fixed locus of the $G$ action correspond to points in the base $\fM_L$ where the whole fiber degenerates to a point, this means points in $\fM_L$ corresponding to a (possibly singular) Lagrangian $L$ such that all holonomies of the Abelian local system are either trivial or ill-defined.\footnote{%
As examples illustrated, a one-cycle in the torus $T^{b_1(L)}$ pinches whenever the corresponding 
abelian holonomy 
parameterizing $S^1\subseteq T^{b_1(L)}$ ceases to make sense. 
For a Lagrangian in $\fD_L$ the whole fiber $T^{b_1(L)}$ collapses to a point, corresponding indeed to a fixed point of the $G$-action.
}
This locus must be a collection of points in $\fM_L$, since the torus fibration is Lagrangian: the shrinking of each fiber places a local constraint on the base coordinates, and the overall number of constraints is equal to the dimension of the base.
We denote the fixed locus by $\fD_L\subset \fM_L$. 
By the $G$-equivariant localization formula for the Euler characteristic (\ref{eq:localization-Euler}) with $G=T^{b_1(L)}$ we deduce that $\chi(\CM_L) = |\fD_L|$. 
Using this into our definition of the BPS invariants (\ref{eq:omega-chi}), we arrive at
\be\label{eq:Omega-localization}
	\Omega(L) 
	=(-1)^{\dim \CM_L} \, |\fD_L| \,.
\ee
The BPS invariant $\Omega(L)$ coincides, up to a sign, with the number of $G$-fixed points.

If $L$ admits an $S^2$-fibration as described in section \ref{sec:S2-fibers}, the moduli space $\fM_L$ can be studied by means of foliations, as explained in Section \ref{sec:foliations}.
In this model, a special Lagrangian $L\in \fM_L$ may map to a leaf of some foliation $\phi_{ij,n}$, or more generally to a set of leaf segments of different foliations, connected by junctions.
In either case the generic leaf is a smooth Lagrangian, but in the whole family of leaves we also usually find certain degenerate leaves.
The question we wish to address is how to find degenerate leaves. More precisely we would like to identify leaves corresponding to points in $\fD_L$, which are \emph{maximally} degenerate.
In the following we assume that, for a generic choice of complex moduli of $\Sigma$, 
only trivalent junctions appear in a system of leaves that represents a single Lagrangian. 

There are two types of mechanisms that produce a degenerate leaf from a smooth one. 
\begin{enumerate}
\item[$i)$]
The first phenomenon occurs when a family of smooth leaves is parameterized by a modulus $r\geq 0$ such that at $r=0$ a local piece of the leaf labeled by $\phi_{ij,0}$ runs into a branch point where $y_i=y_j$. See Figure \ref{fig:leaf-degeneration-1}.
There is typically a cycle in $H_1(L,\IZ)$ corresponding to a lift of the generic leaf of $\phi_{ij,0}$ to a point in $S^2_{ij,0}$, see the blue line in the figure.
The modulus $r$ is a local coordinate on $\fM_L$ corresponding to the `height' of a smooth leaf, measured transversely to the foliation. $r=0$ corresponds to degeneration into a critical (or possibly bi-critical) leaf, where the holonomy around the blue cycle becomes ill-defined.

\begin{figure}[h!]
\begin{center}
\includegraphics[width=0.45\textwidth]{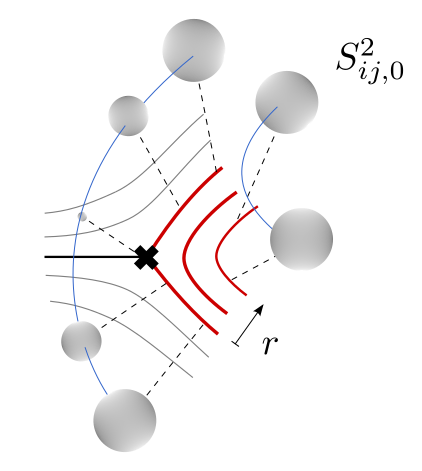}
\caption{The first type of degeneration of a smooth leaf. Red leaves correspond to points in $\fM_L$. At $r=0$ the two-sphere $S^2_{ij,0}$ degenerates over the $ij$ branch point, pinching the $S^1\subseteq T^{b_1(L)}$ parameterizing the holonomy around the blue cycle on $L$.}
\label{fig:leaf-degeneration-1}
\end{center}
\end{figure}

\item[$ii)$]
The second type of degeneration may occur whenever a segment connected by two junctions collapses, and the pair of junctions becomes coincident. See Figure \ref{fig:leaf-degeneration-2}.
If $L$ has a non-contractible cycle that goes through one of the junctions, then that cycle will pinch when the segment shrinks because the two junctions must annihilate. 
For example, the figure shows a blue cycle running above the external segments of types $ij$ and $jk$. In $L$, this loop lifts, above each $x$, to a point $(y(x),u(x))$ in the two-spheres $S^2_{ij,m}$ and $S^2_{jk,n}$ respectively. The lifted paths are joined above the junction, by a path in the triangle in the $\log y$ plane also shown. When the segment shrinks, the triangle collapses and the holonomy around the blue cycle becomes ill-defined.
\end{enumerate}

\begin{figure}[h!]
\begin{center}
\includegraphics[width=0.75\textwidth]{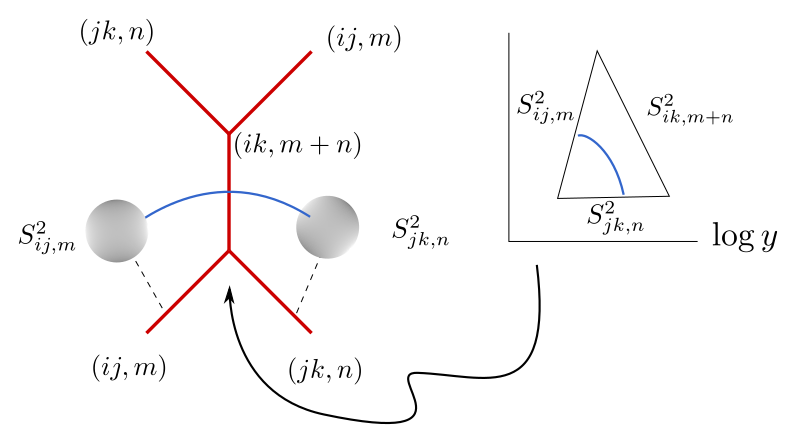}
\caption{The second type of degeneration of a smooth leaf. When the $ik$ segment is non-zero, it supports a junction shown on the right frame as a triangle in the $\log y$ plane. There is a cycle in $H_1(L,\IZ)$ shown in blue that runs through the junction, from a point on $S^2_{ij,m}$ to a point on $S^2_{jk,m}$. When the $ik$ segment shrinks, the junction disappears, and the holonomy supported on the blue cycle becomes ill-defined.}
\label{fig:leaf-degeneration-2}
\end{center}
\end{figure}

We have now explained how to detect degenerate leaves.
What counts for the computation of $\Omega(L)$ are the \emph{maximally} degenerate leaves.
These are Lagrangians with $b_1(L)$ pinching cycles, and they can be found in foliations by looking for leaves (or systems of leaves) that feature $b_1(L)$ degeneration phenomena of the types from Figures \ref{fig:leaf-degeneration-1} and \ref{fig:leaf-degeneration-2}.
Next we illustrate some examples of maximally degenerate leaves, and the corresponding computations of the Euler characteristic.

\paragraph{A remark on degenerations.}
Naively, a third possibile type of degeneration would seem to arise when a segment suspended between a branch point and a junction collapses.
This however would not cause a topological change in the Lagrangian, for the following reason.
As a local model suppose the shrinking segment is of type $ik$, and the other two segments attached at the junction are of types $ij$ and $jk$.
Since the $ik$ segment ends on a branch point, it cannot support a non-contractible cycle in $L$. Any such cycles, if present at all, must run above the $ij$ and $jk$ segments (as in the local model of Figure \ref{fig:leaf-degeneration-2}).
As the $ik$ segment shrinks, the $ij$ segment hits the $ik$ branch point, and when it goes across it finds a branch cut, which turns the $ij$ segment into one of type $jk$. This then continues as the $jk$ segment that was also attached to the junction.
This transition is simply the inverse of the creation of a `string junction' for M2 branes, by Hanany-Witten effect \cite{Hanany:1996ie}, also related to the notion of equivalence for spectral networks discussed in \cite[Section 10.6]{Gaiotto:2012rg}.

\subsection{Some computations of enumerative invariants}

\begin{figure}[h!]
\begin{center}
\includegraphics[width=0.25\textwidth]{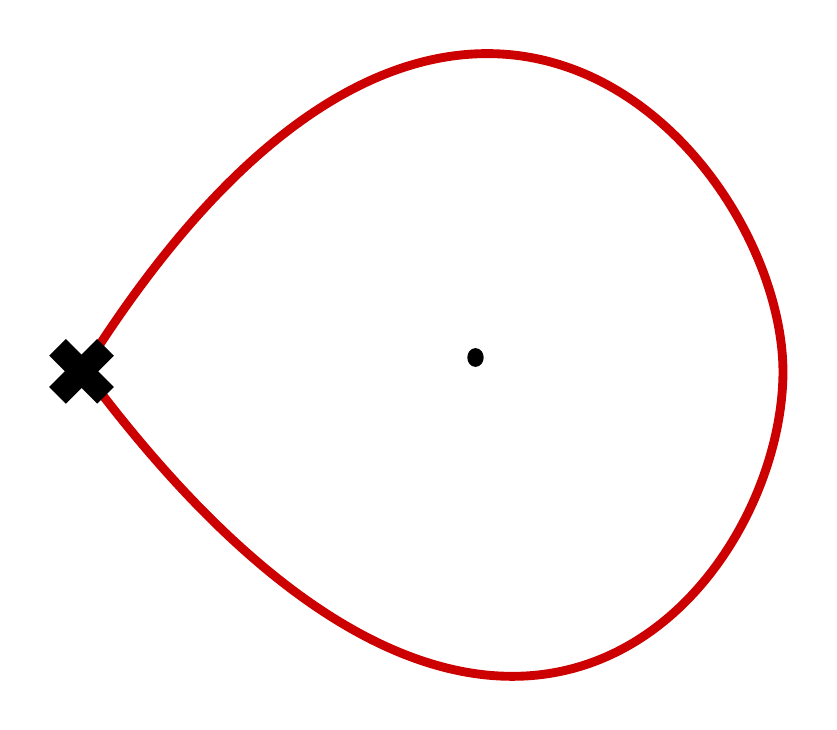}
\caption{The fixed point leaf from the family of Figure \ref{fig:unbounded-circles}.}
\label{fig:unbounded-circles-FP}
\end{center}
\end{figure}

\paragraph{Examples of type $i$.}
The first nontrivial case is unbounded family of compact leaves studied in section \ref{sec:unbounded-circles}. Here $\CM \simeq \IC$ presented as an $S^1$ fibration over $\fM_L\simeq \IR_{\geq 0}$.
The $S^1$ action rotates $\CM_L$ around the origin, corresponding to the fixed point $p\equiv \fD_L$. 
This point is precisely the bi-critical leaf attached to the branch point, the degeneration that occurs is of the first type mentioned above. See Figure \ref{fig:unbounded-circles-FP}.
By counting fixed points we recover the result from (\ref{eq:CP1-BPS-index})
\be
	\Omega(L) = (-1)^1\cdot | \fD_L| = -1\,.
\ee

\begin{figure}[h!]
\begin{center}
\includegraphics[width=0.25\textwidth]{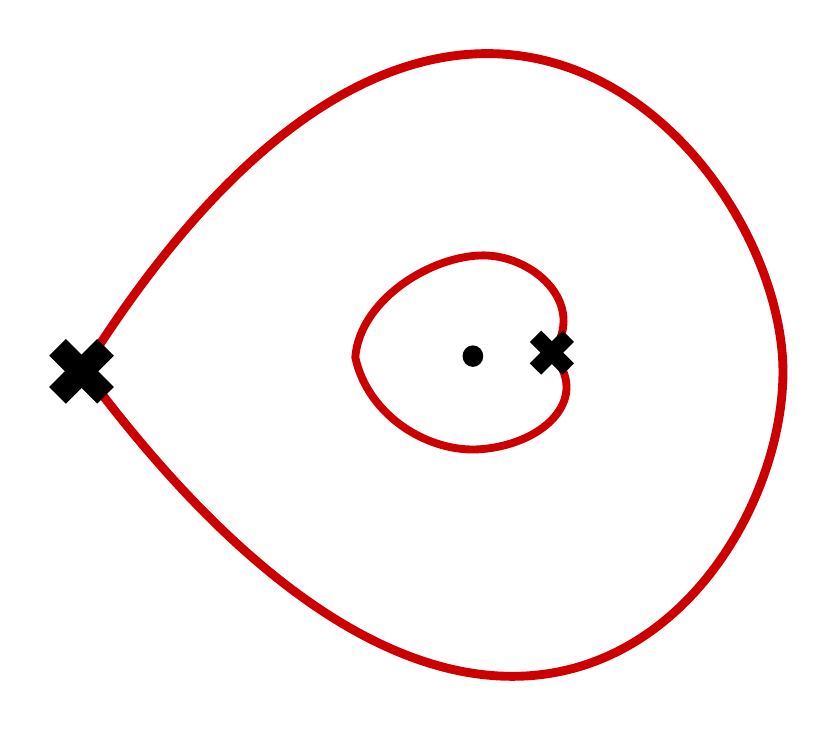}
\caption{The two fixed point leaves from the family of Figure \ref{fig:bounded-circles}.}
\label{fig:bounded-circles-FP}
\end{center}
\end{figure}

A variant of this is the bounded family of compact leaves studied in section \ref{sec:bounded-circles}. Here $\CM \simeq \IP^1$ presented as an $S^1$ fibration over $\fM_L\simeq [0,1]$.
The $S^1$ action rotates $\CM_L$ leaving the north and south poles fixed, corresponding to boundaries of the interval $\partial [0,1] = \{0,1\} \equiv  \fD_L$.
The two fixed points are the two bi-critical leaves attached to the branch points. Again the degeneration that occurs is of the first type mentioned above. See Figure \ref{fig:bounded-circles-FP}.
Again, by counting fixed points we reproduce (\ref{eq:VM-BPS-index}) 
\be
	\Omega(L) = (-1)^1\cdot | \fD_L| = -2\,.
\ee

\begin{figure}[h!]
\begin{center}
\includegraphics[width=0.75\textwidth]{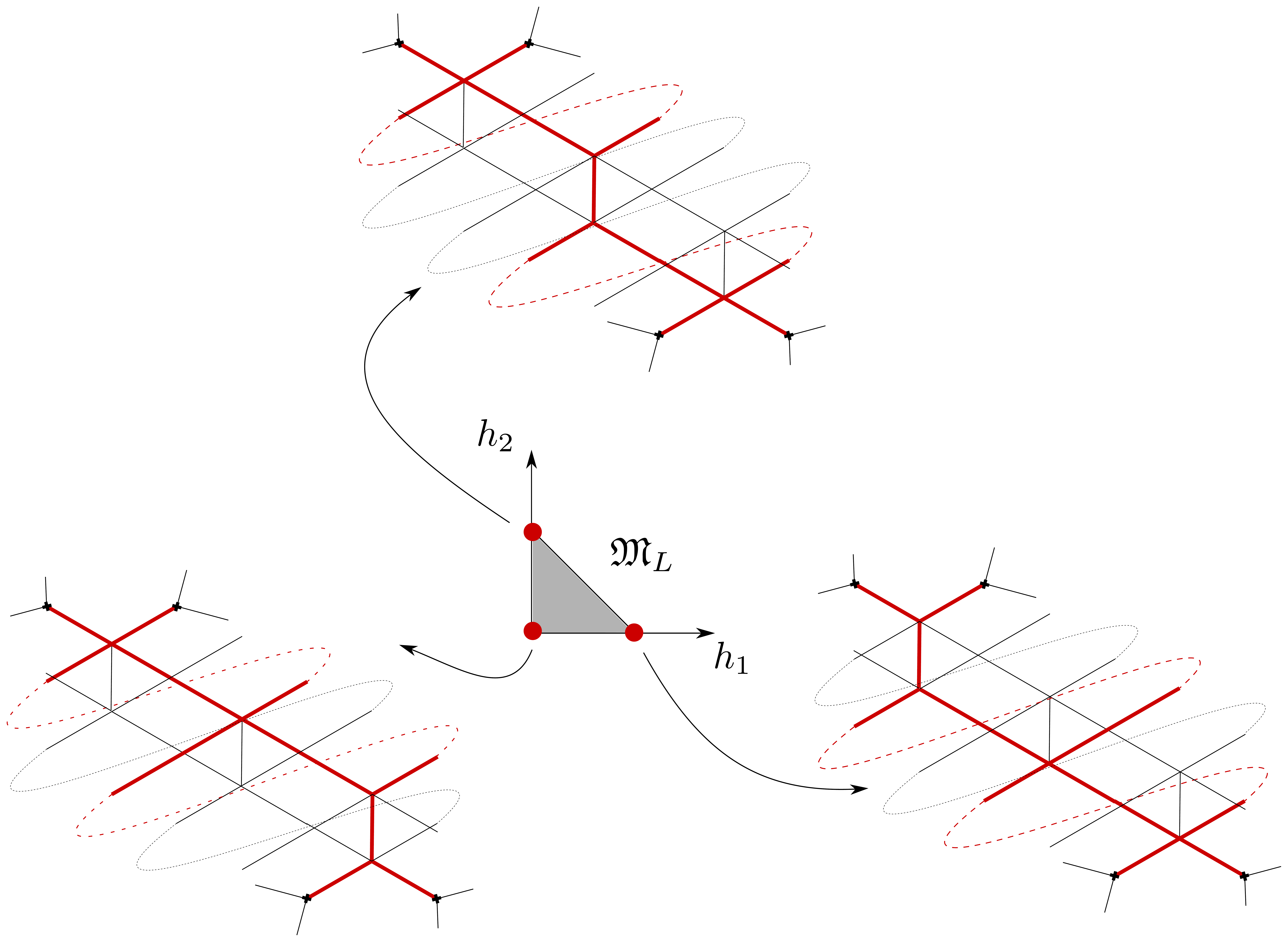}
\caption{The three fixed point leaves from the family of Figure \ref{fig:herds}.}
\label{fig:k-herd-FP}
\end{center}
\end{figure}

\paragraph{Examples of type $ii$.}
The second type of degeneration occurs in the family of foliations with junctions considered in section \ref{sec:sliding-junctions}.
Here $\fM_L$ is a two-simplex $\Delta^2$, and $\CM_L\simeq \IP^2$ is a $T^2$ fibration over it. The set of fixed points $\fD_L$ corresponds to the three vertices of $\Delta^2$, see Figure \ref{fig:herds}.
A segment connecting two junctions shrinks whenever one of the $h_i=0$. This corresponds to one of the codimension-one boundaries of $\fM_L$ in Figure \ref{fig:herds}. 
If only a single edge shrinks, only one cycle shrinks. But $b_1(L)=2$ and a true fixed point requires that both cycles shrink at the same time. So we demand that $h_i=h_j=0$ for $(i,j)\in \{(1,2),(2,3),(1,3)\}$. These three configurations are shown in Figure \ref{fig:k-herd-FP}, and correspond indeed to the vertices of $\Delta^2$.
By counting fixed points we recover the result from  (\ref{eq:3-herd-BPS-index})
\be
	\Omega(L) = (-1)^2\cdot | \fD_L| = 3\,.
\ee

The generalization to $k$-herds is straightforward: there are $k$ internal segments, and fixed points correspond to the $k$ possibilities where one has full length while all others shrink. These are indeed vertices of $\fM_L\simeq \Delta^{k-1}$, corresponding to fixed points of the toric action on $\CM_L$.
Again by counting fixed points we reproduce (\ref{eq:k-herd-BPS-index})
\be
	\Omega(L) = (-1)^{k-1}\cdot | \fD_L| = (-1)^{k-1} k\,.
\ee

\paragraph{The case of SYZ fibers.}
As discussed in sections \ref{sec:SYZ-fibers}-\ref{sec:codimension-one-strata}, only certain codimension-one strata of the moduli space $\fM_{SYZ}$ of special Lagrangian SYZ fibers can be described by leaves of foliations.
This seems to pose an obstruction to computing $\chi(\CM_{SYZ})$, since we are not able to see the full moduli space.
But in fact, thanks to localization this obstacle can sometimes be sidestepped altogether, by shifting the focus to the fixed points of the $G=T^3$ action on $\CM_{SYZ}$.

\begin{figure}[h!]
\begin{center}
\includegraphics[width=0.3\textwidth]{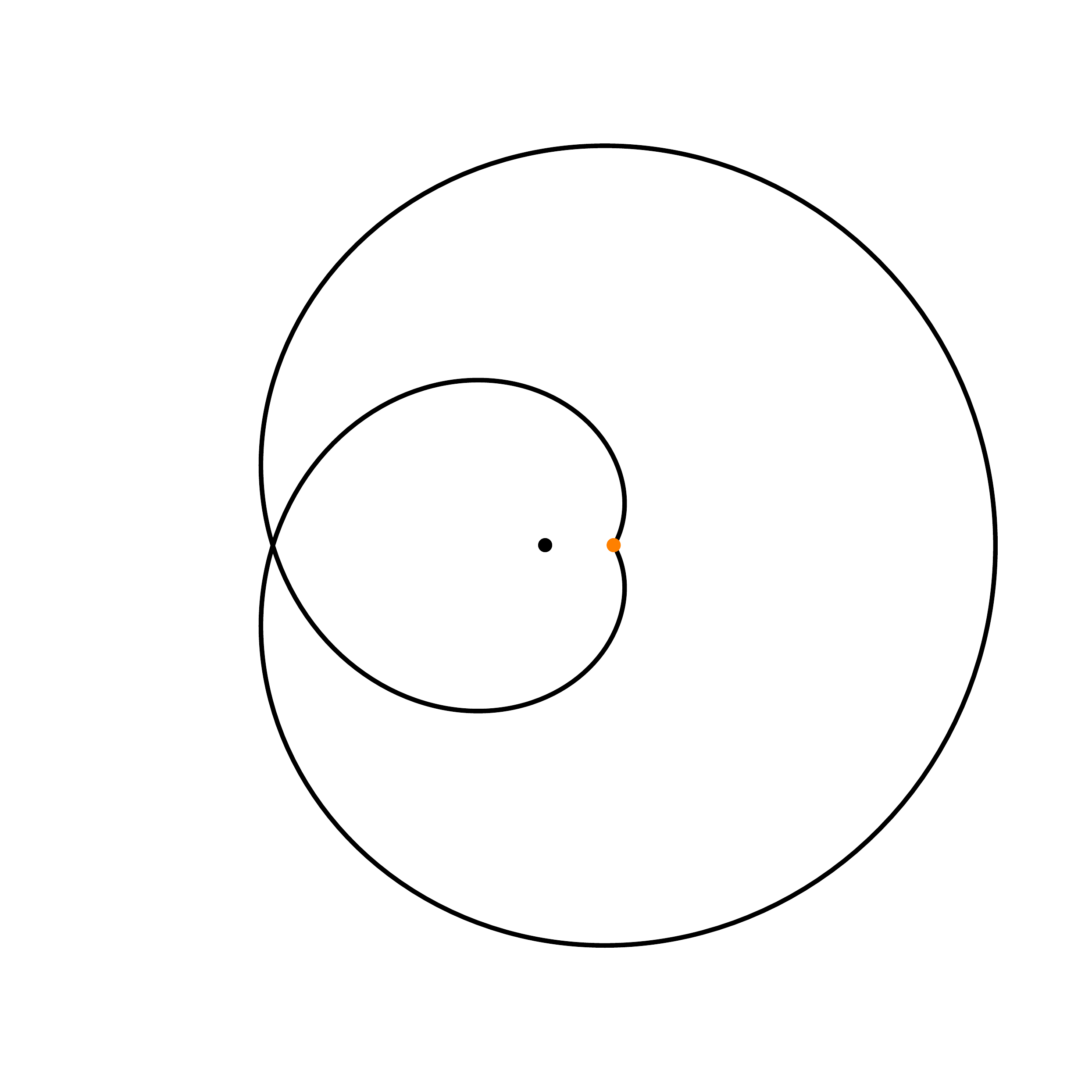}
\caption{The unique fixed point leaf from the family of Figure \ref{fig:circle-decay}.}
\label{fig:C3-D0-fixed-point}
\end{center}
\end{figure}

Fixed points must correspond to codimension-three strata of $\fM_L$. 
If all fixed points belong to the (boundary of the) codimension-one stratum parameterized by foliations, 
localization allows to compute the BPS invariant for the family of special Lagrangian SYZ fibers,
by studying degenerate leaves. 
This was verified in a few cases in \cite{Banerjee:2018syt, Banerjee:2019apt} for moduli spaces of SYZ fibers in mirrors of toric Calabi-Yau threefolds.\footnote{More generally, it is unclear if fixed points of the $T^3$ action on $\CM_{SYZ}$ are always captured by foliations.}
For example, in the case of the Hori-Vafa mirror of $\IC^3$, there is a unique degenerate leaf in the codimension-one stratum studied in Figure \ref{fig:circle-decay}. This is the bi-critical leaf shown in Figure \ref{fig:C3-D0-fixed-point}.
By counting this unique fixed point we recover the correct result
\be
	\Omega(L_{SYZ}) = (-1)^{3}\cdot | \fD_{SYZ}| = -1\,.
\ee

\section{Relation to Spectral and Exponential Networks}\label{sec:relation-to-networks}

In section \ref{eq:eunmerative-inv-A-branes} we have introduced our definition of enumerative invariants for stable $A$-branes, via the Witten index of worldvolume quantum mechanics of D3 branes wrapping special Lagrangians.
Later, in sections \ref{sec:sLag-moduli-foliations}-\ref{sec:some-moduli-spaces} we have applied this definition to the case of special Lagrangians parameterized by leaves of foliations of certain abelian differentials. Then in section \ref{sec:localization} we argued that localization effectively allows us to restrict attention to certain singular leaves of foliations. 
In this section we connect the study of singular leaves to the subject of spectral and exponential networks.

We will argue that the enumerative invariants defined above coincide, in a way that will be made precise below, with the BPS indices computed by non-abelianization for networks.
This has two important consequences.
\begin{enumerate}
\item
First, it implies that the enumerative invariants that we study in this work exhibit jumps over the moduli space of complex structures, which are governed by the Kontsevich-Soibelman wall-crossing formula \cite{Kontsevich:2008fj}.
This follows from the fact that BPS indices computed by networks obey the `$\CK$-wall formula' (see \cite{Gaiotto:2012rg} for spectral networks and \cite{Banerjee:2018syt} for exponential networks).
The compliance with wall-crossing can be taken as evidence that our definition of BPS invariants is a viable candidate for the generalized Donaldson-Thomas invariants considered in \cite{Joyce:2008pc,Kontsevich:2008fj}.
In fact, earlier computations based on exponential networks \cite{Eager:2016yxd, Banerjee:2018syt, Banerjee:2019apt, Banerjee:2020moh, Longhi:2021qvz} have been checked to match with computations of Donaldson-Thomas for $B$-branes on the mirror Calabi-Yau's.
\item
Another reason why the connection with networks is important, is that the latter offer a \emph{systematic} way of computing the BPS index, and therefore our enumerative invariants for $A$-branes. 
On the one hand, we hope that the definition of $\Omega(L)$ introduced in this work may help demistify some of the aura of the BPS indices defined via nonabelianization, which involves a fair deal of definitions and unconventional computations. On the other hand, the framework of nonabelianization serves as a powerful tool for explicit and systematic computations for the enumerative invariants we consider here.
\end{enumerate}

\subsection{BPS index formula from networks}

To pave the way for comparing our BPS invariants with the BPS indices computed by networks, we begin with an executive summary of the latter.
Here we focus entirely on the formula that defines the BPS index $\Omega(\gamma)$, the discussion will not be self-contained. More details can be found in \cite{Gaiotto:2012rg,Banerjee:2018syt} and in reviews included in \cite{KleinLectures, Hollands:2013qza, Hollands:2021itj}.

\subsubsection{Spectral networks}

\begin{figure}[h!]
\begin{center}
\includegraphics[width=0.5\textwidth]{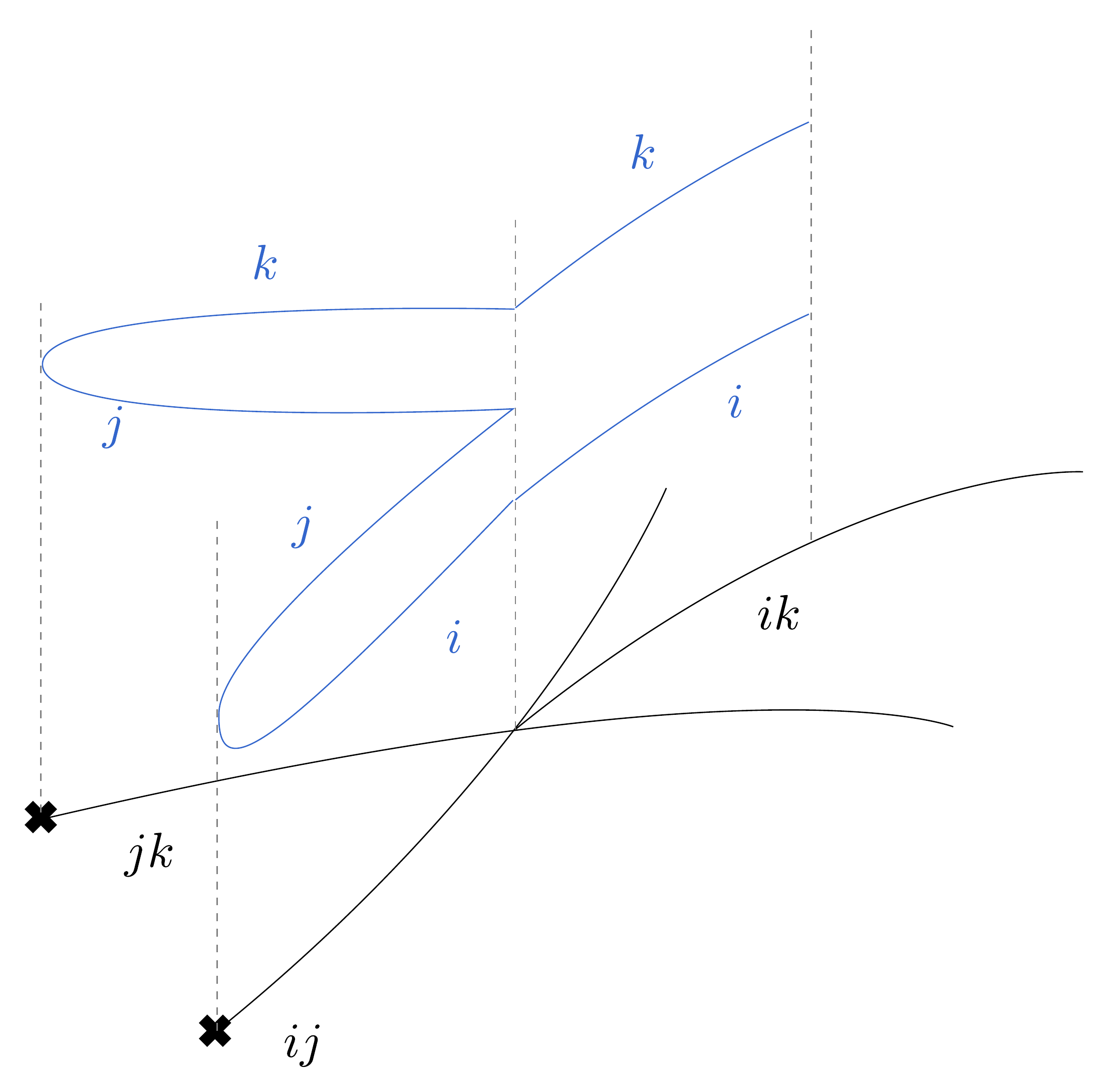}
\caption{A spectral network made of primary critical leaves of types $ij$ and $jk$, intersecting to give a descendant critical leaf of type $ik$. Each trajectory carries combinatorial data of open paths on the covering surface $\Sigma$, shown in blue and running on the respective sheets (e.g. $\lambda_i$ and $\lambda_j$) labeling the trajectory (e.g. $ij$) . Paths on primary trajectories concatenate at intersections to give birth to the paths carried by descendants.}
\label{fig:example-network}
\end{center}
\end{figure}

The definition of networks involves two main pieces of data: the geometric data of trajectories on a Riemann surface $C$, and the combinatorial data of open paths on a covering surface $\Sigma\to C$ that is associated to each trajectory.
We will describe these structures for spectral networks first.
Given a Riemann surface $C$, also known as `UV curve' \cite{Gaiotto:2009we, Gaiotto:2009hg}, consider a ramified covering $\Sigma\to C$ as a curve in $T^*C$ defined by an algebraic equation for the Liouville one-form $\lambda$
\be\label{eq:Sigma-spec-net}
	\lambda^N+\sum_{k=2}^{N} \phi_k \lambda^{N-k} = 0\,.
\ee
Here $\phi_k$ are meromorphic $k$-differentials on $C$, with poles of prescribed degrees at punctures. The curve given here applies to theories of class $\CS$ of type $A_{N-1}$. Extensions to $ADE$ curves and beyond can be found in \cite{Longhi:2016rjt, Ionita:2021tqn}.
After choosing a trivialization for the covering, consisting of a system branch cuts and a global assignment of labels $\lambda_i$ to the sheets $i=1,\dots, N$, one may define foliations $\phi_{ij}$ by 
\be
	\iota_{\partial_t}(\lambda_j-\lambda_i)  \in e^{i\vartheta} \IR^+\,.
\ee
Here $\partial_t$ is the tangent vector to a leaf parameterized by $t\in \IR$. This equation corresponds to (\ref{eq:E-wall}) in the case $\lambda_i = \log y_i\, d\log x$ and $n=0$, with $\zeta=e^{i\vartheta}$.
The geometric data of networks consists of specific leaves of foliations $\phi_{ij}$ for all pairs $i\neq j$.
First of all there are the primary critical leaves, namely those leaves of $\phi_{ij}$ that start from a branch point where $\lambda_i=\lambda_j$ on one side, and flow into a puncture on the other side.
Second, there are descendant critical leaves: when two critical leaves (either primary or descendant) of types $ij$ and $jk$ intersect, one takes the leaf of $\phi_{ik}$ passing through the intersection point. See figure \ref{fig:example-network} for a sketch of a generic spectral network.

The combinatorial data of networks is the assignment to each trajectory of certain open paths on the covering $\Sigma$. 
A critical leaf of type $ij$ may carry open paths that begin at sheet $i$ and end on sheet $j$ at generic points above the trajectory, see Figure \ref{fig:example-network}.
We only keep track of relative homology classes of open paths, denoted $a\in \Gamma_{ij}(z) \equiv H_1^{{\rm rel}}(\Sigma, \lambda_i(z),\lambda_j(z))$ for paths beginning/ending at $\lambda_i(z) / \lambda_j(z)$ for some $z\in C$ on the trajectory.
Open paths are counted by certain integers $\mu(a)\in \IZ$, which end up being $\mu(a)=\pm 1$ for the open path obtained by `lifting' the trajectory to $\Sigma$ and $\mu(a)=0$ for all other classes.\footnote{If the trajectory is a descendant critical leaf, one considers not only its lift, but also the lifts of parent trajectories, all the way to primary ones. The lifts are glued at intersections to give a continuous open path obtained by successive concatenations.}
The sign is determined by additional framing data, see \cite{Gaiotto:2012rg} for a discussion.

The charge of a BPS state is represented by a homology cycle $\gamma\in H_1(\Sigma,\IZ)$.\footnote{More precisely charges are valued in a certain quotient of a sublattice of $H_1(\Sigma,\IZ)$ \cite{Gaiotto:2009hg, Longhi:2016rjt, Ionita:2021tqn}.}
Geometrically the cycle $\gamma$ encodes the data of a Lagrangian cycle $L\in H_3(X,\IZ)$ in a Calabi-Yau threefold described by the hypersurface $\lambda^N+\sum_{k=2}^{N} \phi_k \lambda^{N-k} = uv$ in $T^*C\times \IC^2$. Details of this map have been discussed e.g. in \cite{Klemm:1996bj, Gaiotto:2009hg, Eager:2016yxd, Banerjee:2018syt}. 
One way to think about this map is to consider an $S^2$-fibered special Lagrangian as in section \ref{sec:sLag-moduli-foliations}. 
Then $L$ projects to a path on $\IC^*$, which is here replaced by $C$. Above each $z\in C$ there is a segment in $T^*_zC$ running between $\lambda_i(z)$ and $\lambda_j(z)$, and a circle $uv={\rm const.}$ fibered above the segment, which shrinks at endpoints. This gives an $S^2$ above $z$, and varying the basepoint gives the whole calibrated $L$. Consider the endpoints $\lambda_i(z)$ and $\lambda_j(z)$ at each $z$ on the segment: varying $z$ the endpoints trace arcs on $\Sigma$, which eventually must reconnect together at branch points, or to other arcs above junctions. The overall system of arcs is a closed path on $\Sigma$, whose homology class is $\gamma$. 
We thus have a map 
\be\label{eq:cycle-map}
	L \in H_3(X,\IZ) \quad\longleftrightarrow\quad \gamma\in H_1(\Sigma,\IZ)
\ee
between three-cycles on an appropriate Calabi-Yau and one-cycles on an associated Riemann surface, as first observed in \cite{Klemm:1996bj}.

\begin{figure}[h!]
\begin{center}
\includegraphics[width=0.95\textwidth]{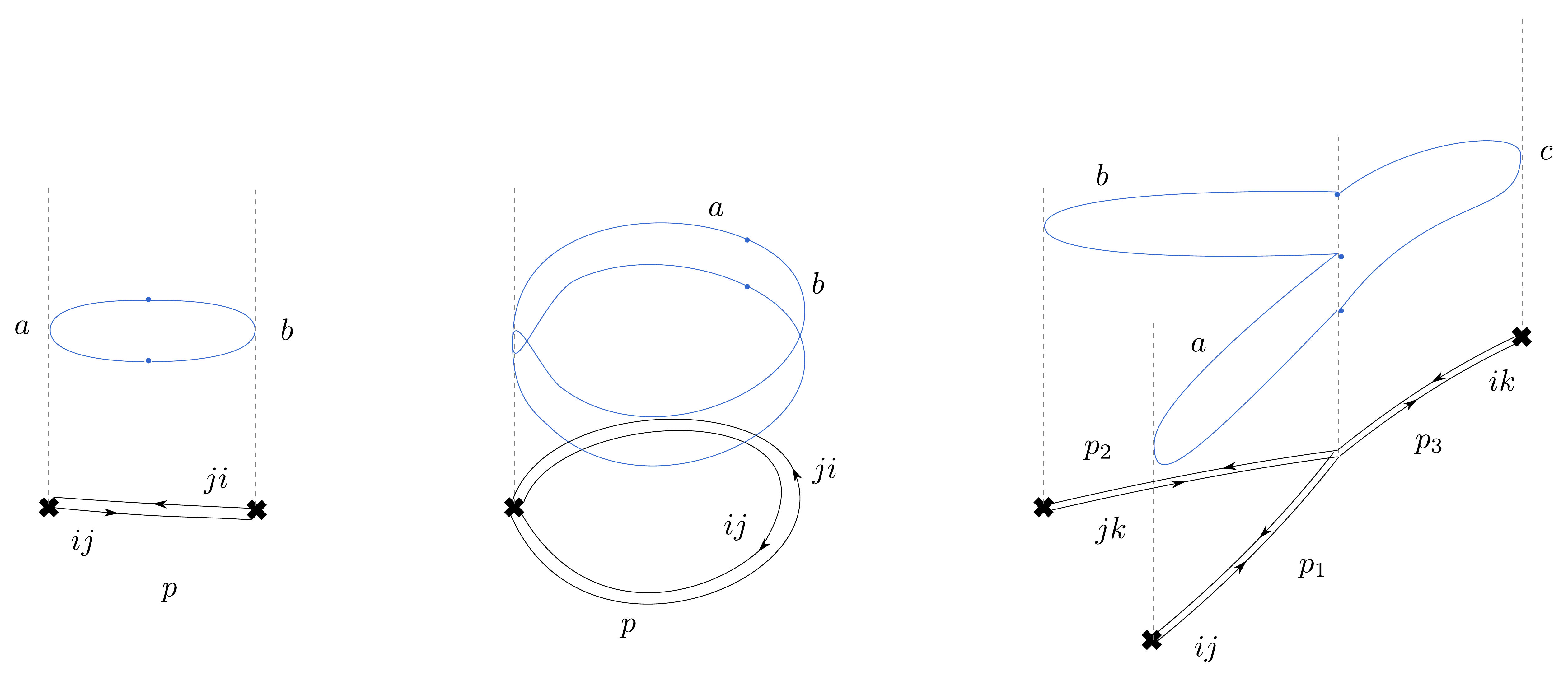}
\caption{Some examples of two-way streets, and the underlying resolution by critical leaves of foliations. Shown in blue are the closed cycles on $\Sigma$ obtained from concatenation of open path combinatorial data.}
\label{fig:two-way-street-examples}
\end{center}
\end{figure}

Now we come to the computation of the BPS index $\Omega(\gamma)$ for a given charge $\gamma$, and a given choice of complex moduli for $\Sigma$.
Let $\vartheta = \arg \oint_\gamma \lambda$ be the phase of the period along a primitive cycle $\gamma$ of the Liouville form pulled back to $\Sigma$. Consider the network at phase $\vartheta$. If all trajectories are non-degenerate then $\Omega(\gamma)=0$.
On the other hand if some trajectories of opposite types (e.g. $ij$ and $ji$) overlap, then there may be stable BPS states with charges proportional to $\gamma$.
One considers only the degenerate trajectories of types $ij$ that overlap with those of opposite types $ji$. We call this a system of two-way streets. A two-way street $p$ of type $ij/ji$ may be attached to branch points of the same type, or to junctions where $ij/ji$ trajectories intersect with trakectories of types $ik/ki$ and $kj/jk$ for some $k\neq i,j$. Some examples are shown in figure \ref{fig:two-way-street-examples}.

Each two-way street of type $ij/ji$ is made of underlying oriented trajectories of types $ij$ and $ji$ (possibly multiple ones for each type). Recall that these carry combinatorial data of open paths.
One builds a generating function of closed paths by considering all possible concatenations of oriented open paths on the underlying trajectories
\be\label{eq:Q-def}
	Q(p) 
	= 1 + \sum_{a\in \Gamma_{ij}}\sum_{b\in \Gamma_{ji}} \mu(a)\mu(b) X_{a\circ b} 
	= 1 + \sum_{n\geq 1} c_{n\gamma} X_{n\gamma}
\ee
Here $a\circ b$ denotes concatenation of $a$ with $b$ at both endpoints, and $X_\gamma$ are formal variables associated with homology cycles, valued in a ring with multiplication rule $X_\gamma X_{\gamma'} = X_{\gamma+\gamma'}$.

It is argued in \cite{Gaiotto:2012rg} that the series $Q(p)$ must factorize as
\be\label{eq:Q-factorization}
	Q(p) = \prod_{n\geq 1}(1 \pm X_{n\gamma})^{\alpha_{n\gamma}(p)}
\ee
for suitable choice of signs, with integer exponents $\alpha_{n\gamma}(p)\in \IZ$.
Using these exponents, one may define a closed cycle on $\Sigma$ by taking the lift $\pi^{-1}:C\to \Sigma$ of formal linear combinations of two-way streets
\be\label{eq:L-path}
	\fL_{n\gamma} = \sum_{p} \alpha_{n\gamma}(p) \,\cdot\, \pi^{-1}(p)\,.
\ee
This is argued to give a closed path, whose homology class is an integer multiple of $n\gamma$
\be\label{eq:BPS-index}
	\Omega(n\gamma) = \frac{[\fL_{n\gamma}]}{n\gamma}\,.
\ee
This is the BPS index for charge $n\gamma$, and a fixed geometry of $\Sigma$. The set of BPS indices for all possible charges $\gamma$ determines the BPS spectrum of a theory. 
Varying the complex moduli, as encoded by the $k$-differentials $\{\phi_k\}_k$, may induce a change in the spectrum of BPS states. It follows from the $\CK$-wall formula of \cite{Gaiotto:2012rg} that BPS indices must change in a controlled way, described by the wall-crossing formula of Kontsevich and Soibelman \cite{Kontsevich:2008fj}.

\subsubsection{Exponential networks}

We briefly comment on how the above definitions evolve in the context of exponential networks, following \cite{Banerjee:2018syt}.
The first difference is in the geometric data: instead of a Riemann surface described by a collection of differentials (\ref{eq:Sigma-spec-net}), one now considers an algebraic curve in $\IC^*\times \IC^*$ described by a polynomial $F(x,y)=0$.
Given this curve, the exponential network consists of a set of trajectories on $\IC^*$ with coordinate $x$. Again the trajectories are described by a differential equation, and consist of critical leaves of a foliation. In this case the differential equation for the critical leaves is precisely (\ref{eq:E-wall}), with $n\in \IZ$ keeping track of logarithmic branching of $\log y_i \frac{dx}{x}$ in the $y$-plane, for a choice of trivialization.
Once again we define primary and descendant critical leaves: the former are sourced at branch points where $y_i(x)=y_j(0)$ and have $n=0$, where as the descendants are sourced at intersections.
Exponential networks feature more types of intersections than spectral networks, in particular when trajectories of types $(ij,m)$ and $(ji,n)$ with $m+n\neq 0$ intersect, there are infinitely many new trajectories generated \cite{Banerjee:2018syt}.
The combinatorial data associated to each trajectory is also more involved: now one considers open paths not on $\Sigma$ but on a logarithmic covering $\tilde \Sigma$, branching over $\Sigma$ around punctures where $y$ runs to $0$ or to $\infty$.\footnote{%
The branch cuts of $\tilde \Sigma \to \Sigma$ are the preimages of the $\log y$ branch cut on the $y$-plane $\IC^*$, see \cite{Banerjee:2018syt} for an extensive discussion.
}
As a result each trajectory carries infinite towers of open paths counted by $\mu(a_N) \in \IZ$ where $N\in \IZ$ denotes a path running from sheet $\log y_i(x) + 2\pi i N$ to sheet $\log y_j(x) + 2\pi i (N+n)$ for a critical leaf of the foliation $\phi_{ij,n}$.
The integers $\mu(a_N)$ are $\pm1$ for the natural lift of the critical leaf (as in the case of spectral networks) irrespective of $N$, and zero otherwise. 

Coming to BPS states, once again one tunes the phase of the foliation to a specific phase, corresponding to a period of $\lambda = \log y \, d\log x$ around a cycle $\gamma\in H_1(\tilde \Sigma,\IZ)$.
We focus on two-way streets, and again consider all possible concatenations of open paths carried by underlying oriented trajectories
\be
	Q_N(p) = 1 + \sum_{a_N\in \Gamma_{ij}}\sum_{b_N\in \Gamma_{ji}} \mu(a_N)\mu(b_N) X_{a_N\circ b_N} 
\ee
It can be shown that the function $Q_N(p)\equiv Q(p)$ does not depend on the choice of logarithmic branch $N\in \IZ$.
Again $Q(p)$ factorizes as in (\ref{eq:Q-factorization}) and one defines closed paths $\fL_\gamma$ on $\tilde\Sigma$ by (\ref{eq:L-path}). The BPS index is computed by (\ref{eq:BPS-index}).

An interesting feature of exponential networks is that every cycle on $\Sigma$ admits an infinite sequence of lifts $\gamma_N$ to $\tilde \Sigma$. 
The cycle $\gamma\in H_1(\Sigma,\IZ)$ may be again mapped to a Lagrangian cycle in $X$ via (\ref{eq:cycle-map}).
The lift to $\gamma_N$ with $N\in \IZ$ is then mapped to the choice of a graded lift for $L$, discussed in section \ref{sec:S2-fibers}.

\subsection{Networks count fixed points}

Recall the map (\ref{eq:cycle-map}) relating Lagrangian cycles in $X$ to one-cycles on $\Sigma$.\footnote{ 
In the case of exponential networks, one may consider cycles on $\tilde \Sigma$. This corresponds to including the data of a graded lift for $L$.}
We would like to argue that the BPS index $\Omega(\gamma)$ computed by networks for a one-cycle $\gamma$ coincides with the BPS invariant defined in (\ref{eq:omega-chi}) for the associated three-cycle $L$
\be\label{eq:Omega-equality}
	\Omega(\gamma) = \Omega(L)\,.
\ee
This will be the main statement we wish to prove in this section. We will not be able to provide a complete proof covering all possible cases. For certain settings, such as spectral networks of type $A_1$, we will be able to come `close' to a complete proof. 
But more generally we will provide an argument that supports (\ref{eq:Omega-equality}) holding for \emph{primitive} cycles.

Localization relates $\Omega(L) = (-1)^{\dim \CM_L} \chi(\CM_L)$ to a simple count of fixed points (\ref{eq:Omega-localization}), up to an overall sign.
Each fixed point of the torus action on $\CM_L$ corresponds to a maximally degenerate leaf (or a system of leaves, in case there are junctions involved).
As argued in section \ref{eq:foliation-fixed-points}, all fixed points can be captured by studying the moduli space $\fM_L$ of special Lagrangian cycles underlying $A$-branes. 
The fixed point locus $\fD_L\subseteq \fM_L$ is a finite collection of maximally degenerate special Lagrangians, represented by degenerate leaves (or system of leaves).
Here we would like to argue that if $\gamma, L$ are primitive cycles in the respective homology lattices, then there is a one-to-one correspondence between the fixed point locus $\fD_L$ and a certain decomposition of the cycle $\fL_\gamma$ defined via networks in (\ref{eq:L-path}). Indeed, comparing (\ref{eq:BPS-index}) with (\ref{eq:Omega-localization}) through (\ref{eq:Omega-equality}) leads to the following claim
\be\label{eq:signed-counts-agree}
	\frac{[\fL_\gamma]}{\gamma} 
	= (-1)^{\dim \CM_L} |\fD_L|\,.
\ee

Recall that $\fL_\gamma$ is the lift of a formal sum of two-way streets $p\subset C$. 
Each $p$ is taken with multiplicity $\alpha_\gamma(p)\in \IZ$, defined by combinatorial data of networks (\ref{eq:Q-factorization}).
Thus $\pi(\fL_\gamma)$ is a system of arcs on $C$, each of which corresponds to a critical leaf of some foliation $\phi_{ij,n}$. According to (\ref{eq:BPS-index}) the overall lift of this system of arcs is in class $\Omega(\gamma)\cdot \gamma$.
In particular, this suggests that one may view $\Lambda_\gamma$ as $|\Omega(\gamma)|$ distinct cycles, each of them in class $\pm \gamma$ (the sign being determined by that of $\Omega(\gamma)$).
Let $\kappa_a$ be a degenerate leaf (or system of leaves) corresponding to a point in $a \in\fD_L$, note that $\pi^{-1}(\kappa_a) = \gamma$ by construction, via the map (\ref{eq:cycle-map}).
Our goal will be to show that 
\be\label{eq:fixed-points-L-path}
	\fL_\gamma =  (-1)^{\dim \CM_L}  \, \sum_{a \in \fD_L} \pi^{-1}(\kappa_a)\,.
\ee
This equation implies (\ref{eq:signed-counts-agree}), which is recovered by passing to homology and dividing each side by $\gamma$. By extension it also implies (\ref{eq:Omega-equality}).
On the other hand, acting on each side of (\ref{eq:fixed-points-L-path}) by the projection map $\pi:\Sigma\to C$ gives an equality between formal sums of critical leaves of foliations on the left hand side, and degenerate leaves on the right hand side.
We will proceed to show that (\ref{eq:fixed-points-L-path}) holds in examples of increasing complexity.

\subsubsection{$A_1$ spectral networks}

\begin{figure}[h!]
\begin{center}
\includegraphics[width=0.5\textwidth]{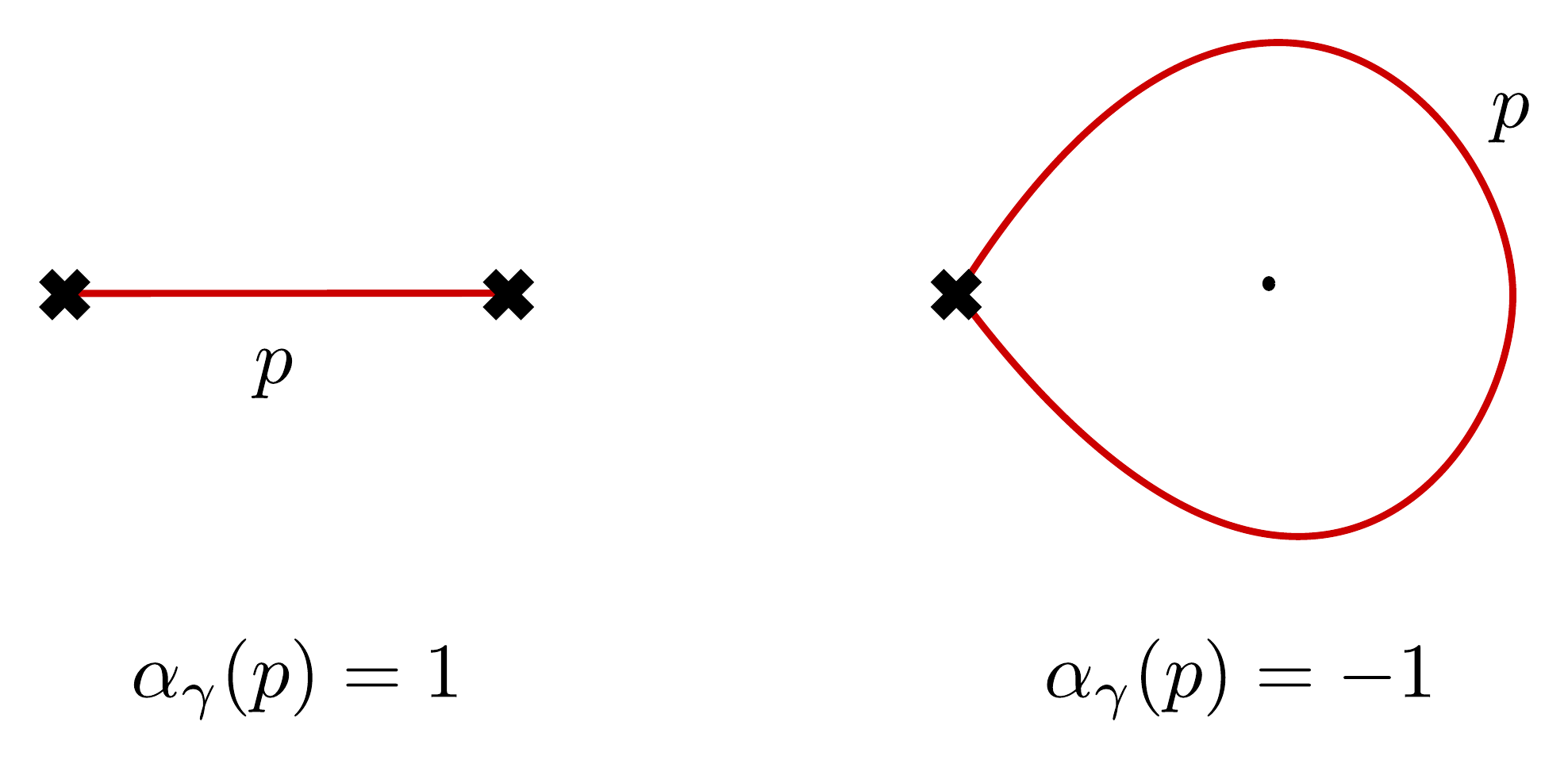}
\caption{Two-way streets of $A_1$ spectral networks, with values of $\alpha_\gamma(p)$ indicated.}
\label{fig:A1-saddles-alpha}
\end{center}
\end{figure}

In the case of $A_1$ theories of class $S$, the spectral curve $\Sigma$ is described by $\lambda^2+\phi_2(z)$ in $T^*C$ where $z\in C$ is a local coordinate on the underlying Riemann surface.
It is known that there are only two types of two-way streets that can appear, corresponding to saddles of quadratic differentials \cite{Gaiotto:2009hg, 2013arXiv1302.7030B, Strebel}.\footnote{We are assuming a generic choice of $\phi_2$. For a non-generic choice the rings may break up into multiple concatenations of saddles, see e.g. \cite{Hollands:2013qza, Longhi:2016wtv, Gabella:2017hpz, Fluder:2019dpf} for discussions of these cases.}
The first type is a saddle with endpoints on distinct branch points, as in figure \ref{fig:A1-saddles-alpha}. In this case $Q(p)=1+X_\gamma$ and therefore $\alpha_\gamma(p)=1$. 
Since $[\pi^{-1}(p)]=\gamma$, the contribution of this saddle to $\Omega(\gamma)$ is $+1$.
The second type is a saddle with both endpoints attached to the same branch point, also shown in figure.
In this case $Q(p)=(1-X_\gamma)^{-1}$, therefore $\alpha_\gamma(p)=-1$.

Let us compare this with degenerate leaves of foliations. 
If a foliation of type $\phi_{ij,0}$ has a bi-critical leaf, this has trivial moduli space $\fM_L=\{{\rm pt}\}$. Therefore its contribution to $\Omega(L) = \chi(\CM_L)=1$, matching the contribution of the first type of saddle for $A_1$ networks.
Next we consider compact circular leaves that come in families.
Recall from section \ref{eq:foliation-fixed-points} that when we consider a family of compact leaves there are two types of degenerations. In the case of $A_1$ networks there are no junctions, therefore only type $(i)$ degenerations are possible. This corresponds to the circular leaf hitting a branch point, compare Figures \ref{fig:unbounded-circles} and \ref{fig:unbounded-circles-FP}. The moduli space for these leaves is one-dimensional $\dim_\IR \fM_L = \dim_\IC\CM_L =1$, parameterizing the `height' of the generic compact leaf away from the branch ploint.
Then $\fM_L$ may be either homeomorphic to the half-line or to an interval, corresponding to having one or two fixed points.
Each of the fixed points corresponds, in fact, to a \emph{critical} leaf of the foliation -- namely to the two-way streets of a spectral network.
The contribution of each fixed point to $\Omega(L)$ is $(-1)^{\dim_\IC\CM_L} = -1$, precisely matching the contribution of the circular two-way streets to $\Omega(\gamma)$.

Thus for $A_1$ networks we are able to argue that two-way streets always correspond to the boundaries of $\fM_L$
\be\label{eq:A1-LHS}
	\pi(\fL_\gamma)
	\quad\longleftrightarrow\quad
	\partial\fM_L\,.
\ee
The moduli space of special Lagrangians in this case is either a point, the half-line, or an interval. In each of these cases we have argued that $\partial \fM_L$ corresponds to fixed points
\be\label{eq:A1-RHS}
	\partial\fM_L
	\quad\longleftrightarrow\quad
	\fD_L\,.
\ee
To establish the validity of (\ref{eq:fixed-points-L-path}) we first
invoke the correspondence between special Lagrangian fixed points in $\fD_L$ 
and degenerate leaves of foliations discussed in section \ref{eq:foliation-fixed-points},
and then act by the lift map $\pi^{-1}:C\to \Sigma$ on both the left-hand side of (\ref{eq:A1-LHS}) and the right-hand side of (\ref{eq:A1-RHS}).
To match signs we observe that $(-1)^{\dim \CM_L}$ agrees with ${\rm sgn}\,\Omega(\gamma)={\rm sgn}\,\alpha_\gamma(p)$ for each case taken individually, see Figure \ref{fig:A1-saddles-alpha}.

As explained earlier, establishing (\ref{eq:fixed-points-L-path}) implies (\ref{eq:signed-counts-agree}), which is equivalent to the desired statement (\ref{eq:Omega-equality}).

\subsubsection{Higher rank spectral networks}

Moving on to spectral networks of higher-rank class $S[A_r]$ theories, the novelty is that sytems of two-way streets $p$ making up $\pi(\fL_\gamma)$ may now include junctions.
This implies that there are infinitely many possible types of `generalized saddles', unlike in the $A_1$ case where we had only two.
Nevertheless we would still like to argue that, in general, the overall collection of generalized saddles defining $\pi(\fL_\gamma)$ captures all fixed points in $\CM_L$. In other words the system of two-way streets $\pi(\Lambda_\gamma)$ can again be viewed as a collection of $|\Omega(\gamma)|$ degenerate leaves of foliations, each of which corresponds to a maximally degenerate special Lagrangian $L\in \fD_L$.

\begin{figure}[h!]
\begin{center}
\includegraphics[width=0.35\textwidth]{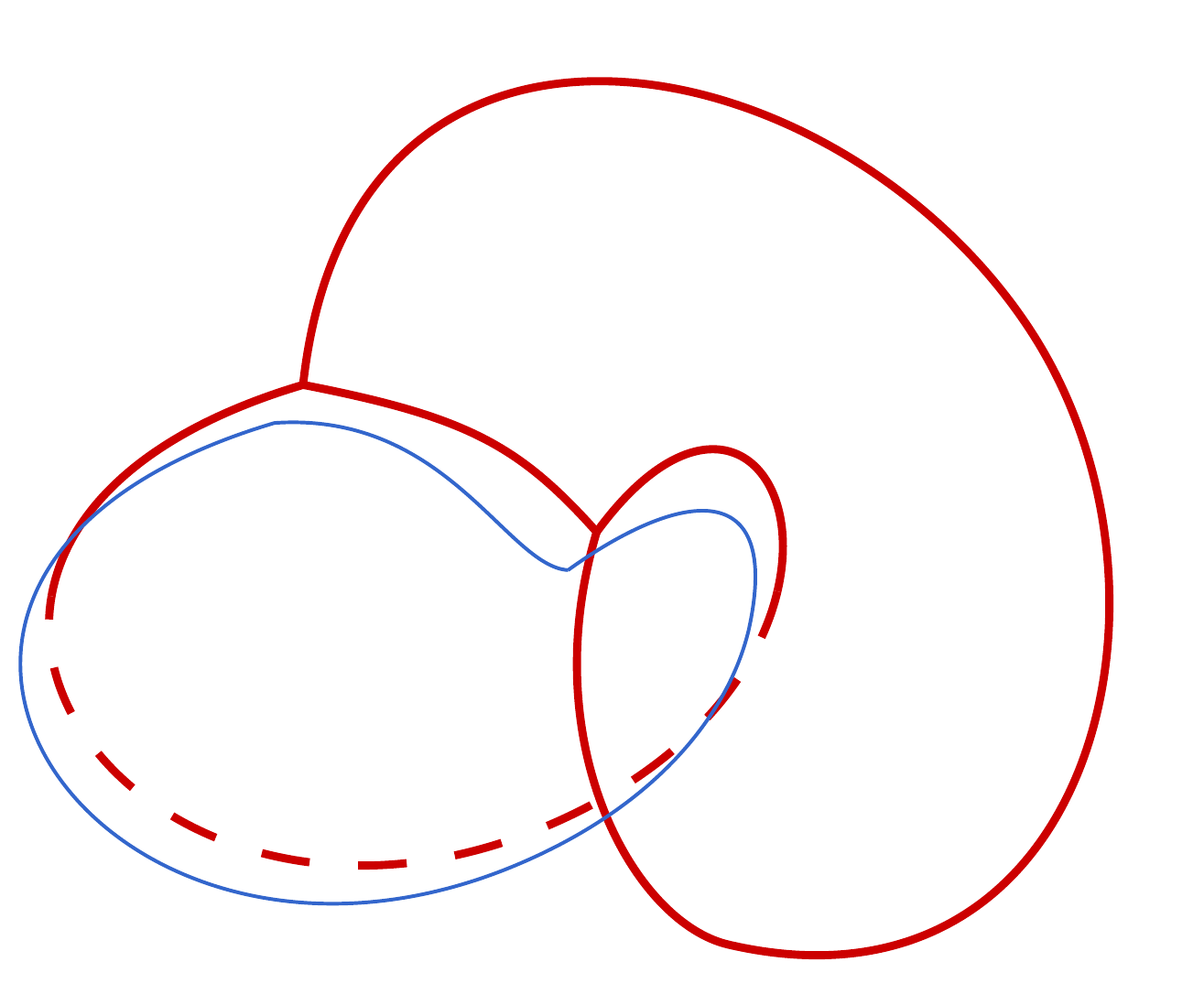}
\caption{A system of compact leaves (in red) connected only by junctions must wrap at least one nontrivial cycles on $C$. As a consequence $b_1(L)>0$ must count at least one generator (in blue). }
\label{fig:compact-leaf-generic}
\end{center}
\end{figure}

Consider a generic system of compact leaves, corresponding to a generic point in $\fM_L$. 
Suppose the generic leaf involves only junctions but no branch points, as sketched in figure \ref{fig:compact-leaf-generic}.
What we would like to check is that any fixed point of the torus action, corresponding to maximally degenerate points $\fD_L\subset \fM_L$, must attach to a branch point. 
If this were not the case, there would be contributions to $\chi(\CM_L)$ that cannot be seen by networks, because all two-way streets are generalized critical leaves, which must be anchored to branch points.
Recall from section \ref{eq:foliation-fixed-points} that there are two types of degeneration: what we would like to rule out is that $\fD_L$ cointains fixed points obtained only through degenerations of type $(ii)$. 
We argue this by \emph{reductio ad absurdum}.
Take a generic $L\in \fM_L$, and tune moduli to achieve a sequence ot type $(ii)$ reductions, eventually reaching $L'\in \fD_L$.  
Since $L$ is not anchored to a branch point, and since we only invoked moves of type $(ii)$, neither is $L'$ attached to any branch point.
In order for $L'$ to be topologically nontrivial, the underlying system of (degenerate) compact leaves must wrap around a noncontractible cycle in $C$. But then $b_1(L)>0$ and by McLean's theorem \cite{mclean1998deformations} it follows that $L'$ must still have some nontrivial modulus, therefore $L'\notin \fD_L$. Thus every degenerate Lagrangian in $\fD_L$ must correspond to degenerate leaves attached to at least one branch point.
This in turn means that the degenerate leaves correspond to two-way streets of a spectral network, and therefore should be captured by $\pi(\fL_\gamma)$. The argument we gave can be clearly sharpened further, but this would require introducing systematic definitions and considerably more work. We leave this as a sketch of proof, and proceed with the next part of the argument.

We have argued that any torus fixed point, or degenerate Lagrangian in $\fD_L$, must be captured by the collection of 2-way streets $\pi(\fL_\gamma)$.
Next one needs to ensure that $\pi(\fL_\gamma)$ gets contributions \emph{only} from torus fixed points, and not from other points in $\fM_L$. 
A generic Lagrangian in $\fM_L$, if anchored to branch points, must involve junctions that can `slide', as in the degeneration of type $(ii)$.
We say that a junction is critical if all three leaves attaching to it are critical leaves, in the generalized sense (either primary critical leaves, or descendant critical leaves).
Critical junctions cannot slide, because critical leaves are rigid: they are solutions of a first-order differential equation with fixed boundary conditions. Since $\pi(\fL_\gamma)$ is made of critical leaves, it only contains critical junctions, which therefore cannot slide. This shows that $\pi(\fL_\gamma)$ only contains contributions from special Lagrangians in $\fD_L$ and not from the complement in $\fM_L$.

Finally we argue that integer multiplicities $\alpha_\gamma(p)$ attached to two-way streets $p$ of $\pi(\fL_\gamma)$ produce the correct signed counts of fixed points. This follows by taking an expansion in $X_\gamma$ of (\ref{eq:Q-def}) and its factorization (\ref{eq:Q-factorization})
\be
\begin{split}
	Q(p) 
	& = 1 +  \sum_{a,b \,|\, a\circ b= \gamma} \mu(a)\mu(b) X_{a\circ b} + O(X_{2\gamma})\\
	& = 1 \pm \alpha_\gamma(p) \, X_\gamma + O(X_{2\gamma})
\end{split}
\ee
which shows how $\alpha_\gamma(p)$ counts all possible concatenations of open paths $a,b$ in class $\gamma=a\circ b$. Recall that $a,b$ are nothing but lifts of the underlying oriented critical leaves that make up the two-way street $p$ (up to a sign).
Also recall that $a,b$ may run only above $p$ is this attached to a branch point, or they may arise as concatenations of lifted critical leaves, if there are junctions.
This means that $\pm\alpha_\gamma(p)$ counts how many times $p$ appears in a system of compact critical leaves whose lift to $\Sigma$ is a closed cycle in class $\gamma$. As we argued above, a system of critical leaves of this type must correspond to a fixed point Lagrangian $L\in \fD_L$, therefore $\pm\alpha_\gamma(p)$ counts how many times the leaf-segment corresponding  to $p$ appears in the overall count of degenerate Lagrangians in $\fD_L$.

To summarize we have argued (admittedly, to a lesser degree of rigour compareed to the $A_1$ case) that 
all torus fixed points in $\CM_L$ correspond to (systems of) compact leaves given by two-way streets of $\pi(\fL_\gamma)$, and that $\pi(\fL_\gamma)$ only gets contributions from fixed points $\fD_L\subset\fM_L$. We illustrate these statements with the example of $k$-herds

\begin{figure}[h!]
\begin{center}
\includegraphics[width=0.85\textwidth]{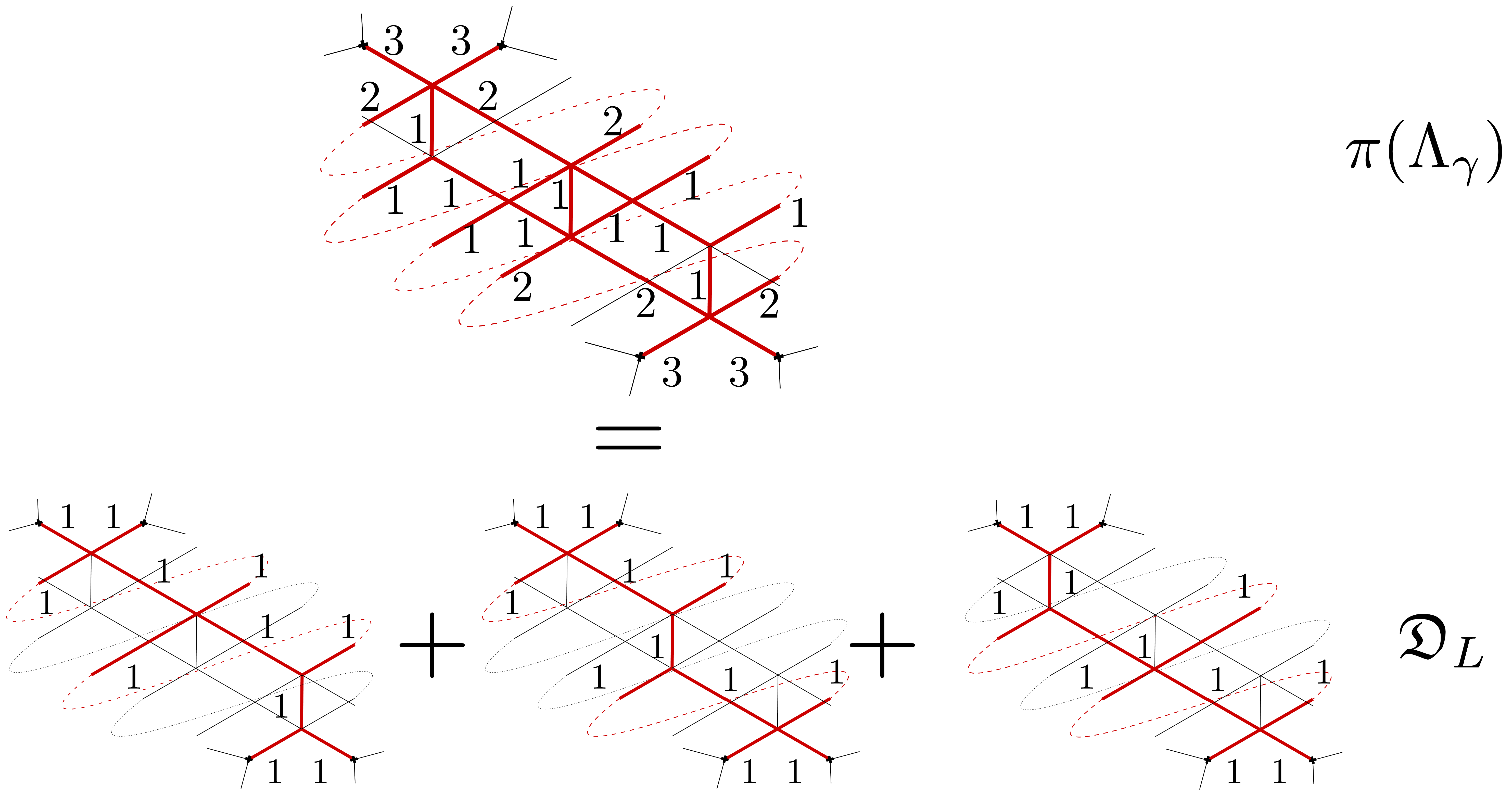}
\caption{The two-way streets in $\pi(\fL_\gamma)$ for the $3$-herd, with respective multiplicities $\alpha_\gamma(p)$ computed by spectral networks. $\pi(\fL_\gamma)$ corresponds exactly to the sum of the three fixed points in $\fD_L$ identified earlier in Figure \ref{fig:k-herd-FP}.}
\label{fig:3-herd-alpha-gamma}
\end{center}
\end{figure}

\paragraph{Example.}
To illustrate these points we consider the example of the $3$-herd shown in Figure \ref{fig:3-herd-alpha-gamma}.
A full analysis of the combinatorial data on two-way streets of $k$-herds can be found in \cite{Galakhov:2013oja}.
Here we focus on $k=3$, the generalization to generic $k$ is straightforward. The integers $\alpha_\gamma(p)$ are encoded by equations  (3.2)-(3.3) in \cite{Galakhov:2013oja}. We report them in Figure \ref{fig:3-herd-alpha-gamma}, also see \cite[Appendix B.2]{Galakhov:2014xba}.
The top part of the figure shows $\pi(\Lambda_\gamma)$, with each 2-way street $p$ of the network labeled by the corresponding integer $\alpha_\gamma(p)$. The bottom part shows the decomposition of $\pi(\Lambda_\gamma)$ into three pieces, corresponding to systems of leaves that represent degenerate Lagrangians $\kappa_a\in \fD_L$ for $a=1,2,3$. Since the dimension of the moduli spaces is even, it is straightforward to check that this data obeys (\ref{eq:signed-counts-agree}).

\begin{comment}
As we already stressed, we only argue the equality (\ref{eq:Omega-equality}) for primitive cycles (respectively primitive $\gamma\in H_1(\Sigma,\IZ)$ and primitive $L\in H_3(X,\IZ)$).
To see the kind of subtleties arising with non-primitive cycles we may again consider the example of the $3$-herd.
In this case it is known from direct computation with spectral networks that $\Omega(n\gamma)$ is nontrivial not only for $n=1$ but also for higher values of $n>1$. Indeed $\Omega(n\gamma) = 3, -6, 18, -84, 465, \dots$, see \cite[Table 1]{Galakhov:2013oja}.
One may ask whether $\Omega(n\gamma)$ for $n>1$ is again, up to a sign, the Euler characteristic of a certain moduli space of $A$-branes, namely
\be\label{eq:naive-non-primitive}
	\Omega(n\gamma)\ \mathop{=}^{?} \ (-1)^{\dim \CM_{nL}} \chi(\CM_{nL})\,.
\ee
To address this naive guess, one should first understand the moduli space $\CM_{nL}$, and this requires some care.
For instance consider the case $n=2$, and let us ask what would the generic special Lagrangian in class $[2L]$ look like.
It seems unlikely that this could be just two copies of the underlying once-around Lagrangian $L$ lying on top of each other.
In fact, recall the unbounded family of circular compact leaves of figure \ref{fig:A1-saddles-alpha}: in that case we know from spectral networks that $\Omega(n\gamma) = -\delta_{n,1}$, therefore taking a Lagrangian wrapping twice the generic compact leaf should result in an empty moduli space.
Instead it seems likely that one should take two copies of $L$, perhaps at different points in $\fM_L$, and glue them together to obtain a new special Lagrangian in class $[2L]$ with a larger number of moduli.
The expected number of moduli can be deduced by comparing with the appropriate BPS quiver, recall comment~\ref{comm:DT-quivers}.
As observed in \cite[eq. (8.8)]{Galakhov:2013oja} the dimension $\dim_\IR\fM_{2L} = \dim_\IC\CM_{2L}$ is expected to be $(k-2)n^2+1$ for a $k$-herd. This means we expect a $5$-dimensional moduli space for the generic calibrated cycle in class $[2L]$. Indeed the Poincar\'e polynomial for the state of charge $2\gamma$ corresponds to a character $\chi_{5/2}(y)$ of $SU(2)$ of spin $5/2$ \cite[Appendix A.2]{Galakhov:2013oja}. 
By McLean's theorem this suggests that we should glue two copies of $L$ in such a way that $b_1(2L) = 5$.
In fact it is not too hard to find a way to do that. Something more interesting happens for $n=3$: in this case the naive dimension of the quiver moduli space is $10$, however we know that the Poincar\'e polynomial of the true moduli space, as predicted by wall-crossing or spectral networks `with spin' \cite{Galakhov:2014xba}, should be $\Omega(3\gamma,y) = \chi_{3}(y)+\chi_{5}(y)$ the sum of a spin-$3$ and a spin-$5$ character. This raises the possibility that the naive guess (\ref{eq:naive-non-primitive}) needs to be corrected: instead of the Euler characteristic of a single moduli space, there may be weighted contributions from different types of moduli spaces when the charge vector is non-primitive. A similar phenomenon indeed plays a role in the definition of Donaldson-Thomas invariants for $B$-branes \cite{2005math......7523B}. 
Clearly the question of whether, or how, (\ref{eq:naive-non-primitive}) should be adapted to the non-primitive case is very interesting, we leave it to future work at the moment.
\end{comment}

\subsubsection{Exponential networks}

The more general case of exponential networks is covered by similar arguments as higher-rank spectral networks.
We mention two technical novelties here. The first one is the introduction of a nontrivial logarithmic index $n$ (not necessarily zero) for the types of foliations $\phi_{ij,n}$ involved.
The second novelty is the presence of new types of junctions, involving leaves of $\phi_{ij,m}$ and $\phi_{ji,n}$ with $m+n\neq 0$.\footnote{These junctions are absent for spectral networks, since for those $m=n=0$.}

Modulo these minor differences, one may repeat all steps adopted for higher-rank spectral networks to argue once again that $\fL_\gamma$ must all torus fixed points, and only those. This statement is understood in the sense that two-way streets $p$ of $\fL_\gamma$ correspond to generalized critical leaves of foliations, which correspond in turn to the degenerate compact Lagrangians in $\fD_L\subset \fM_L$.
The integer multiplicities $\alpha_\gamma(p)$ attached to two-way streets $p$ of $\pi(\fL_\gamma)$ are then expected to encode the correct signed counts of fixed points.

\bibliography{biblio}{}
\bibliographystyle{JHEP}

\end{document}